\documentclass[11pt,aps,tightenlines,nofootinbib,longbibliography,superscriptaddress, floatfix]{revtex4-1}

\usepackage{graphicx,graphics,subfigure}
\usepackage{amsfonts,amsmath,amssymb,mathrsfs,bm,amsthm,nccmath}

\usepackage{array}
\usepackage{tabu}
\usepackage{dcolumn}
\usepackage{float}
\usepackage{url}
\usepackage{ragged2e}
\usepackage{mathtools} 
\usepackage{multirow} 
\usepackage{slashed}
\usepackage[usenames,dvipsnames]{color}
\usepackage{soul}
\usepackage{url}
\usepackage[colorlinks = true,
            linkcolor = blue,
            urlcolor  = blue,
            citecolor = blue,
            anchorcolor = blue]{hyperref}
\usepackage{physics}
\usepackage{makecell}
\usepackage{units}
\usepackage{upgreek}

\usepackage[total={6.5in,9in},top=1in,headsep=0.1in,headheight=1in]{geometry}
\usepackage[utf8]{inputenc}
\usepackage{charter}
\usepackage{fullpage}

\def\eV{\,\rm eV \,}
\def\bea{\begin{eqnarray}}
\def\eea{\end{eqnarray}}

\definecolor{Zpurple}{RGB}{119, 50, 168}

\definecolor{cardinal}{RGB}{140, 21, 21}

\newcommand\snowmass{
\begin{center}
  \rule[-0.2in]{\hsize}{0.01in}\\
  \rule{\hsize}{0.01in}\\
  \vskip 0.1in
  Submitted to the Proceedings of the US Community Study\\
  on the Future of Particle Physics (Snowmass 2021)\\
  \rule{\hsize}{0.01in}\\
  \rule[+0.2in]{\hsize}{0.01in}\\[-2em]
\end{center}
}

\usepackage[firstpage=true]{background}
\backgroundsetup{contents={\parbox{6.5in}{\snowmass}}, scale=1,placement=top,opacity=1,color=black,position={3.25in,1.2in}}

\usepackage{fancyhdr}
\fancypagestyle{plain}{%
  \fancyhf{}%
  \fancyhead[C]{}
  \fancyfoot[C]{\thepage}
}

\usepackage{lipsum}

\begin{document}

\title{Snowmass 2021 White Paper\\
New Horizons: Scalar and Vector  Ultralight Dark Matter}

\begin{abstract}
\end{abstract}

\author{D.~Antypas}
\affiliation{Johannes Gutenberg University, Mainz, Mainz, Germany}
\affiliation{Helmholtz Institute (GSI) Mainz, Mainz, Germany}

\author{A.~Banerjee}
\affiliation{Department of Particle Physics and Astrophysics, Weizmann Institute of Science, Rehovot, Israel}

\author{C.~Bartram}
\affiliation{University of Washington, Seattle, WA, USA}

\author{M.~Baryakhtar}
\affiliation{University of Washington, Seattle, WA, USA}

\author{J.~Betz}
\affiliation{Department of Physics and Astronomy, University of Delaware, Newark, DE, USA}

\author{J.~J.~Bollinger}
\affiliation{National Institute of Standards and Technology, Boulder, CO, USA}

\author{C.~Boutan}
\affiliation{Pacific Northwest National Laboratory, Richland, WA, USA}

\author{D.~Bowring}
\affiliation{Fermi National Accelerator Laboratory, Batavia, IL, USA}

\author{D.~Budker}
\affiliation{Helmholtz Institute (GSI) Mainz, Mainz, Germany}
\affiliation{Johannes Gutenberg University, Mainz, Mainz, Germany}
\affiliation{University of California, Berkeley, Berkeley, CA, USA}

\author{D.~Carney}
\affiliation{Physics Division, Lawrence Berkeley National Laboratory, Berkeley, CA, USA}

% ...

\include{macros}

\maketitle


\newpage
\vspace*{7cm}
\begin{center}
{\large{\bf{Abstract}}}
\end{center}

The last decade has seen unprecedented effort in dark matter model building at all mass scales coupled with the design of numerous new detection strategies. 
Transformative advances in quantum technologies have led to a plethora of new high-precision quantum sensors and dark matter detection strategies for ultralight ($<10\,$eV) bosonic dark matter that can be described by an oscillating classical, largely coherent field.
This white paper focuses on searches for wavelike scalar and vector dark matter candidates. 

\newpage 

\tableofcontents

\newpage 

\section{Executive summary}

\textbf{Dark matter puzzle} --- A large number of astrophysical and cosmological measurements at many different scales \cite{Ber05,Fen10,Gor14} suggest that more than 80\% of all matter in the Universe is invisible, nonluminous dark matter (DM) which is not explained by the standard model (SM). 
Understanding the nature of dark matter is one of the biggest fundamental problems in modern science.  
Solving this problem will not only reveal the composition of the Universe, but can also offer insights into the cosmology of the early Universe, uncover new physical laws, and potentially lead to the discovery of other fundamental forces \cite{Safronova:2018RMP}. The vast range of possible dark matter masses and strengths of interactions have  prompted searches for 
scenarios where potential DM particles naturally arise in theories aimed at solving other problems of fundamental physics, such as the hierarchy problem or the strong-CP (C-charge, P-parity) problem. 
Such models allow one to restrict the range of dark matter masses and provide a clear blueprint and parameter space target goal for DM detection.

The past decade has seen unprecedented effort in dark matter model building at all mass scales coupled with the design of numerous new detector types. It is the goal of this (as well as other Snowmass white papers) to present and highlight  recent advances as well as show the promising prospects for the future.
In particular, transformative advances in quantum technologies have led to a plethora of new high-precision quantum devices joining the search for light and ultralight  dark matter.

\textbf{Ultralight bosonic dark matter} --- Within a broad class of models, dark matter can be composed of bosonic fields associated with ultralight ($m_{\phi} \lesssim 10~\eV$) particles and generally classified  by their spin and intrinsic parity (scalar, pseudoscalar, vector)~\cite{Safronova:2018RMP,Preskill:1982cy, Abbott:1982af,Dine:1982ah,Piazza:2010ye,Goodsell:2009xc,Arias:2012az,Arvanitaki:2014faa,Stadnik:2015DM-laser,graham2016dark}. 
We note that, for this mass range and based on measurements of the galactic halo 
DM density, these candidates are necessarily bosonic and feature typical occupation numbers larger than 1 \cite{bookDM}. The key idea is then that such ultralight dark matter (UDM) particles behave in a ``wave-like’’ manner and their phenomenology is described by an oscillating classical field:
$\phi(t) \approx \phi_{0} \cos(m_{\phi}t)$, where $\phi_0\sim\sqrt{2\rho_{\rm DM}}/m_{\phi}$ is the field oscillation amplitude and $\rho_{\rm DM}$ is the local DM density. ``Fuzzy'' dark matter, in the lowest mass range $m_\phi \lesssim 10^{-18}~\eV$, has been a subject of intense astrophysical studies due to its effect on large-scale structures and other astrophysical signatures; see Section~\ref{astro}. 
\begin{figure}
    \centering
    \includegraphics[width = \textwidth]{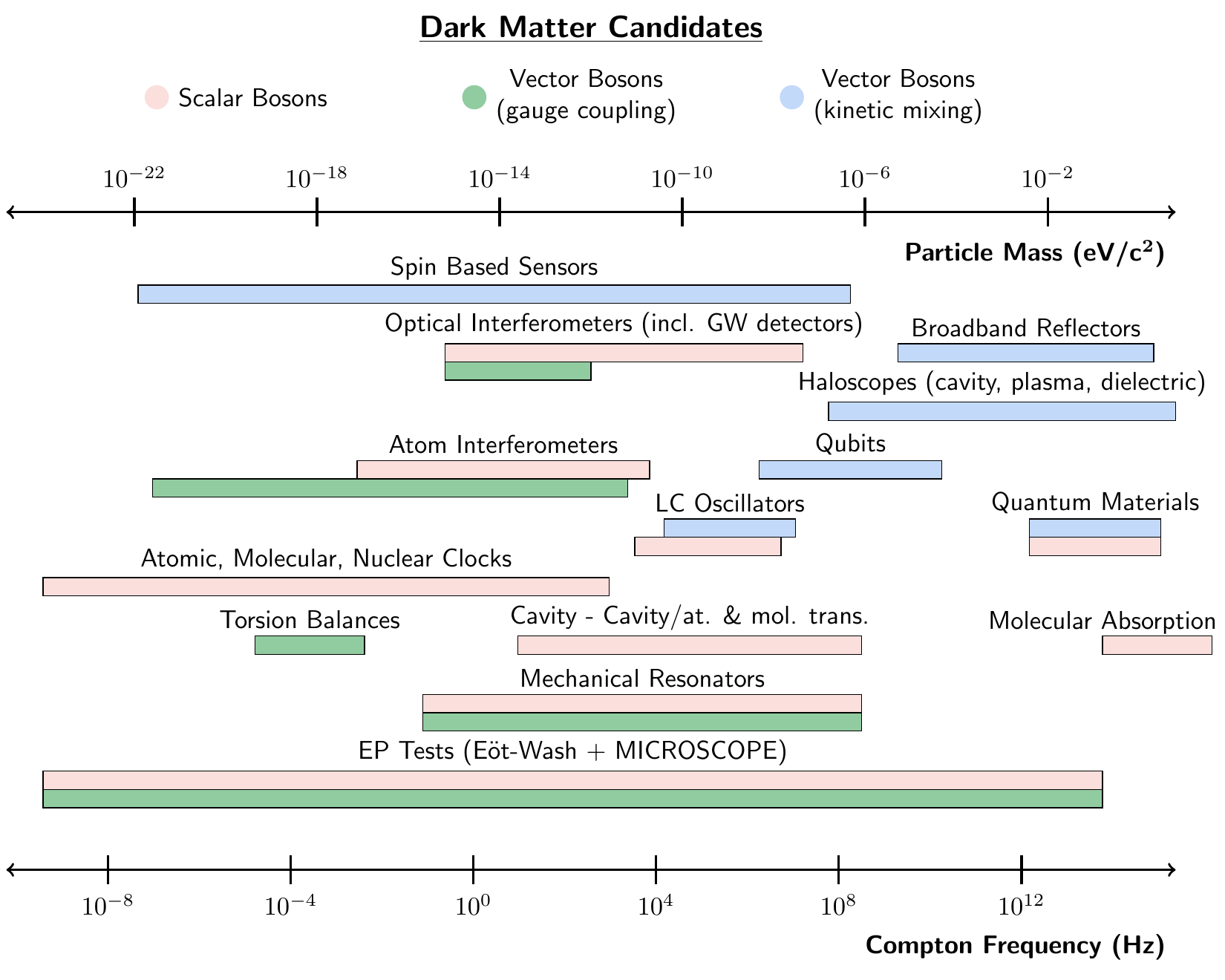}
    \caption{
    Summary of current and future laboratory direct-detection experiments to set constraints on scalar and vector dark matter. These are complementary to cosmological and astrophysical probes, both in mass and the coupling strength to the SM, see Fig.~\ref{det1}. 
     ``Qubits'' includes Rydberg atoms, trapped ions, and superconducting transmon qubits. ``Torsion Balances'' refers specifically to direct DM searches, as opposed to more traditional equivalence principle (EP) violation tests. ``GW'' refers to gravitational wave detectors. 
     }
         \label{det}
\end{figure}
Such DM candidates act as coherent entities on the scale of individual detectors or networks of detectors, leading to a new detection paradigm. UDM fields may cause precession of nuclear or electron spins, drive currents in electromagnetic systems, produce photons, or induce equivalence-principle-violating accelerations of matter. They may also modulate the values of the fundamental ``constants'' of nature, which would in turn induce changes in atomic transition frequencies  and local gravitational field and affect the length of macroscopic bodies.

\textbf{New paradigm:~Quantum technologies for dark matter detection} --- The unprecedented progress in controllable quantum systems  
and other precision measurement technologies  has profound implications on our ability to detect such ultralight (wavelike) dark matter.  
In the past ten years, precision searches for UDM with quantum technologies have emerged as a vibrant research area, with many promising new proposals joining several ongoing experiments. In fact, a key impact of the emergent second quantum revolution should be on fundamental physics, i.e., using quantum entanglement to discover new phenomena. 

The CF2 Snowmass Wavelike Dark Matter group, which did not even exist in the 2013 Snowmass process, has received 85 Letters of Intent.  These LOIs naturally divide into the axion and axion-like particle (ALP) searches and searches for other wavelike dark matter candidates. Hence, one Cosmic Frontier (CF2) white paper \cite{CF21} discusses searches for axions and ALPs (i.e., pseudo-scalars), while this white paper focuses on searches for scalar and vector dark matter candidates.
Two other Snowmass white papers in the Instrumentation Frontier (IF1) focus on quantum sensors \cite{IF11,IF12} that are used for the purpose of UDM detection discussed in this work.

\begin{figure}
    \centering
    \includegraphics[width = \textwidth]{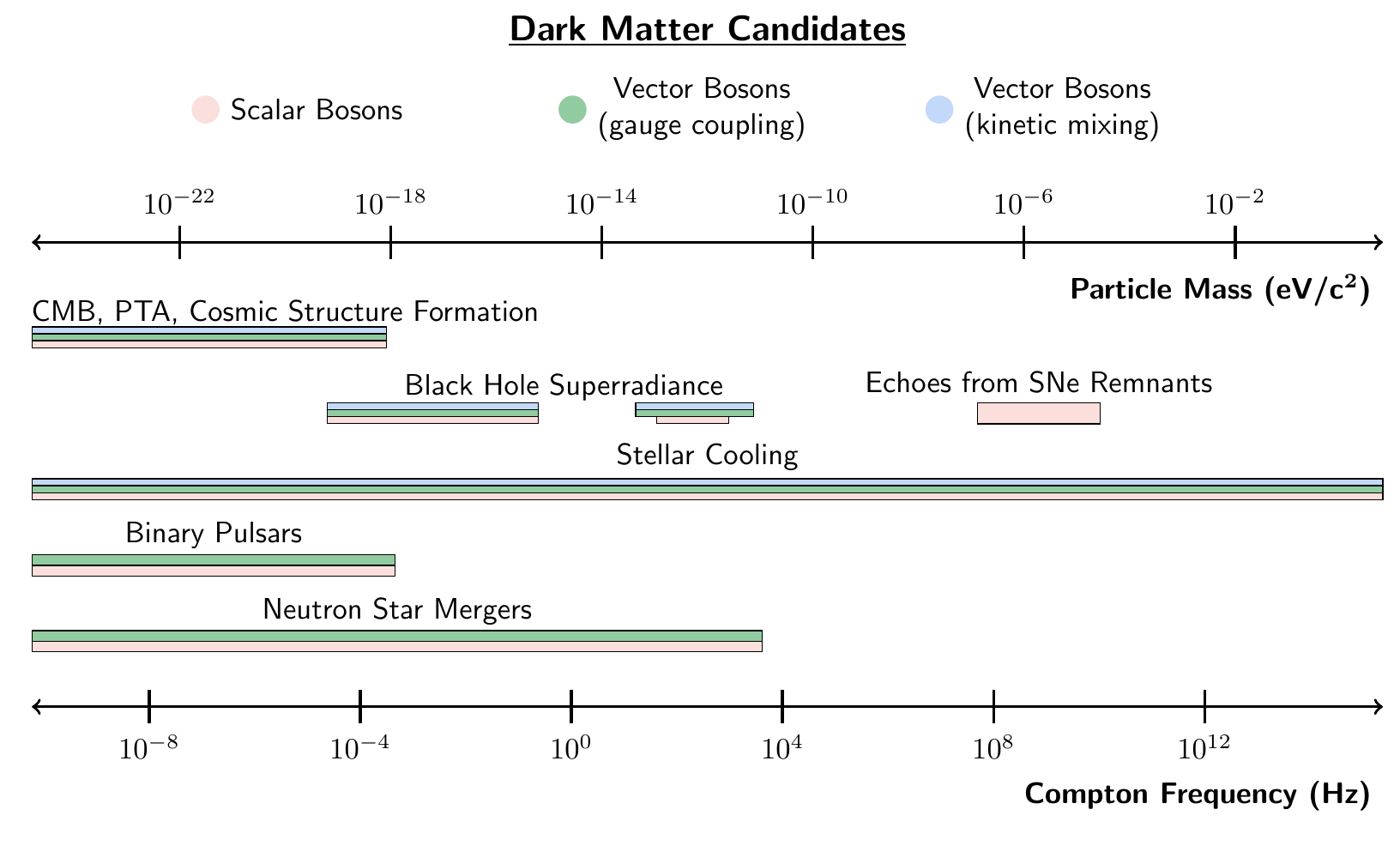}
    \caption{Summary of current and future constraints on scalar and vector dark matter using cosmological and astrophysical measurements, see Section~\ref{astro} for details. CMB: cosmic microwave background; PTA: pulsar timing arrays; SNe: supernovae.
    }
    \label{det1}
\end{figure}

We have only begun the exploration of quantum technologies in this field, and improvements of many orders of magnitude in 
the sensitivity to UDM are expected  
over the next decade. Just in the few years since the previous Snowmass process, a 
wide range of new experiments have emerged. Nearly the entire effort described in this paper is less than 10 years old and still in its infancy, with rapid improvements of many orders of magnitude expected over the next few years. 
We show the diversity of new experiments and their coverage of UDM mass ranges in Fig.~\ref{det}. Cosmological and astrophysical probes illustrated by Fig.~\ref{det1} can be complementary to laboratory searches, both in mass and coupling strength to the SM.

\begin{center}
{\textbf{\large{Roadmap to discovery}}}
\end{center}
Several interconnected directions have to be pursued to maximize scalar and vector UDM detection capabilities:
\begin{enumerate}
    \item
    \textbf{Initiate/maintain extensive theory program} in scalar and vector UDM  that will include model building, study of  production mechanisms and distinctive astrophysical signatures, identification and study of specific UDM candidates such as the relaxion and others, develop models of dark matter distributions of UDM candidates that can affect direct detection (such as compact objects and objects with non-trivial topology), study potential UDM production in extreme astrophysical events, and other relevant topics.
    
    \item
    \textbf{Enable development of strong collaborations between particle physics and quantum science fields} including atomic, molecular, optical, nuclear, and solid state physics, metrology, and quantum information science. Such collaborations will identify new detection signals and strategies, lead to the development of new quantum technologies dedicated to dark matter detection, design specific experiments including those enabled by networks of quantum detectors, analyse the experiments, create procedures for signal verification by multiple types of detectors, apply quantum metrology strategies for the extraction of weak signals to dark matter detection and measurement beyond the standard quantum limit, and explore extended quantum coherence as a new resource for the detection of new physics.
    
    \item
    \textbf{Initiate robust detector R\&D program} on the detection of scalar and vector UDM with a broad range of technologies described in this review and novel detection strategies and ensure construction of detectors that are projected to improve sensitivities to UDM by many orders of magnitude. Example projects could include the development and continued support for the MAGIS program (both MAGIS-100 and its future upgrades) and nuclear clocks.
    \end{enumerate} 
 A critically important topic to discuss is the technology development that accompanies the best, state-of-the-art laboratory science breakthroughs. To make a real breakthrough, we need to employ the best laboratory quantum sensors, for example, by establishing a network of (different types of) optical clocks (as proposed in 
 \cite{2014QuantClock}). This effort will necessitate a large-scale R\&D program, but it can be coordinated with other programs, which can leverage each other's resources. The technological maturation of this effort is going to be important for many other emerging scientific experiments and commercial-use potentials. We should also emphasize the importance of international collaborations in this field. No single player can truly make this profound discovery (one can take LIGO-Virgo-KAGRA as a good example). Verification of signals across experiments will be key.

\newpage 

\section{Scalar and vector ultralight dark matter models and interactions}
\label{theory}

\subsection{Scalar ultralight dark matter}

Ultralight scalar bosons are an attractive candidate to explain the observed DM. 
The standard picture of ultralight scalar DM is that of a coherently oscillating field $\phi$ with the mass of the underlying particle $m_{\phi}$ and present-day oscillation amplitude $\phi_{0}$. Then the contribution of the coherently-oscillating scalar to the cosmological energy density can be written as\footnote{We use natural units, where $\hbar=c=1$, unless explicitly stated otherwise.} 
\begin{align}
    \label{eq:Omega}
    \Omega_{\phi} = \frac{\rho_{\phi}}{\rho_{c}} = \frac{\frac{1}{2} m^{2}_{\phi}\bar{\phi}_{0}^{2}}{3 H^{2}_{0} M^{2}_{\rm Pl}} = 0.28 \left(\frac{0.67}{h}\right)^{2} \left(\frac{m_{\phi}}{10^{-11}\eV}\right)^{2} \left(\frac{\bar{\phi}_{0}}{4.5 \times 10^{5} \eV}\right)^{2} \, , 
\end{align}
where $\rho_{c}$ is the critical (average total) energy density today, $H_{0} = 100~h~{\rm km/s~Mpc}^{-1} = 2.1~h~\times10^{-33}\,{\rm eV}$ 
is the present-day Hubble parameter, with $h \approx 0.7$ being the dimensionless Hubble constant, and $M_{\rm Pl} = (8 \pi G)^{-1/2} = 2.4\times 10^{27} \eV$ is the reduced Planck mass, with $G$ being the universal gravitational constant. 
Note that in virialized DM halos (including the one we reside in), the local density is $\mathcal{O}(10^{5}) - \mathcal{O}(10^{6})$ times higher than the mean cosmological density, and hence locally $\phi_{0}$ will be a factor of $\mathcal{O}(10^{3})$ times larger than the mean cosmological amplitude $\bar{\phi}_{0}$. 

Searches over many decades of particle masses are well motivated. 
Assuming all of the dark matter comes from a single light scalar, experimental searches are motivated between $10^{-21}\,{\rm eV} \lesssim m_{\phi} \lesssim 10 \eV$, where the lower bound comes from cosmological and astrophysical constraints \cite{2021PhRvL.126g1302R,Bozek:2014uqa,Armengaud:2017nkf,2017PhRvL.119c1302I,Zhang:2017chj,Kobayashi:2017jcf,Bar:2018acw} (although ultralight scalars not being all the DM is also well-motivated and allowed at many masses) discussed in Section~\ref{astro} and the upper bound comes from the assumption that the light scalar is a classical field. We note that, for this mass range and based on measurements of the galactic halo velocity distribution, these DM candidates are necessarily bosonic and feature typical occupation numbers larger than 1 \cite{bookDM}.
The latter condition requires a large number of scalar particles within a single coherence volume, i.e., $n_\phi \lambda_\textrm{coh}^{3} \gg 1$, where $n_\phi$ is the number density of scalar particles and $\lambda_\textrm{coh}$ is the coherence length of the classical field (see Sec.~\ref{sec2} for discussion of coherence).

The present-day amplitude of the oscillation can be sourced by the misalignment of the scalar field in the early universe, through either of the two mechanisms below.

\textit{Standard Misalignment Mechanism} -- The standard scenario of generating a nonzero energy density for light scalar fields is through the misalignment mechanism originally proposed for the production of axions~\cite{Preskill:1982cy,Abbott:1982af,Dine:1982ah} (see also~\cite{Nelson:2011sf,Arias:2012az} for more general scalars and vectors).  The evolution starts from a (random) initial field value of $\phi$ (often homogenized over the entire observable Universe during inflation) displaced from the minimum of the potential.
Once the Hubble rate becomes comparable to the mass of the scalar field $(3H = m_{\phi})$, the scalar field starts to oscillate, and the energy density redshifts like non-relativistic matter assuming a quadratic potential for the scalar field.

\textit{Thermal Misalignment Mechanism} -- If the scalar field couples to fermions (or bosons) in the thermal plasma, then the one-loop thermal potential of the fermions (or bosons) gives a temperature-dependent thermal potential for the scalar field. This thermal potential can then misalign the scalar field away from its zero-temperature minimum, and hence source the amplitude of the oscillations. Since the thermal potential for the light scalar field depends on the mass of the fermion or boson the light scalar couples to, one can predict the dark matter density as a function of the light scalar coupling and the fermion or boson mass. As examples of thermal misalignment, a light scalar dark matter coupling to muons is given in \cite{Batell:2021ofv} and coupling to Higgs is given in \cite{Piazza:2010ye}.  Another variation of scalar dark matter, discussed below in the context of relaxion, is when the displacement of the dark matter from its minimum is associated with reheating the theory above the electroweak scale, resulting with effective "disappearance" of the, Higgs-dependent, backraction potential~\cite{Banerjee:2018xmn}.

A scalar field $\phi$ can couple to the SM fields in a variety of ways. 
Generally, the simplest possibility involves linear-in-$\phi$ interactions (in the notation of, e.g., Refs.~\cite{Damour:2010EP-A,Arvanitaki:2014faa,Hees:2018DM_EP}): 
\begin{align}
\label{scalar-field_Lagrangian_alternative}
 &\mathcal{L}^{\textrm{lin}}_\textrm{int} = \kappa \phi \left\{ \left[ \frac{d_e F_{\mu\nu}F^{\mu\nu}}{4} - d_{m_e} m_e \bar{\psi}_e \psi_e \right]
 - \left[ \frac{d_g \beta_3 G^a_{\mu\nu}G^{a\mu\nu}}{2 g_3} + \sum_{q=u,d,s} \left(  d_{m_q}  +  \gamma_m d_g \right)  m_q \bar{\psi}_q \psi_q \right]
 \right\}  \, , 
\end{align}
where $F$ is the electromagnetic field tensor, $\psi_e$ denotes the electron field, $G$ is the gluonic field tensor, $\psi_q$ denote the quark fields (at low energies, it suffices to consider only the up, down and strange quarks), $\beta_3 (g_3)$ is the beta function for the running of the strong coupling constant $g_3$, while $\gamma_m$ is the anomalous dimension giving the energy-running of the masses of the strongly-coupled fermions. 
The factor $\kappa = (\sqrt{2}M_\textrm{Pl})^{-1}$ is essentially the inverse of the reduced Planck mass $M_\textrm{Pl}$. 
The dimensionless parameters $d_i$ encode the strengths of the interactions between the scalar field and SM fields relative to the strength of gravity.

When experiments are conducted at energies well below the QCD scale of $\Lambda_\textrm{QCD} \approx 250~\textrm{MeV}$, 
it is possible to treat nucleons as the degrees of freedom instead of quarks and gluons. 
In this case, Eq.\,(\ref{scalar-field_Lagrangian_alternative}) can be rewritten in the following form: 
\begin{align}
    \label{linear_scalar_interactions}
    \mathcal{L}_\textrm{int}^\textrm{lin} &= \frac{g_\gamma \phi F_{\mu\nu} F^{\mu\nu}}{4} - \sum_{\psi=e,p,n} g_\psi \phi \bar{\psi} \psi
    \, ,
\end{align}
where we have used a common alternative notation, which is related to the notation in Eq.\,(\ref{scalar-field_Lagrangian_alternative}) via the identifications $g_\gamma \leftrightarrow \kappa d_e, g_e \leftrightarrow \kappa m_e d_{m_e}, \textrm{etc.}$ 
Besides the notations for the interaction parameters appearing in Eqs.\,(\ref{scalar-field_Lagrangian_alternative}) and (\ref{linear_scalar_interactions}), 
another often used convention in the literature involves the new-physics energy scales explicitly via the identifications $g_\gamma \leftrightarrow 1/\Lambda_\gamma, g_e \leftrightarrow m_e/\Lambda_e, \textrm{etc.}$ (see, for example, Refs.\,\cite{Stadnik:2015DM-laser,Stadnik:2015DM_VFCs,Derevianko:2014TDM}). 

The linear couplings in Eqs.\,(\ref{scalar-field_Lagrangian_alternative}) and (\ref{linear_scalar_interactions}) can be generated, for example, via the super-renormalizable interaction $\phi H^\dagger H$ between the light scalar field $\phi$ and the Higgs field $H$ \cite{Piazza:2010ye}. 
Linear couplings, however, may be absent, for example, as a result of an underlying $Z_2$ symmetry (invariance under the $\phi \to -\phi$ operation). 
In this case, the simplest possibility would instead involve quadratic-in-$\phi$ interactions: 
\begin{equation}
    \label{quadratic_scalar_interactions}
    \mathcal{L}_\textrm{int}^\textrm{quad} = \frac{g'_\gamma \phi^2 F_{\mu\nu} F^{\mu\nu}}{4} - \sum_{\psi=e,p,n} g'_\psi \phi^2 \bar{\psi} \psi
    \, . 
\end{equation}

The linear interactions in Eq.\,(\ref{linear_scalar_interactions}) effectively alter the electromagnetic fine-structure constant $\alpha$ and fermion masses according to
\begin{equation}
    \label{variations_FCs_linear}
    \alpha \to \frac{\alpha}{1 - g_\gamma \phi} \approx \alpha (1 + g_\gamma \phi) \, , ~~~ m_\psi \to m_\psi + g_\psi \phi \, , 
\end{equation}
while the quadratic interactions in Eq.\,(\ref{quadratic_scalar_interactions}) effectively alter the constants according to
\begin{equation}
    \label{variations_FCs_quadratic}
    \alpha \to \frac{\alpha}{1 - g'_\gamma \phi^2} \approx \alpha (1 + g'_\gamma \phi^2) \, , ~~~ m_\psi \to m_\psi + g'_\psi \phi^2 \, . 
\end{equation}
As a result, the coupling to the SM fields of the oscillating scalar UDM field $\phi \approx \phi_0 \cos(m_\phi t)$ with the linear interactions in Eq.\,(\ref{linear_scalar_interactions}) or quadratic interactions in Eq.\,(\ref{quadratic_scalar_interactions}) results in apparent oscillation of fundamental constants. 
These oscillations occur at about the Compton DM frequency $m_\phi$ for a theory with linear coupling or at twice that frequency for a theory with quadratic coupling \cite{Arvanitaki:2014faa,Stadnik:2015DM-laser,Stadnik:2015DM_VFCs}. 
Such oscillations can affect a variety of physical quantities, such as energy levels in atoms \cite{Arvanitaki:2014faa,Stadnik:2015DM-laser} and lengths of solids \cite{Stadnik:2015DM-laser,Stadnik:2016DM-cavity}. 
Gradients in an oscillating scalar DM field (including motional gradients) induce time-varying equivalence-principle-violating forces on bodies \cite{graham2016dark,Hees:2018DM_EP}. 
In models of scalar-field DM with $\phi^2$ interactions, a number of additional effects are possible. 
Scalar-field DM with $\phi^2$ interactions induces variations of the fundamental constants that are correlated with differences in the local DM density, potentially affecting astrophysical and cosmological phenomena via temporal drifts of the fundamental constants \cite{Stadnik:2015DM_VFCs}.  
Additionally, scalar-field DM with the quadratic interactions in Eq.\,(\ref{quadratic_scalar_interactions}) experiences screening in the vicinity of dense bodies such as Earth \cite{Hees:2018DM_EP} (for a discussion of screening in other scalar-field models with $\phi^2$ interactions, see Refs.~\cite{Damour-Esposito:1993Nonperturb,Olive:2008Enviro,Hinterbichler:2010Screening,Stadnik:2020TDM_enviro,Stadnik:2021comment}). 

Furthermore, scalar coupling to the SM fields leads to a Yukawa-type interaction between matter, in addition to the gravitational interaction (sometimes referred to as a ``fifth force"). 
A non-universal coupling of a scalar to SM fields leads to an equivalence-principle (EP) violating acceleration between two bodies~\cite{Kaplan:2000hh,Damour:2010EP-B,Damour:2010EP-A}, whereas a universal coupling (when the scalar field couples to the total mass-energy of a body) leads to deviations from the inverse square law~\cite{Fischbach:1996eq}. Thus, experiments searching for violation of the EP and/or the inverse square law can constrain the existence of a light scalar~\cite{Fischbach:1996eq,Berge:2017ovy,Hees:2018DM_EP,Tino:2020nla,Oswald:2021vtc}. 

Table~\ref{tab1-1} gives a summary of possible tree-level couplings of ultralight scalars, as well as corresponding experimental and astrophysical probes that are discussed in Sections~\ref{astro} and~\ref{searches}.

A scalar field that interacts non-gravitationally with SM fields is generically expected to receive positive quantum corrections to its mass. 
This poses challenges in the construction of models with naturally ultralight scalar fields. 

In the past few years, a number of concrete proposals have been put forward for constructing a natural model of light DM which has scalar interactions with the SM. 
These proposals can be divided into two classes: i) where the DM mass is protected by an approximate scale-invariance symmetry~\cite{Arvanitaki:2014faa}, and ii) where it is protected by an approximate shift symmetry that is broken, together with CP~\cite{Flacke:2016szy}, by two sequestered sectors~\cite{Banerjee:2018xmn} (inspired by the relaxion paradigm~\cite{Graham:2015cka}), or protected by a $Z_N$ symmetry~\cite{Brzeminski:2020uhm}.  
These two classes of models are qualitatively different; yet, in both frameworks, the DM couples to the SM fields either due to the fact that their coupling breaks scale invariance~(see for instance~\cite{Goldberger:2007zk}) or via mixing with the Higgs~\cite{Flacke:2016szy,Banerjee:2020kww}.
A light scalar field can also arise in theories with more than four dimensions, and in particular in string theory with the dilaton and the moduli fields (see, for example, \cite{Damour:1994zq, Gasperini:2001pc} and references therein).

\textbf{Relaxion DM} -- One example of a light scalar DM candidate is the coherently oscillating relaxion DM discussed in~\cite{Banerjee:2018xmn}.  
This is motivated by the relaxion mechanism, which ameliorates the Higgs naturalness problem by dynamically selecting the electroweak (EW) scale~\cite{Graham:2015cka}. In this framework, during the inflationary phase of the universe, the EW scale is selected by an evolution of an axion-like field, referred to as the relaxion. 
Compared to conventional models for the EW hierarchy problem, the relaxation mechanism includes one infrared degree of freedom, the relaxion, 
which couples feebly to SM particles via its mixing with the Higgs field due to the presence of CP violation~\cite{Gupta:2015uea,Choi:2016luu,Flacke:2016szy,Banerjee:2020kww}. 

\begin{table}
\begin{tabular}{ |c|c|c|c| } 
\hline
\textbf{DM} & \textbf{Couplings} & \textbf{Experimental and astrophysical probes} \\
\hline
\multirow{17}{1em}{$\phi$} & Gravity & CMB, Matter power spectrum (Ly$\alpha$f, Halo Mass Function)\\&& Galactic rotation curves, Black hole superradiance \\ \cline{2-3}
&  & EP violation and fifth-force searches \\
& Electromagnetism $\left(\phi F_{\mu \nu} F^{\mu \nu}\right)$ &All optical and microwave clocks, Optical cavities\\
&&Optical and atom interferometers (including GW detectors)   \\
&&LC oscillators, Cosmic distance measurement \\ && Stellar observations, DM stimulated emission \\\cline{2-3} &  & EP violation and fifth-force searches\\ 
 & & Stellar observations \\
& Electrons  $\left(\phi \bar{e} e\right)$ & Microwave and molecular clocks, Optical cavities\\ && Optical and atom interferometers (including GW detectors) \\&&Mechanical resonators, Molecular absorption\\ \cline{2-3}
& Muons  $\left(\phi \bar{\mu} \mu\right)$ & $g_{\mu} -2 $, Stellar observations, Neutron star mergers \\ \cline{2-3}
& Gluons $\left(\phi G_{\mu \nu}G^{\mu \nu}\right)$ & Microwave, molecular, and nuclear clocks\\ && EP violation and fifth-force searches \\ \cline{2-3}
& Quarks $\left(\phi \bar{q}q\right)$ / Nucleons $\left(\phi \bar{N}N\right)$ & Microwave, molecular, and nuclear clocks\\
 &  & EP violation and fifth-force searches\\ 
\hline
\multirow{12}{1em}{$A'_\mu$}&Gravity & CMB, Matter power spectrum (Ly$\alpha$f, Halo Mass Function) \\
&&Black hole superradiance  \\
\cline{2-3}
& Kinetic Mixing & Coulomb's law, Light-shining-through-a-wall, CMB\\
&  & Stellar observations, Resonant cavities, LC circuits  \\ & & Quantum materials, Molecular absorption, Magnetometers \\ & & Broadband reflectors, Plasma haloscopes, Dielectric haloscopes\\
\cline{2-3} &  & EP violation and fifth-force searches \\ 
 & Minimal Gauge Coupling $(B, L , B-L)$ & Stellar observations \\
&  &  Optical and atom interferometers (including GW detectors) \\ & & Molecular absorption\\ & & Mechanical resonators\\
\cline{2-3}
&  Minimal Gauge Coupling $(L_{\mu} - L_{\tau})$ &  Neutron star mergers \\
\hline
\end{tabular}
\caption{Experimental and astrophysical probes for scalar and vector fields coupling to the SM particles through a particular portal. For the probes, we only consider the corresponding tree-level coupling and not any radiatively induced couplings. Ly$\alpha$f stands for Lyman-alpha forest.} \label{tab1-1}
\end{table}

The relaxion potential consists of two parts: one for scanning the Higgs mass, and the other for providing feedback to the relaxion evolution as a function of the Higgs vacuum expectation value (VEV), referred to as the ``backreaction'' potential.  During inflation, the relaxion scans the EW Higgs mass, and eventually settles to $\dot{\phi}\simeq 0$ at one of the local minima of its potential~\cite{Banerjee:2020kww}. 
If the universe is reheated with temperature above the EW scale\footnote{The requirement of a high reheating temperature, motivated by a large class of inflation models (see, e.g., Ref.~\cite{PhysRevD.40.1753} and references
therein), is generic and also needed in most models that explain the observed baryon abundance (see, e.g., Ref~\cite{Bodeker:2020ghk} and references therein).}, the EW symmetry is restored, and the backreaction potential disappears. As a result, the relaxion field begins to evolve again, until the backreaction potential reappears at some temperature. If the relaxion is trapped in a nearby minimum, the relaxion field is displaced from its local minimum with a certain misalignment angle, and consequently it starts to oscillate when the Hubble scale drops below its mass. 

This coherently-oscillating relaxion field produced via dynamical misalignment contributes to the DM in the present universe~\cite{Banerjee:2018xmn}. 
Depending on the trapping mechanism, the relaxion can be DM in the mass range $10^{-11}\,{\rm eV} \lesssim m_{\phi}\lesssim 1\,{\rm eV}$~\cite{Banerjee:2021oeu}. 
Due to the presence of relaxion-Higgs couplings, coherently-oscillating relaxion DM effectively leads to temporal oscillation of the Higgs VEV, and thus in turn causes apparent temporal oscillation of various fundamental constants, which can be probed by a variety of precision experiments (see, e.g., \cite{Arvanitaki:2014faa,Stadnik:2016DM-cavity,Stadnik:2016DM-clock,Safronova:2018RMP,Antypas:2019DM_atom-cavity,Grote:2019DM-LIFO}). 
However, due to the rather high mass range noted above, the relaxion DM model seems to prefer high oscillation frequencies which pose both challenging and
exciting targets for a variety of precision experiments \cite{Safronova:2018RMP,Antypas:2019DM_atom-cavity,Grote:2019DM-LIFO,Aharony:2021DM_atom-cavity}. 

Note that the physical relaxion is not a CP eigenstate, and this allows the relaxion to have both scalar- and pseudoscalar-type couplings to the SM particles (see e.g. \cite{Flacke:2016szy}). 
Due to the axion-like (pseudoscalar) coupling to the SM particles, axion DM searches such as ADMX, CASPEr, GNOME, etc. can also be applied to the relaxion DM scenario. Some such experiments can probe various spin-dependent effects induced by axions and/or ALPs (see for instance~\cite{graham2016dark} and references therein). 
Additionally, the exchange of a non-CP-eigenstate particle such as the relaxion leads to CP-violating forces between fermions or macroscopic bodies that can be probed in various ways (see, e.g., Ref.~\cite{Stadnik:2018axion-EDM} and references therein). 

We finish this part by mentioning that the scalar interaction with the standard model fields is leading to naturalness issues.
Assuming that the underlying theory is natural we can relate the scalar mass to its coupling to the matter fields and the cutoff of the theory, $\Lambda_{\rm Nat}$, where new degrees of freedom are required to be added, for instance for the electron coupling, $d_{m_e}$ defined in Eq.~\eqref{scalar-field_Lagrangian_alternative},
\begin{equation}
d_{m_e}\lesssim {4\pi m_\phi\over \kappa m_e \Lambda_{\rm Nat}} \,.
\end{equation}
Clearly the natural boundary for $d_{m_e}$ inversely proportional to $\Lambda_{\rm Nat}$, and thus has considerable theoretical, model dependence.
To demonstrate this we can consider two scenarios, one is conservative assuming that the  new degrees of freedom directly interact with the standard model fields for which we require
$\Lambda_{\rm Nat}=10\,$TeV, while in the second, which is more aggressive, we can assume mirror models~\cite{Chacko:2005pe,Burdman:2006tz,Craig:2014aea,Kats:2017ojr}, where the new degrees of freedom, which can be thought as dark electrons, only couple to $\phi$ with coupling of the order of $d_{m_e}$, and are therefore hidden from us. In this case we can assume that $\Lambda_{\rm Nat}=m_e$, allowing a much bigger coupling. 
We end this discussion pointing out that in some of the models discussed above, for instance the relaxion, the coupling is generically driven to unnatural values by the relaxation mechanism~\cite{Banerjee:2020kww}.

\subsection{Vector ultralight dark matter}
\label{Sec:vector_DM_theory}
Another possibility for UDM is a naturally light gauge boson, denoted as the ``dark photon" (DP). It naturally appears in models
with compactified extra dimensions, such as string-inspired models. Its mass
can be naturally light \cite{Goodsell:2009xc}, thus it is a good ultralight dark matter candidate.
Much like the case of scalar UDM, ultralight DPs can be described by a coherently-oscillating condensate, using a classical field of the form $A'_\mu(t, \boldsymbol{x}) \simeq A'_\mu(t) = A'_{\mu,0}\cos(m_{\gamma'} t)$ over a characteristic length scale of $\lambda_{\rm coh} \propto m_\phi^{-1}$ (see Section \ref{sec2} for details).
If the DP dark matter (DPDM) reproduces the total observed DM density, then $A'_{\mu,0}$ can be written as $A'_{\mu,0}=\sqrt{2\rho_{\rm DM}}/m_{\gamma'}$. 
For DPs to act as cold DM, they need to be decoupled from the SM thermal bath so that they can effectively cool, becoming non-relativistic before matter-radiation equality. 
The relic abundance of DPDM may be produced by the misalignment mechanism associated with the inflationary epoch~\cite{Nelson:2011sf,Arias:2012az,Graham:2015rva}, as described in the previous section. 
Other non-thermal production mechanism include production of DPDM through parametric resonance~\cite{Dror:2018pdh}, tachyonic instability developed via the rolling of the axion (or ALPs)~\cite{Agrawal:2018vin, Co:2018lka, Co:2021rhi} or the inflaton~\cite{Bastero-Gil:2018uel}, and decay of global cosmic strings~\cite{Long:2019lwl}.

There are several possibilities for the DP to couple with SM particles. Generally, the interactions arise either through the kinetic mixing term, or through a direct gauge coupling. For the kinetically-mixed DP, the Lagrangian can be written as
\begin{equation}
{\cal L} \supset -\frac{1}{4}(F_{\mu\nu}F^{\mu\nu}+F'_{\mu\nu}F'^{\mu\nu}-2\epsilon F_{\mu\nu}F'^{\mu\nu})+\frac{1}{2}m_{\gamma'}^2 A'_\mu A'^\mu-e A_\mu j^\mu_\textrm{EM} \, , 
\label{qa}
\end{equation}
where $\epsilon$ denotes the strength of the kinetic mixing (sometimes referred to as the ``mixing parameter").
We note that $A_{\mu}$ is the ordinary electromagnetic field and $A'_{\mu}$ is the DP gauge potential. In addition, $F_{\mu\nu}=\partial_\mu A_\nu-\partial_\nu A_\mu$ and $F'_{\mu\nu}=\partial_\mu A'_\nu-\partial_\nu A'_\mu$ are the field strength of the SM photon and DP, respectively. 
In order to make the kinetic terms canonical, one needs to perform a field redefinition. In the so-called mass basis, the DP couples to the electromagnetic current, $j^\mu_\textrm{EM}$, with a coupling proportional to the mixing parameter $\epsilon$, 
\bea
\mathcal{L} \supset -e A_{\mu} J^{\mu}_\textrm{EM}- \epsilon e A'_{\mu} J^{\mu}_\textrm{EM} \, . 
\eea
A similar possibility involves a DP that mixes with the $Z$-boson instead of the SM photon~\cite{Galison:1983pa}.

In another large class of models, the DP couples to the SM through direct gauge couplings. Generically, the Lagrangian can be written as
\begin{equation}
\label{vector_gauge_coupling}
{\cal L} \supset -\frac{1}{4}(F_{\mu\nu}F^{\mu\nu}+F'_{\mu\nu}F'^{\mu\nu})+\frac{1}{2}m_{\gamma'}^2 A'_\mu A'^\mu-e A_\mu j^\mu_\textrm{EM} - g A'_\mu j^\mu_\textrm{SM} \, . 
\end{equation}
Here $j^\mu_\textrm{SM}$ is not necessarily the electromagnetic current. For example, if one gauges the baryon number $B$ in the SM and assumes the DP to be the gauge boson of such a gauge group, then $j^\mu_\textrm{SM}$ should be identified as the baryon current. In this case, a gauged $U(1)_{B}$ counts the total baryon number of a body. Other popular choices include the $B-L$ current, which effectively counts the number of neutrons associated with a charge-neutral body~\cite{Pierce:2018DM-LIFO_vector} and/or $U(1)_{L_e-L_\tau}$ (see \cite{Fayet:2016nyc,Bauer:2018onh} and references therein for a detailed discussion).

The couplings of a DP to SM particles can lead to various interesting experimental signatures. The existence of the DP leads to an additional force between SM particles, and therefore can potentially be detected in experiments testing the EP. In addition, if the mass of the DP is nonzero, it may also lead to a deviation from Coulomb's Law, namely the $1/r^2$ scaling of the Coulomb force. The DP can also be produced in various astrophysical environments (see also Section~\ref{astro}), such as in the cores of stellar objects or during supernova explosions. Such production processes provide addition cooling channels and hence may lead to deviations from observations. If the DP plays the role of the dark matter, it may also cause displacements of objects that are charged under the DP field.  Ultralight DP detection strategies are discussed  in Section~\ref{searches}.

Table~\ref{tab1-1} gives a summary of possible tree-level couplings of ultralight vectors, as well as corresponding experimental and astrophysical probes that are discussed in Sections~\ref{astro} and~\ref{searches}.

\section{Ultralight dark matter properties for detection considerations}
\label{sec2}

Several properties of the wavelike DM distribution stand out as important for detection: local persistent density and spectral shape, coherence qualities, and prevalence of transient structures. 

\textbf{Wave-like nature of UDM ---} DM is considered to be wavelike where the values of its state occupation numbers $N_{\rm dB}=n_\phi \lambda_{\rm coh}^3$ greatly exceed unity. This transition is expected to occur, for our local Galactic region, when 
\begin{equation}
    m_\phi \lesssim 30\,{\rm eV}
    \left(\frac{250\,{\rm km/s}}{\langle v^2\rangle^{1/2}}\right)^{3/4}
    \left(\frac{\rho_{\rm DM}}{0.4\,{\rm GeV/cm}^3}\right)^{1/4},
\end{equation}
assuming that the UDM candidate comprises the majority of the DM that is gravitationally bound in a near-thermal halo distribution \cite{Hui:2021tkt}. The root-mean-square speed of the DM is given here by $\langle v^2\rangle^{1/2}$ (see below).

\textbf{Spatio-temporal coherence of bosonic UDM ---} The spatio-temporal coherence scales represent the observed typical length and time scales over which the UDM field has a near-constant amplitude and phase. 
The coherence time is governed by the spread in the UDM kinetic energies and is given by $\tau_\textrm{coh} \sim 2\pi / \Delta E_\phi$, while the coherence length is governed by the spread in the UDM momenta and is given by $\lambda_\textrm{coh} \sim 2 \pi / \Delta p_\phi$. 
UDM bosons that are primarily bound to the Milky Way halo are expected to have 
a root-mean-square speed relative to the Galactic center of $\langle v^2\rangle^{1/2}_\textrm{Galaxy} \sim 10^{-3} c \approx 300$~km/s, which is slightly higher than the orbital speed of the Sun about the Galactic center. 
We can hence estimate the coherence time on Earth to be $\tau_\textrm{coh} \sim 4 \pi/ (m_{\phi} \langle v^2\rangle_\textrm{Earth}) \sim 10^6 T_\textrm{osc}$, where $T_\textrm{osc} \approx 2\pi / m_{\phi}$ is the period of oscillation (Compton timescale). 
The coherence length is given by $\lambda_\textrm{coh} \sim 2 \pi / (m_{\phi} \langle v^2\rangle^{1/2}_\textrm{Earth}) \sim 10^{3} \lambda_\textrm{c}$, where $\lambda_\textrm{c} = 2 \pi/(m_{\phi} c)$ is the Compton wavelength. 

The issue of coherence in UDM searches has received attention recently in the context of searches in the lower mass range when the coherence timescale exceeds the duration of experiments \cite{Centers:2021Stochastic}, as well as searches in the higher mass range when the detector size exceeds the coherence length scale \cite{Brun:2019lyf,universe8010005}. 
The amplitude of an oscillating UDM field varies stochastically on time and length scales which exceed the coherence time and coherence length, respectively \cite{Derevianko:2018Stochastic,Foster:2018Stochastic}. 
In the standard halo model, the UDM field amplitude is expected to follow a Rayleigh-type distribution, while the UDM particle velocities are expected to follow a Maxwell-Boltzmann-type distribution~\cite{Evans:2018bqy}. 

In experiments probing UDM in the temporally incoherent regime, a range of different UDM field amplitudes is expected to be sampled.
Thus, terrestrial searches whose integration time is much longer than the coherence time of the target candidate will be exposed to many density fluctuations, converging to the expected local density on average (excluding transients). Searches for lighter candidates, where observation may only occur over a single or a few coherence times, may be strongly biased by the strength of the experienced fluctuation.
Therefore, in experiments that probe UDM in the temporally coherent regime, the stochastic nature of the UDM field should still be taken into account due to only partial sampling of the distribution of UDM field amplitudes \cite{Centers:2021Stochastic}.

\textbf{Compact dark matter objects and halos ---} 
Structures formed from ultralight scalars and vectors can arise during the history of the Universe when local relative density fluctuations grow to $\mathcal{O}(1)$, either due to gravitational instability (i.e., gravity dominates over the gradient energy) or tachyonic instability (i.e., due to attractive self-interactions). 
Once decoupled from the Hubble flow, a coherent state of the ultralight bosonic field could form as a Bose-Einstein condensate given enough time to dynamically relax (see e.g.~\cite{Levkov:2018kau,Kirkpatrick:2020fwd,Kirkpatrick:2021wwz,Chen:2021oot}).
A compact object of this type is independent of background dark matter and its density differs from the surrounding density. 
If the scalar DM forms a compact object, the amplitude of the DM oscillation becomes $\phi_0=\sqrt{2\rho_*}/m_\phi$, which is determined by the density $\rho_*$ of the compact object.

In the presence of gravity alone, a free scalar field can support itself against gravitational collapse through the repulsive pressure associated with gradients in the scalar field, below a critical value of the total bound mass \cite{Kaup:1968zz,Ruffini:1969qy}. When formed from spin-0 particles, these configurations have come to be known as boson stars, axion stars, relaxion stars, or solitons~\cite{Colpi:1986ye,Eby:2014fya,Eby:2015hsq,Jain:2021pnk,Banerjee:2019epw} (see Ref.~\cite{Visinelli:2021uve} for a recent review). Similar objects can form from spin-1 fields, which are known as Proca stars or vector boson stars~\cite{Brito:2015pxa,SalazarLandea:2016bys,Sanchis-Gual:2017bhw,Zhang:2021xxa}. As long as self-interactions are weak, the balance of gravitational and gradient forces implies that the boson star radius is inversely proportional to its mass.

The size and mass of boson stars varies over many orders of magnitude, depending on the mass $m_\phi$ of the UDM boson. On galactic scales, light scalar and vector fields of mass $m_\phi\sim 10^{-22}\;\mathrm{eV}$ give rise to subhalo structures with size of $\mathcal{O}(\mathrm{kpc})$, which can form highly dense central cores in galaxies~\cite{Schive:2014,Schwabe:2016,Mocz:2017,Nori:2018,Mocz:2019, Amin:2022pzv}. On stellar scales, light scalar and vector fields of mass $m_\phi\sim 10^{-10}\;\mathrm{eV}$ (or above) could form (sub-)solar-mass bound structures. In principle, these objects can be freely moving throughout the Galaxy, giving rise to unique lensing signals~\cite{Kolb:1995bu, Fairbairn:2017sil, Dai:2019lud, Croon:2020ouk}.
They may also undergo transient encounters with Earth, enhancing the observed DM density for a finite time.
However, this is only viable for $m_\phi \gtrsim 10^{-8}\,\eV$; when $m_\phi\lesssim 10^{-8}\,$eV, it is the case that either the boson-star energy density is large but the encounter rate with Earth is too small for terrestrial experiments on human timescales, or the rate is high enough but the boson-star energy density becomes even smaller than that of the background DM~\cite{Banerjee:2019epw}. 

On the other hand, when a light scalar field has repulsive self-interactions, the mass profile could largely deviate from that above~\cite{Colpi:1986ye}, which can enhance their density. Gravitational waves from binaries of such compact objects can lead to gravitational-wave signals that are sensitive to the specific form of the scalar potential \cite{Croon:2018ybs}.

Large density enhancements can be achieved if the scalar field becomes bound to an external gravitational source, rather than supported by its self-gravity. This configuration is referred to as a \emph{bound scalar halo} (or ``relaxion halo'' in case of relaxion DM) \cite{Banerjee:2019epw}. If such a halo is hosted by the Sun (Earth) it is called the Solar (Earth) halo (see also \cite{Anderson:2020rdk}). The strongest direct constraints on the density $\rho_*$ in this scenario arise from measurements of solar system ephemerides (for a Solar Halo) \cite{Pitjev:2013sfa}, or from comparison of low-orbit satellites to lunar-laser ranging observations (for an Earth Halo) \cite{Adler:2008rq}. In addition to a potential density enhancement, the scalars bound in such objects are colder (have lower velocity dispersion) than the background DM, which implies a longer coherence time which can be exploited in experimental searches.
The existence of such objects can be probed by atomic and optical precision measurements on Earth~\cite{Banerjee:2019epw,Grote:2019DM-LIFO} or in space~\cite{Tsai:2021lly}, and also by experiments looking for ALP DM as discussed in~\cite{Banerjee:2019xuy}.

In the Solar system, planetary and asteroidal data provide strong constraints on dark sector particles, including model-independent bounds on DM \cite{Pitjev:2013sfa}, as well as long-range forces mediated by ultralight bosons \cite{Poddar:2020exe,Tsai:2021irw}. These model-independent bounds are especially strong for DM bound to Earth and the Sun, and in turn affect the direct-detection signatures.

\textbf{Topological solitons ---}
Ultralight bosonic fields may also form solitonic objects of topological nature. 
Topological solitons (also known as topological defects) may arise in a variety of dimensionalities, namely zero-dimensional monopoles \cite{Hooft:1974Monopole,Polyakov:1974Monopole}, one-dimensional strings \cite{Abrikosov:1957String,Nielsen:1973String} and two-dimensional domain walls \cite{Zeldovich:1974DW}. 
Such objects may be produced during a cosmological phase transition \cite{Vilenkin:1985Defects_review}. 
Monopoles, being nearly pressureless, are a good candidate to explain the observed DM. 
However, the equations of state for canonical strings and domain walls differ significantly from the non-relativistic equation of state. 
Additionally, canonical strings and domain walls would be expected to induce large anisotropies in the CMB temperature spectrum that are ruled out by observations \cite{Press:1989Walls,Urrestilla:2008Strings}, and so these types of objects may not comprise the entirety of the DM. 
However, decays of strings and domain walls in the early universe can provide a sizeable contribution to the present-day bosonic UDM abundance (see, e.g., Refs.~\cite{Nagasawa:1994Axion_Strings-DWs,Chang:1998Axion_Strings-DWs,Hagmann:2010Axions_Strings-DWs,Hiramatsu:2011Axions_Strings-DWs,Hiramatsu:2012Axions_Strings-DWs,Gelmini:2021yzu,OHare:2021zrq} and references therein).

\section{Cosmological and astrophysical probes}
\label{astro}

Important constraints on scalars and dark vectors come from cosmological and astrophysical probes. The cosmological and astrophysical probes are complementary to each other and to terrestrial experiments.

\begin{figure}
\begin{center}
\includegraphics[width=1.0\textwidth]{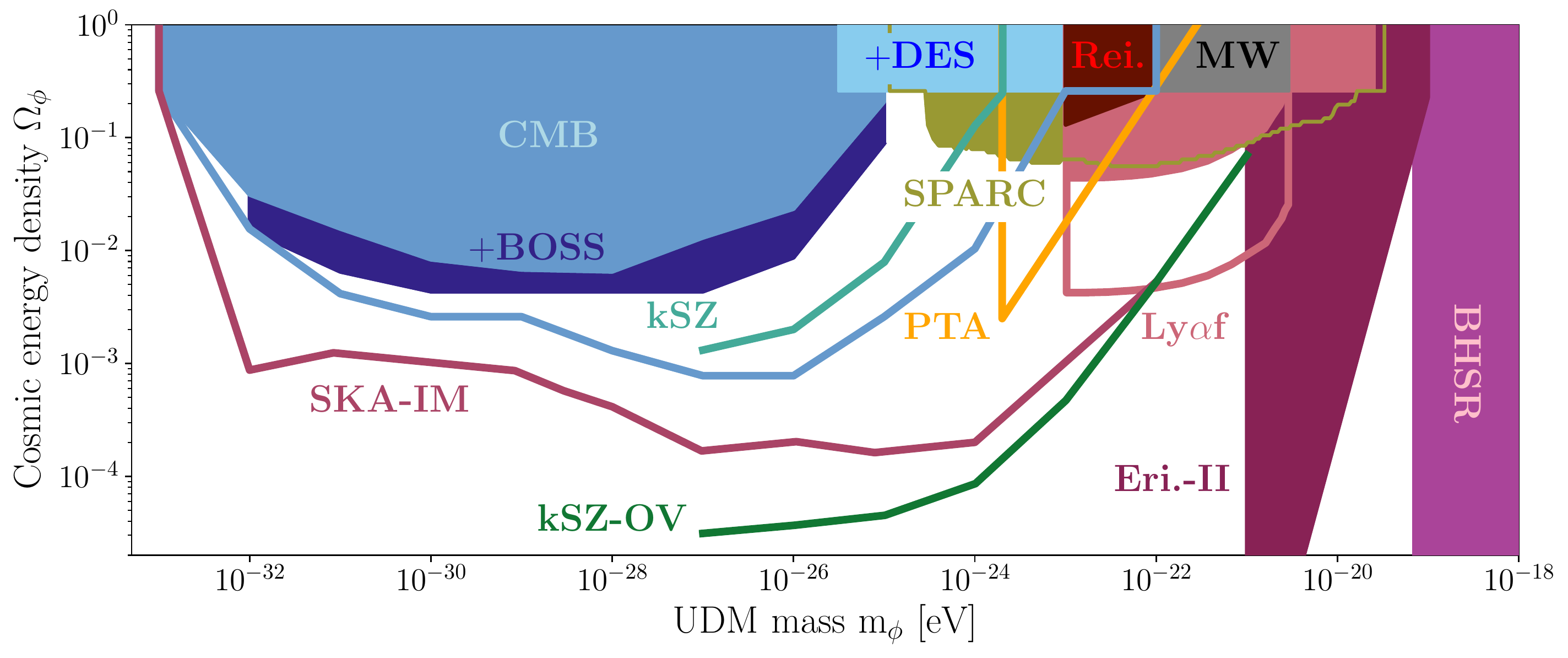}
\caption{Current (in solid) and projected (thick lines) 95\% c.l. exclusions on ultralight scalars from astrophysical probes, for cosmic energy density \(\Omega_\phi\) (where \(\Omega_\phi = 1\) is the critical density) as a function of mass \(m_\phi\). Existing cosmic microwave background (CMB) bounds come from \textit{Planck} \cite{Hlozek:2017zzf,Poulin:2018dzj}, while the thick blue line shows the forecast sensitivity for CMB-S4 \cite{Hlozek:2016lzm}. The kinetic Sunyaev-Zeldovich mean pairwise velocity (kSZ) measurement in CMB-S4 and DESI will also probe heavier UDM, while the Ostriker-Vishniac (kSZ-OV) signal in a future CMB-HD experiment could probe a sub-dominant contribution of UDM up to \(m_\phi \sim 10^{-21}\,\mathrm{eV}\) \cite{2021arXiv210913268F}. Galaxy surveys are already powerful probes of UDM: a joint analysis of \textit{Planck} CMB anisotropies and galaxy clustering multipoles from the Baryon Oscillation Spectroscopic Survey (+BOSS) sets the strongest bounds on the UDM energy density for \(m_\phi < 10^{-25}\,\mathrm{eV}\) \cite{2022JCAP...01..049L}; combining \textit{Planck} and galaxy weak lensing shear from the Dark Energy Survey (+DES) sets \(m_\phi > 10^{-23}\,\mathrm{eV}\) for UDM being all the DM \cite{2021arXiv211101199D}. The strongest bound from the Lyman-alpha forest (Ly\(\alpha\)f) sets \(m_\phi > 2 \times 10^{-20}\,\mathrm{eV}\) for UDM being all the DM \cite{2021PhRvL.126g1302R}, while \cite{Kobayashi:2017jcf} considered a sub-dominant contribution to the DM. UDM is also excluded from galaxy rotation curves in the SPARC database \cite{Bar:2021kti}; a combination of the high-redshift UV luminosity function and the optical depth to reionization (Rei.) \cite{Bozek:2014uqa}; and the Milky Way sub-halo mass function (MW) \cite{Nadler200800022}. The central star cluster in the dwarf galaxy Eridanus-II (Eri.-II) can be used to exclude \(m_\phi \sim 10^{-20}\,\mathrm{eV}\) \cite{2019PhRvL.123e1103M}, while black hole superradiance (BHSR) probes \(m_\phi \sim 10^{-18}\,\mathrm{eV}\) (although there are differences between analyses) \cite{Arvanitaki:2010sy,Arvanitaki:2014wva,Brito:2014wla,Brito:2015oca,Arvanitaki:2016qwi,Brito:2017zvb,Cardoso:2018tly,Caputo:2021efm}. Pulsar timing array (PTA) residuals are sensitive to gravitational potential oscillations for \(m_\phi \sim 10^{-23}\,\mathrm{eV}\). A future intensity mapping survey from the Square Kilometre Array (SKA-IM) could probe up to \(m_\phi \sim 10^{-22}\,\mathrm{eV}\) and \(\Omega_\phi \sim 10^{-4}\). For \(m_\phi < 10^{-30}\,\mathrm{eV}\), the ultralight scalars behave like dark energy and do not contribute to the dark matter density. 
}
\label{fig:astro}
\end{center}
\end{figure}

Cosmological measurements probe ultralight dark matter particles of very low masses\footnote{In the astrophysical literature, UDM is often used interchangeably with ultralight scalar/bosonic/axion dark matter; in the mass range $m_\phi<10^{-18}\,\mathrm{eV}$, it is often referred to as ``fuzzy dark matter'' owing to large-scale astrophysical wave-like phenomena.} (\(m_\phi < 10^{-18}\,\mathrm{eV}\)).
Such low masses are sufficiently light that wave-like behavior would manifest on astrophysical scales 
($\gtrsim$ pc) and could explain possible tensions within the standard cold dark matter (CDM) cosmological model \cite{2021PhRvD.104l3011B}; in particular a mass of \(\sim 10^{-22}\,\mathrm{eV}\) is often invoked to address the so-called cold dark matter ``small-scale crisis'' \cite{2017ARA&A..55..343B}. 
Much of the constraining power of astrophysical systems
derives from the imprint of the UDM Jeans suppression scale in the growth of cosmic structure: \(\lambda_\mathrm{Jeans} = 9.4\,(1 + z)^\frac{1}{4} \left(\frac{\Omega_\phi h^2}{0.12}\right)^\frac{1}{4} \left(\frac{m_\phi}{10^{-26}\,\mathrm{eV}}\right)^{-\frac{1}{2}} \mathrm{Mpc}\), where \(\Omega_\phi h^2\) is the physical UDM energy density and \(z\) is redshift \cite{Hu:2000ke,2017PhRvD..95d3541H}. The very lightest ultralight particles (\(m_\phi \lesssim 10^{-30}\,\mathrm{eV}\)) behave like dark energy for some time after matter-radiation equality, and so are heavily constrained by the integrated Sachs-Wolfe effect in the cosmic microwave background \cite{Marsh:2015xka}. Figure\,\ref{fig:astro} shows the current and projected experimental status.\footnote{An up-to-date version of Fig.~\ref{fig:astro} is maintained at \url{https://keirkwame.github.io/DM_limits}.} Unless otherwise stated, the bounds and forecasts given in this section refer to ultralight scalars with \(m_\phi > 10^{-27}\,\mathrm{eV}\). Nonetheless, any limits arising from non-relativistic physics, i.e., those set by matter clustering after recombination, can be applied to vectors (and indeed tensors), if the self-interactions are weak enough. This complementarity has been under-appreciated to-date in the astrophysical literature and we advocate for a wider understanding of how cosmological data can be used to search for ultralight vectors.

Other astrophysical probes are additionally effective at intermediate to large masses $10^{-13}\,\mathrm{eV} \lesssim m_{\phi} \lesssim  10\,\mathrm{eV} $, and are complementary to laboratory searches and cosmological probes. Black hole superradiance currently probes light scalars and vectors in the mass range ($10^{-13}\,\mathrm{eV} \lesssim m_{\phi} \lesssim  10^{-11}\,\mathrm{eV} $). Assuming the light scalar or vector couples to muons, neutron star inspirals can dominantly constrain $m_{\phi} \lesssim 10^{-11}\,\mathrm{eV}$. Stellar cooling constraints can be the most stringent ones in different mass ranges, depending on the spin, coupling and other properties (e.g. the mass generation mechanism) of the light particles.

Below we describe the cosmological and astrophysical probes that constrain ultralight scalar and vector dark matter.

\textbf{Cosmic microwave background} -- UDM would alter the standard picture of cosmic expansion and perturbation growth \cite{Hu:2000ke,Amendola:2005ad,Hwang:2009js,Marsh:2010wq,Chavanis:2011uv,Park:2012ru,Marsh:2013taa,Hlozek:2014lca,Suarez:2015fga,Marsh:2015xka,Urena-Lopez:2015gur,Emami:2016mrt,Hlozek:2016lzm,Cedeno:2017sou,Hlozek:2017zzf}, allowing \textit{Planck} cosmic microwave background (CMB) data to bound the UDM energy density fraction $\Omega_\phi \lesssim 10^{-2}$ for $10^{-32}\,{\rm eV}\lesssim m_\phi \lesssim 10^{-26}\,{\rm eV}$ \cite{Hlozek:2017zzf,Poulin:2018dzj}. Future experiments at the Simons Observatory (SO) \cite{Ade:2018sbj} and in CMB Stage-4 (CMB-S4) \cite{Abazajian:2016yjj} (with map noise levels of $6~\mu{\rm K}$-${\rm arcmin}$ and $1~{\mu}{\rm K}$-${\rm arcmin}$ respectively) will achieve sensitive measurements of primary CMB anisotropies, and reduce CMB lensing noise by a factor of $\sim 20$ \cite{Dvorkin:2022bsc}. Consequently, CMB-S4 should improve UDM sensitivity to $\Omega_\phi \sim 10^{-3}$ at the most constrained $m_\phi$ values and probe up to $m_\phi \sim 10^{-23}\,{\rm eV}$ \cite{Hlozek:2016lzm}. Proposed small-scale experiments like CMB-HD \cite{Sehgal:2019ewc,CMB-HD-Snowmass} could measure the lensing power spectrum to $L\sim 40,000$, allowing the CMB to distinguish between particle CDM (e.g. WIMPs) and wave-like UDM in the $m_\phi \sim 10^{-22}\,{\rm eV}$ regime relevant to challenges coming from observations at the Milky Way-scale to the standard cosmological model \cite{Marsh:2013ywa,Marsh:2016vgj,2017ARA&A..55..343B}. In order to prepare for future ambitious observations like CMB-HD, we advocate that predictions for linear theory observables should be reexamined to clarify the relationship between UDM field and fluid descriptions \cite{Suarez:2011yf,Cembranos:2015oya,Suarez:2015fga,Urena-Lopez:2015gur,Urena-Lopez:2015gur,Leong:2018opi,Cookmeyer:2019rna,2022arXiv220110238P}. Adding information from secondary CMB anisotropies in the kinetic Sunyaev-Zeldovich effect could further increase sensitivity and also probe up to \(m_\phi \sim 10^{-22}\,\mathrm{eV}\) \cite{2021arXiv210913268F, 2019AAS...23334915F, Mueller:2014dba}. Electromagnetic UDM couplings will also yield CMB spectral distortion signatures \cite{Lee:2014rpa,Sigl:2018fba,Mukherjee:2018oeb,Mukherjee:2018zzg,Fedderke:2019ajk}.

UDM would source isocurvature perturbations \cite{Axenides:1983hj,Turner:1983sj,Lyth:1991ub,Fox:2004kb,Hertzberg:2008wr,Komatsu:2008hk,Langlois:2010xc} (if the symmetry breaking that sets the relic density occurs during inflation) with amplitude set by the inflationary Hubble parameter $H_{\rm I}$ and mass \(m_\phi\). $H_{I}$ sets the amplitude of the inflationary gravitational wave background, detectable through CMB B-mode polarization \cite{Seljak:1996gy,Kamionkowski:1996ks} (a driver for future experiments like SO \cite{Galitzki:2018avl}, CMB-S4 \cite{Abazajian:2016yjj}, LiteBIRD \cite{2018JLTP..193.1048S}), allowing sensitivity to 
$H_{\rm I}\gtrsim 10^{13.3}~{\rm GeV}$ and $\Omega_\phi \sim 0.01$ with CMB-S4 \cite{Abazajian:2016yjj,Hlozek:2017zzf} (assuming $\mathcal{O}(1)$ misalignment). Cosmological probes of $\Omega_\phi$ and the inflationary energy scale \cite{Visinelli:2009zm,Marsh:2013taa,Marsh:2014qoa,Visinelli:2014twa,Visinelli:2017imh} are thus complementary. If $10^{-25}~{\rm eV}\lesssim m_\phi \lesssim 10^{-24}~{\rm eV}$, current data allow a $\sim 10\%$ contribution of ultralight bosons to the DM along with $\sim 1\%$ contributions of isocurvature and tensors \cite{Hlozek:2016lzm}.

\textbf{Clustering in the high-redshift Universe} -- The Lyman-$\alpha$ forest of neutral hydrogen absorption sourced in the intergalactic medium (IGM) traces the underlying dark matter distribution at high redshift (up to \(z \sim 5\)) \cite{1998ApJ...495...44C, 2019MNRAS.489.2247C}, probing small scales (\(<\,\mathrm{Mpc}\)) in the quasi-linear matter power spectrum. Because the clustering of the Lyman-alpha is sensitive to only mildly over-dense structures, the main effect of UDM is the suppression of clustering below Jeans wavenumber at matter-radiation equality. The interference pattern and soliton cores that are important for galactic structure do not affect these results \cite{Nori2019, Li2019}. As the UDM Jeans wavenumber \(k \propto m_\phi^{1/2}\), accessing smaller scales increases sensitivity to heavier UDM \cite{Hu:2000ke}. The challenge is to marginalize over the astrophysical state of the IGM, leveraging recent advances in modeling the IGM and cosmic reionization \cite{2017ApJ...837..106O, 2019MNRAS.486.4075O, 2019MNRAS.490.3177W, Davies2016, Keating2020, 2022MNRAS.509.6119M}, methods for statistical inference (e.g., emulators) \cite{2020RogersPRD, 2019JCAP...02..031R, 2019JCAP...02..050B, 2019ApJ...872...13W}, and understanding of systematic effects in the data \cite{2022MNRAS.509.2423W, 2020MNRAS.495.1825C, 2018MNRAS.474.3032R}. The strongest current bound is \(m_\phi > 2 \times 10^{-20}\,\mathrm{eV}\) at 95\% credibility for the entirety of the dark matter being ultralight \cite{2021PhRvL.126g1302R,Kobayashi:2017jcf,2017PhRvL.119c1302I,Armengaud:2017nkf}. The high-redshift clustering also puts strong constraints on the UDM mass in the post-inflation scenario, where the typical signal is enhancement of small-scale clustering due to the emergence of isocurvature density fluctuations \cite{Irsic2020}. Dark matter bounds from the Lyman-alpha forest are currently limited by statistics; increasing the number of observations with upcoming high-resolution spectrographs (e.g., ESPRESSO \cite{2014arXiv1401.5918P}, high-resolution spectrographs on future Extremely Large Telescopes) could increase sensitivity. Independent measurements of IGM properties by upcoming spectrographic surveys (DESI \cite{2016arXiv161100036D}, WEAVE \cite{2014SPIE.9147E..0LD}, PFS \cite{2014PASJ...66R...1T}, 4MOST \cite{2012SPIE.8446E..0TD}) could help further to break UDM/IGM degeneracies, allowing high-$z$ measurements of the power spectrum to smaller scales, thus probing larger UDM masses. We advocate for improved theoretical modeling and simulations of the impact of reionization on the IGM \cite{2015MNRAS.453.2943C}. At higher $z$, intensity mapping of lines like the neutral hydrogen 21 cm transition by surveys like HIRAX \cite{2016SPIE.9906E..5XN} and the Square Kilometre Array (SKA) \cite{2020PASA...37....2W} could detect a $\sim 10\%$ contribution of UDM to the dark matter for $m_\phi \lesssim 10^{-22}~{\rm eV}$ \cite{Bauer:2020zsj,Lidz2018}. Further, measurement of the velocity acoustic oscillation feature in the 21 cm power spectrum with experiments like HERA \cite{2022ApJ...924...51A} could probe \(m_\phi \sim 10^{-18}\,\mathrm{eV}\) \cite{2021arXiv211206943H}.

\textbf{Clustering in the low-redshift Universe} -- The effect of UDM on the matter power spectrum will also manifest in the properties of individual galaxies and the large-scale galaxy distribution \cite{Hlozek:2014lca}. Galaxy clustering bounds on the UDM energy density from the Baryon Oscillation Spectroscopic Survey are already competitive with \textit{Planck} and combining datasets improves the constraint by up to a factor of five over the CMB alone, e.g., physical UDM energy density \(\Omega_\phi h^2 < 0.0002\) for \(10^{-30}\,\mathrm{eV} \leq m_\phi \leq 10^{-27}\,\mathrm{eV}\) \cite{2022JCAP...01..049L}. The gravitational weak lensing of galaxies by intervening matter is sensitive to smaller, non-linear scales in the matter power spectrum. A combined analysis of \textit{Planck} and galaxy shear data from the Dark Energy Survey sets a bound of \(m_\phi > 10^{-23}\,\mathrm{eV}\) if UDM constitutes the entirety of the dark matter  \cite{2021arXiv211101199D}. The Rubin Observatory Legacy Survey of Space and Time (LSST) will vastly improve upon current constraints with measurements of galaxy clustering and weak lensing \cite{Drlica-Wagner190201055,2019BAAS...51c.207B}. Future space-based missions, e.g., \textit{Euclid} and Roman Space Telescope, will improve sensitivity to the matter power spectrum by an order of magnitude, improving sensitivity to UDM. If the effect of small-scale astrophysical feedback processes is accurately modeled, \textit{Euclid} weak lensing could probe \(m_\phi \sim 10^{-20}\,\mathrm{eV}\) \cite{2021arXiv211101199D}.

\textbf{Galaxy/halo mass function} -- UDM suppresses high-redshift galaxy formation and thus delays cosmic reionization. A combination of the Hubble Space Telescope Ultra Deep Field UV luminosity function at redshifts \(6 < z < 10\) and the CMB optical depth to reionization \cite{Bozek:2014uqa,2015ApJ...803...34B,2015PhRvD..91b3518S} exclude UDM being more than half of the dark matter with masses \(m_\phi \leq 10^{-23}\,\mathrm{eV}\). Future James Webb Space Telescope measurements would probe the canonical UDM mass of \(10^{-22}\,\mathrm{eV}\) \cite{Bozek:2014uqa}. The UDM suppression of the matter power spectrum also translates into a low-mass cutoff in the halo/subhalo abundance. Milky Way (MW) satellite galaxies detected by the Dark Energy Survey and Pan-STARRS impose a lower limit on the UDM mass $m_\phi > 2.9\times 10^{-21}\,\mathrm{eV}$ at $95\%$ confidence \cite{Nadler200800022}. Subhalos inferred from sub-structure in the strong gravitational lensing of quasars \cite{Hsueh:2019ynk,Gilman:2019nap} and stellar stream perturbations observed by \emph{Gaia} and Pan-STARRS \cite{Bonaca:2019, Banik:2019cza,Banik:2019smi} translate to a competitive lower limit $m_\phi > 2.1\times 10^{-21}\,\mathrm{eV}$ \cite{Schutz200105503}. Future surveys like \textit{Rubin} LSST are expected to discover hundreds of new MW satellites, thousands of new strong lenses, and to deliver precise stellar stream measurements, improving UDM mass sensitivity to $\sim 10^{-19}\,\mathrm{eV}$ \cite{Drlica-Wagner190201055}. Analytic arguments and simulations suggest that UDM forms soliton cores in dark matter halos, with imprints on  galactic rotation curves \cite{Lesgourgues:2002hk,Marsh:2015wka,Calabrese:2016hmp,Bar:2018acw,Bar:2019bqz,Bar:2019ifz,Bar:2021kti}. Cored density profiles of faint dwarf galaxies have been interpreted as signatures of UDM \cite{Deng:2018}, although it is unclear if the UDM masses that solve the ``core--cusp problem'' (\(m_\phi \sim 10^{-22}\,\mathrm{eV}\)) \cite{2017ARA&A..55..343B} are compatible with other constraints \cite{2021PhRvL.126g1302R,2019PhRvL.123e1103M,Safarzadeh:2020,Nadler200800022}.

\textbf{UDM field oscillations} -- If UDM constitutes all or a significant fraction of the dark matter, the MW gravitational potential will oscillate on timescales \(t \sim 6.6\,\frac{10^{-23}\,\mathrm{eV}}{m_\phi}\,\mathrm{yr}\)~\cite{Khmelnitsky:2014pulsar}. This is accessible with pulsar timing arrays, e.g., NanoGRAV \cite{2020JCAP...09..036K}, Parkes Pulsar Timing Array \cite{2018PhRvD..98j2002P}, with pulsar timing residuals \(\delta t \sim 20\,\left(\frac{m_\phi}{10^{-23}\,\mathrm{eV}}\right)^{-3} \left(\frac{\Omega_\phi}{0.3}\right)\mathrm{ns}\). The recent detection of a stochastic background by NANOGrav \cite{2021ApJS..252....4A,2021ApJS..252....5A,2020ApJ...905L..34A} can be associated with a wide-band spectrum of ultralight scalar and vector (and tensor) dark matter \cite{2021arXiv211215593S}. Temporal oscillations would disrupt the old star cluster in the ultra-faint dwarf galaxy Eridanus-II. The survival of such a cluster implies \(m_\phi \gtrsim 6 \times 10^{-20}\,\mathrm{eV}\), although limitations in modeling allow UDM masses between \(10^{-21}\,\mathrm{eV}\) and \(10^{-20}\,\mathrm{eV}\) \cite{2019PhRvL.123e1103M}.

\textbf{Black hole superradiance} --  Ultralight bosons can trigger a superradiant instability in the presence of a spinning black hole (BH), depleting the rotational energy of the BH into macroscopic bound ``clouds'' of ultralight bosons \cite{Arvanitaki:2009fg,Arvanitaki:2010sy}. BH superradiance is most efficient when the Compton wavelength of the particle is comparable to the BH radius, and is parametrically faster for vector bosons as compared to scalar bosons \cite{Baryakhtar:2017ngi,Siemonsen:2019ebd}. The process relies on the gravitational interactions of the particles, and is less efficient for particles with large self-interactions, interactions with the Standard Model, or in the presence of multiple boson fields~\cite{Baryakhtar:2020gao,Fukuda:2019ewf, Roy:2021uye}, providing excellent complementarity to laboratory searches. There are two main observable consequences: quasi-monochromatic gravitational radiation, and gaps at large spin in the BH spin-mass distribution~\cite{Arvanitaki:2010sy,Arvanitaki:2014wva,Brito:2014wla,Brito:2015oca,Arvanitaki:2016qwi,Brito:2017zvb}. Current gravitational wave searches with LIGO/Virgo data are starting to be sensitive to scalar masses $m_\phi \in \left[2\times 10^{-13},\,10^{-12} \right]$ eV \cite{Tsukada:2018mbp,Sun:2019mqb,Palomba:2019vxe,Zhu:2020tht,LIGOScientific:2021jlr} and vector masses in the range $m_V \in \left[4\times 10^{-14},\,10^{-12} \right]$ eV \cite{Tsukada:2020lgt}.  Constraints from lack of spindown of BHs disfavor particle masses in the range $m_\phi \in\left[ 10^{-13},\,6\times 10^{-12} \right]$ eV for weakly-coupled scalars, \cite{Baryakhtar:2020gao,Ng:2020ruv} and $m_V \in\left[5\times 10^{-14},\,2\times 10^{-11} \right]$ eV for gravitationally-coupled vectors \cite{Baryakhtar:2017ngi}. In the future, observations of supermassive black holes and low-frequency gravitational waves will be sensitive to boson masses in the range $\sim 10^{-19}- 10^{-15}$ eV \cite{Arvanitaki:2014wva,Brito:2017zvb,Cardoso:2018tly}. For further discussion and references, see \cite{CF03xtreme,CF0705}.

\textbf{Neutron star binaries and mergers} -- Muons are naturally present in large quantities inside neutron stars. If a light boson couples to muons, the additional Yukawa force and the radiation of the light boson during neutron star merger can significantly modify the neutron star inspiral, thus modifying the gravitational wave spectrum of the merger observed by LIGO. Such neutron star inspirals have the ability to probe a light boson (dark photon) coupled to muons for $m_{\phi} (m_{A'}) \lesssim 10^{-11} \eV$ with a coupling strength of $g_{\mu} (g_{A'}) \gtrsim 10^{-21}$ \cite{Dror:2019uea}. In 
neutron star binaries, the presence of new vector-mediated forces, taken broadly, can induce an anomalously fast decay of the orbital period due to the emission of dipole radiation, and we refer to studies of constraints on gauged ${\rm U}(1)_{L_\mu-L_\tau}$~\cite{Dror:2019uea} and
${\rm U}(1)_{B}$~\cite{Berryman:2022zic} models.

\textbf{Stellar cooling and observations} -- Scalars and vectors that couple to the SM can be copiously produced in the cores of stars, transporting the energy from the stars and resulting in anomalous cooling of stars \cite{Grifols:1986fc,Hoffmann:1987et,Grifols:1988fv,Raffelt:1996wa,An:2013yfc,Redondo:2013lna,Hardy:2016kme}. In the stellar plasma, both transverse and longitudinal electromagnetic excitations exist and can have unusual dispersion relations. Dark photons that mix with the SM photon are produced resonantly when the dark photon mass is comparable to the plasma frequency of the SM photons, with lower masses suppressed due to the small SM photon--dark photon mixing (depending on the origin of the dark photon mass~\cite{An:2013yua}) and heavier masses experiencing Boltzmann suppression~\cite{An:2013yfc,Redondo:2013lna}. Scalars can mix with the longitudinal mode of the in-medium photon and be resonantly produced to arbitrarily low masses since, unlike the dark photon, its mixing with the longitudinal mode is not suppressed for lower masses~\cite{Hardy:2016kme}. This is the case also for pseudoscalars with CP-violating interactions~\cite{OHare:2020wah}.

Dark photons with masses $m_{A'} \lesssim {\rm keV}$ are better constrained by the Sun~\cite{An:2013yfc,Redondo:2013lna}. Solar data global fits based on helioseismology and solar neutrino observations are a particularly powerful approach~\cite{Vinyoles:2015aba}. Dark photons from the Sun can also potentially be observed with dark matter experiments~\cite{An:2013yua,An:2020bxd}.
The strongest constraints for scalar masses $m_{\phi} \lesssim {\rm keV}$ come from red giant (RG) cores before helium ignition and stars on the horizontal branch (HB) during the helium burning phase~\cite{Hardy:2016kme}. In RG cores, any cooling mechanism that dominates over neutrino cooling delays the onset of helium ignition, contrary to observations. For stars on the HB, any additional cooling mechanism contracts the star core, resulting in faster helium burning and hence shorter than observed lifetimes for the helium burning phase~\cite{Raffelt:1996wa}. 

Similar bounds apply also for other couplings. Astrophysical bounds on pseudoscalars (axions and axion-like particles) with a photon coupling can be translated, mutatis mutandis, to scalars with a coupling $\phi F_{\mu \nu} F^{\mu \nu}$~\cite{Raffelt:1996wa}. Scalars with coupling to muons can be constrained through HB stars and supernovae~\cite{Caputo:2021rux}.

\textbf{Other astrophysical probes} -- Novel astrophysical observables realized in recent years further tighten the constraints on the UDM-SM coupling strength. Recent studies of axion DM stimulated decay can be easily adapted to any bosonic UDM that admits two-photon decay~\cite{Buen-Abad:2021qvj,Sun:2021oqp}, which amplifies the UDM decay signal through Bose enhancement and can set constraints with SKA and FAST in the mass range of $m_\phi \in \left (5\times 10^{-7},\,10^{-4} \right )$~eV. In addition, photon-to-scalar oscillations inside magnetic fields in the IGM and intracluster medium (ICM) during the photon propagation leads to redshift-dependent modification of type Ia supernovae magnitude, as well as cluster-dependent modification of the Sunyaev-Zeldovich effect and X-ray surface brightness of the galaxy clusters. These modifications are bounded by the corresponding cosmic distance measurement data sets. Although originally modeled for the photon-axion system~\cite{Buen-Abad:2020zbd}, they can be easily adapted to constrain the scalar-photon coupling in the mass range of $m_\phi \lesssim 10^{-12}$~eV. The discovery of low-metallicity and cold gas medium in dwarf galaxies can as well set strong constraints on the kinetic mixing parameter of dark photons and the photon coupling of axions~\cite{Wadekar:2019mpc, 2021arXiv211108025W}. 

\textbf{Numerical simulations of cosmic UDM structure} -- Simulation codes for UDM structure formation \cite{Schive:2014,Schwabe:2016,Mocz:2017,Nori:2018,Mocz:2019} provide a tool to run numerical experiments to compare competing dark matter models. As future astrophysical probes cover a larger region of the mass -- energy density parameter space, with higher fidelity and smaller, non-linear scales, we advocate for further theoretical research into the accuracy of different simulation approaches (see below) and to consider the impact of more sophisticated particle physics models, e.g., with self-interactions. This effort must consider the complex interplay of UDM physics and multi-scale, multi-epoch astrophysics. Further, we advocate for research into robust approaches to comparing numerically-expensive simulations to data in the placing of accurate UDM bounds, e.g., using machine learning models called emulators \cite{2020RogersPRD, 2019JCAP...02..031R, 2019JCAP...02..050B, 2019ApJ...872...13W}. As for observational efforts, astrophysical simulation efforts have to-date mostly focused on the ultralight scalar setting. Although it is anticipated that the general behavior will also apply for ultralight vectors, we again advocate for a future focus on the details of higher-spin candidates. In particular, scalar simulations cannot capture the polarisation of vectors, while the early-Universe cosmology and relativistic physics are expected to differ.

The cosmic distribution of UDM is potentially rich due to the candidates' unique nature as highly degenerate Bose fluids. The early standard halo model based on a non-singular or lowered isothermal sphere produced the first conservative estimates for the distribution's line shape and the response to daily and annual modulations from the motion of a Earth-bound detector \cite{PhysRevD.42.3572}. Another model for the Bose DM halo was also created during this period, producing line shapes containing more fine structure and sensitivity to the path of a detector \cite{Sikivie1998}. UDM ($m_\phi \le 10^{-20}\eV$) has been simulated via a Schr\"odinger-Poisson (SP)/Gross-Pitaevskii (GP) equation in small cosmological volumes capable of holding a cluster of field dwarf galaxies ($10^{9-11}~M_{\odot}$), with semi-analytic techniques extrapolating the results to larger halo and particle masses \cite{Schive:2014,2020ApJ...889...88L,2020PhRvD.101h3518V,2021arXiv211009145S}. Numerical models of higher-mass UDM (with expected coherence lengths much less than a parsec) cannot be resolved in modern zoom-in simulations of MW-size halos, necessitating the use of \(N\)-body simulations with pressureless DM \cite{Lentz_2017}. More recent work has found that Bose DM infall may be modified from the SP/GP model, even above the coherence scale.  Preliminary \(N\)-body simulations have shown novel halo distributions and the potential for fine structure \cite{10.1093/mnras/stz488,10.1093/mnras/staa557,ManyBodyDMloi}.

Some models of higher-mass Bose DM predict the formation of non-linear solitons, strings and domain walls that appear after inflation. Domain walls and strings are expected to decay once the candidate acquires mass, but point-like fluctuations and solitons may become bound by their gravity and form boson stars with surrounding mini-halos of sub-solar mass \cite{2019JCAP...04..012V,2020PhRvL.124p1103B,OHare:2021zrq,2021arXiv210805368B}. The mass bound into mini-halos can be a significant fraction of the candidate's cosmic energy density \cite{2020PhRvL.125d1301E}. These mini-halos are too small to be resolved in a numerical simulation of a MW-mass galaxy, and their rate of survival to the present from tidal stripping due to stars, planets and other compact objects is unsettled \cite{TidalStreams2016,OHare:2017yze,2017JETP..125..434D,PhysRevD.104.063038}. It is expected that a residual number of compact mini-halos and boson stars, as well as ultra-cold flows of tidally stripped matter, will be present in our local Galactic region, where we
note ~\cite{Gardner:2021ntg} for a review of the broader
possibilities. The former will be seen as transient objects in terrestrial searches. The latter will be visible in the local DM distribution, but with the number and location of flows uncertain \cite{TidalStreams2016,OHare:2017yze,2017JETP..125..434D,PhysRevD.104.063038,OHare:2018trr,OHare:2019qxc}.

\section{Searches for scalar and vector ultralight dark matter}
\label{searches}

Revolutionary advances in the control of quantum systems have enabled precise manipulation and interrogation of ultracold ions, atoms, and molecules, and brought forth a wide variety of ultra-precise quantum sensors \cite{2021QST}.
The unprecedented progress in accuracy has had profound implications for experimental tests of fundamental physics postulates and searches for new particles and forces~\cite{Safronova:2018RMP}. 
This section discusses current and proposed searches for ultralight scalar and vector DM with atomic, molecular, and nuclear clocks (\ref{Sec:ULDM_clocks}), atom interferometers (\ref{Sec:3_AI}), optical cavities (\ref{Sec:ULDM_cavities+LIFO}),
 optical interferometers (\ref{Sec:LIFO-GW_detectors}), torsion balances (\ref{Sec:3_TB}),    mechanical  resonators (\ref{Sec:3_MR}), a variety of ALP detectors (\ref{sec:ALP}), magnetometer networks (\ref{Sec:3_MN}), Rydberg atoms, superconducting qubits, LC oscillators, trapped ions and others~(\ref{Sec:3_Other}). All of the data are presented in summary plots in Section \ref{summary}.

\subsection{Atomic, molecular, and nuclear clocks} 
\label{Sec:ULDM_clocks}

 Optical atomic clocks have improved by more than three orders of magnitude in precision in less than 15~years~\cite{ludlow_optical_2015}, reaching a fractional frequency precision below 10$^{-18}$ \cite{2019Alclock} and enabling new types of searches for scalar ultralight DM.
 
\subsubsection{UDM detection signal in quantum clocks\footnote{We use ``quantum'' clocks to encompass all atomic, molecular, and nuclear clocks.}} The coupling of scalar DM to the SM leads to oscillations of fundamental constants, as described in Section~\ref{theory}. 
If the fundamental constants, such as the fine-structure constant $\alpha$ or proton-to-electron mass ratio,
are space or time varying, then so are atomic, molecular, and nuclear spectra, as well as the clock frequencies. Therefore, the variation of fundamental constants would change the clock tick rate and imply dependence on the location,
time, or type of clock (since
 frequencies of different clocks depend differently on fundamental constants)~\cite{Safronova:2018RMP}.
Such an oscillation signal would be detectable with atomic clocks for a large range of DM masses ($m  \lesssim 10^{-13}$ eV) and interaction strengths.
Clock DM searches are naturally broadband, with mass range depending on the total measurement time and specifics of the clock operation protocols (see \cite{Tsai:2021lly} for details). 

A scalar UDM signal would appear as an oscillation in the ratio of two frequencies with different sensitivities to the fundamental constants. For periods longer than an experiment cycle-time, the oscillations would arise in a discrete Fourier transform or power spectrum of the data \cite{Arvanitaki:2014faa,Stadnik:2015DM-laser,Kennedy:2020DM_atom-cavity}. A UDM detection signal would be a peak at the DM Compton frequency (for linear couplings) with a specific asymmetric lineshape
\cite{Arvanitaki:2014faa,Centers:2021Stochastic,Derevianko:2018Stochastic}. For periods shorter than an experiment cycle-time, narrow-band dynamic decoupling \cite{Aharony:2021DM_atom-cavity} can be used to enhance clock sensitivity to UDM of specific masses.  In addition to comparing atomic, molecular, or nuclear frequencies, UDM searches can also be carried out using a single clock by comparing the frequency of the quantum clock to the frequency of the local oscillator (typically an optical cavity) \cite{Stadnik:2016DM-cavity,Wcislo:2016TDM_clock-cavity,Kennedy:2020DM_atom-cavity}, as described in Section~\ref{Sec:clocks_cav}. 

Transient changes in fundamental constants that are potentially detectable with networks of clocks may be induced by the passage of DM objects with large spatial extent, such as stable topological defects (Section~\ref{sec2}). See Section \ref{networks} for a discussion of searches with networks of detectors. 
In summary, quantum clocks can map small fractional variations of  fundamental constants  (of any cause or type, e.g., temporal, spatial, slow-drift, oscillatory, gravity-potential dependent, transient or other) onto fractional frequency deviations that are measured with outstanding precision, enabling scalar UDM searches. 

\textbf{Types of quantum clocks and their sensitivity to UDM-SM interactions} -- All presently operating atomic clocks are based either on transitions between hyperfine substates of the ground state of the atom (microwave clocks: frequencies of a few GHz)  or transitions between different electronic levels (optical clocks: frequencies of  $0.4-1.1\times10^{15}$~Hz) \cite{ludlow_optical_2015}. The frequencies of optical clock based on atomic transitions are primarily sensitive to variation of $\alpha$, less so of other constants. Therefore, optical atomic clocks can probe the $\phi F^{\mu \nu}F_{\mu \nu}$ SM-DM coupling and the corresponding quadratic coupling that are defined in Eqs.\,(\ref{scalar-field_Lagrangian_alternative}) -- (\ref{quadratic_scalar_interactions}).

The degree of sensitivity is  parametrized by dimensionless sensitivity factors $K$
that translate the fractional accuracy of the ratio of frequencies $\nu$ to the fractional accuracy in the variation of the fundamental constant. For example, for the fine-structure constant,
\begin{equation}
\frac{\partial}{\partial t} \textrm{ln}\frac{\nu_2}{\nu_1}=(K_2-K_1)\frac{1}{\alpha}\frac{\partial \alpha}{\partial t}\,,
\label{K}
\end{equation}
where indices 1 and 2 refer to clocks 1 and 2, respectively.
 The  values of $K$ of all presently operating and proposed atomic clocks  have been calculated from first principles with high precision \cite{FlaDzu09}. 
 In comparison, $K$ tends to increase for states with similar electronic configurations for atoms with heavier nuclei, and details of electronic structure can lead to significantly larger enhancement factors. The Yb$^+$ clock based on an electric octupole transition that excites an electron from the closed $4f$ shell has the largest (magnitude) sensitivity factor $K=-6$ \cite{FlaDzu09} among all presently operating clocks. It should be noted that using transitions between nearly degenerate levels enhances the sensitivity in terms of the usual clock-sensitivity parameter, $\delta \nu/\nu$. However, for the cases of close degeneracy, this does not, in fact, necessarily translate into enhancement of sensitivity to variation of constants; see, for example,~\cite{Nguyen2004}.
 
Other types of clocks described below can probe different combinations of couplings in Eqs.\,(\ref{scalar-field_Lagrangian_alternative}) -- (\ref{quadratic_scalar_interactions}). 
Microwave clocks are sensitive to variation of $\alpha$ and $\mu=m_p/m_e$ (with a sensitivity factor of $K = 1$). There is also a small sensitivity of microwave clocks to $m_q/\Lambda_\textrm{QCD}$ \cite{Flambaum:2006quarks}, where $m_q$ is average light quark mass and $\Lambda_\textrm{QCD}$ is the QCD scale.

In addition to presently operating clocks, a number of new clocks are being developed \cite{2019MSreview}, based on molecules and molecular ions \cite{Patraeaba0453,2021Hanneke,ZelevinskyKondovNPhys19_MolecularClock}, highly charged ions (HCIs) \cite{kozlov_highly_2018,micke_coherent_2020}, and the $^{229}$Th nucleus \cite{Peik2021}.
Molecular clocks provide sensitivity to 
$\mu$
variation and are projected to reach $10^{-18}$ uncertainties \cite{2021Hanneke}.  
The nuclear clock is highly sensitive to the hadronic sector couplings, with possible $K=10^4$ sensitivity to the variation of $m_q/\Lambda_\textrm{QCD}$ \cite{2020nuc}; see below for further details.

\textbf{Clock stability and uncertainty} -- The state-of-the-art clocks are characterised by stability and uncertainty \cite{ludlow_optical_2015,2013ClockReview,2022QSNET}. The  \emph{uncertainty} of an atomic clock describes how well one understands the physical processes that shift the measured frequency from its unperturbed natural value.  \emph{Stability} is the precision with which we can measure a quantity,  usually determined as a function of averaging time, since noise is reduced through averaging over many noise processes, and the precision increases with repeated measurements \cite{ludlow_optical_2015,2013ClockReview}. Depending on the signal and  noise, reaching a $1\times 10^{-18}$ fractional uncertainty requires averaging times in the range of  $10^3$ to $10^6$ seconds.  Short-term stability is particularly important for the detection of UDM with masses above $10^{-15}$~eV, where the DM oscillation frequencies become comparable with the common extent of a single clock probe time. 

While the fractional accuracy of microwave clocks reaches its technical limit of about $10^{-16}$ uncertainty~\cite{2018CsClock}, both stability and uncertainty of  optical clocks are expected to continue to rapidly improve. 
The advances demonstrated in Ref. \citep{2022JunSr}  enable optical atomic coherence of 37\,s and expected
single-clock stability of $3.1 \times 10^{-18}$ at 1\,s using macroscopic samples. 

\subsubsection{Present UDM search limits from atomic clocks} 
The most recent limits on the drift of the fine-structure constant come from 
the comparison of two optical clocks based on the $^2\rm{S}_{1/2} (F=0) \rightarrow {}^2\rm{D}_{3/2} (F=2)$ electric quadrupole (E2) and the $^2\rm{S}_{1/2} (F=0) \rightarrow {}^2\rm{F}_{7/2} (F=3)$ electric octupole (E3) transition of $^{171}$Yb$^+$ \cite{2021YbclockAlpha}, and measurement of the ratio of these two frequencies to the frequency of the Cs microwave clock. Repeated measurements  over several years are analysed for potential violations of local position invariance. These measurements  improved 
the limits for fractional temporal variations of $\alpha$ to $1.0(1.1)\times10^{-18}$/yr and of the proton-to-electron mass ratio $\mu$ to $-8(36)\times10^{-18}$/yr, an improvement by factors of about 20 and 2 (respectively). Using the annual variation of the Sun's gravitational potential at Earth $U$, Ref.~\cite{2021YbclockAlpha}  improved limits for a potential coupling of both constants to gravity, $(c^2/\alpha)(d\alpha/dU)=14(11)\times 10^{-9}$ and $(c^2/\mu)(d\mu/dU)=7(45)\times10^{-8}$.

The UDM limits from the atomic clocks and precision spectroscopy  are included in the combined plots  in Section~\ref{summary}. These include re-analyses of $\alpha$ drift data for Dy/Dy \cite{PhysRevLett.115.011802}, Rb/Cs microwave clocks~\cite{PhysRevLett.117.061301}, and  Al$^+$/Hg$^+$ optical clocks~\cite{Beloy2021}. New experiments included clock-comparison experiments with Yb/Al$^+$ and Yb/Sr  clock pairs limits~\cite{Beloy2021} and comparison of hydrogen maser and strontium optical clock with a cryogenic crystalline silicon cavity (H/Si and Sr/Si) ~\cite{Kennedy:2020DM_atom-cavity}.
Further discussion of the clock-cavity experiments is given in Section~\ref{Sec:clocks_cav}.

\subsubsection{Future improvements of DM clock searches} Several directions are being pursued to drastically improve the reach of clock  experiments for DM detection:
\begin{itemize}
\item Significant improvement of the current clocks~\cite{sanner_optical_2019}, that is expected to rapidly evolve in the next decade;
\item Development of clock networks at the new level of precision~\cite{roberts_search_2020,2022QSNET};
\item Development of new atomic clocks based on highly charged ions (HCI) that have much higher sensitivities to the variation of $\alpha$~\cite{kozlov_highly_2018,micke_coherent_2020}, see Section~\ref{HCI};
\item Development of a nuclear clock that is based on a nuclear rather than atomic transition~\cite{Peik2021}, see Sec.\,\ref{Sec:Nuclock};
\item Development and implementation of new clock-comparison schemes specifically designed to improve the reach of oscillatory and transient dark matter searches~\cite{Aharony:2021DM_atom-cavity};
\item Development of molecular clocks~\cite{Patraeaba0453,2021Hanneke,ZelevinskyKondovNPhys19_MolecularClock}, see section \ref{MolClocks};
\item Use of quantum entanglement to measure beyond the standard quantum limit \cite{2020PietSqueezing, pedrozo-penafiel_entanglement_2020}.
\end{itemize}

The experimental effort is  complemented by the development of high-precision atomic theory~\cite{Ir} and particle physics model building~\cite{Graham:2015cka, Flacke:2016szy,Banerjee:2020kww}.

\subsubsection{State-of-the-art and next-generation molecular DM clock searches} 
\label{MolClocks}
Molecular vibrational and rotational dynamics are directly sensitive to $\mu$-variations, without any coupling to nuclear parameters, and present an alternative route to DM searches.  To date, the most stringent laboratory measurement of $\mu$-variation in a molecule was performed in KRb, assembled from ultracold atoms via photoassociation, finding that any linear fractional drift in $\mu$ must be smaller than $10^{-14}$ per year~\cite{Kobayashi2019}. This experiment utilized a near-exact energy match between a highly excited vibrational state in the ground electronic manifold and a relatively low-lying vibrational state in an excited electronic manifold. Due to the dramatically different potential energy curves governing the motion of the nuclei in these two states, the absolute sensitivity of the transition energy to $\mu$ is far larger than for a generic molecular microwave transition. Besides utilizing enhanced sensitivity factors of transitions between accidentally near-degenerate states, molecular experiments in a clock configuration can take advantage of their ultrahigh projected precision.  An initial demonstration of a vibrational molecular clock based on Sr$_2$ molecules at microkelvin temperatures yielded spectroscopic quality factors approaching $10^{12}$, with molecule-light coherence times of $\sim100$\,ms \cite{ZelevinskyKondovNPhys19_MolecularClock,ZelevinskyLeungPRL20_MClockRabi100ms}.  The possibility to probe vibrational states across the entire ground-state electronic potential \cite{ZelevinskyLeungNJP21_STIRAP} can allow one to use molecular clock transitions with different sensitivities to $\mu$-variations in a self-referenced configurations.  This success opens the door to tighter constraints on $\mu$-variations than what was previously achieved with molecular systems.  A highly competitive approach is being pursued for the O$_2^+$ molecular ion \cite{Carollo2019,2021Hanneke,wolf_prospect_2020}, where the large molecular binding energies, and therefore high absolute sensitivities to $\mu$, combine with a promising level of quantum control that parallels that for atomic ions. An alternative method is possible in polyatomic molecules, where two vibrational states arising from distinct vibrational modes may be accidentally near-degenerate. As a result, their energy difference may depend on $\mu$ at the scale of a typical vibrational transition, but can be probed with more technically convenient microwave sources. The linear triatomic molecule SrOH, which has been previously laser-cooled~\cite{Kozyryev2017c}, possesses a low-lying pair of near-degenerate vibrational states. A trap of SrOH molecules is projected to offer a large sensitivity to changes in $\mu$ and a high degree of control over systematic errors~\cite{Kozyryev2021}.

\subsubsection{Development of HCI clocks and perspectives towards higher frequencies}
\label{HCI}
HCI are attractive candidates for the development of novel atomic clocks with high sensitivity to both dark matter and variation of $\alpha$. Several proposals based on HCI optical clocks, their fundamental physics applications, and experimental progress towards HCI high-precision spectroscopy were reviewed in~\cite{kozlov_highly_2018}. Recent development of HCI cooling, trapping, and quantum logic techniques are enabling rapid progress in the development of HCI clocks~\cite{micke_coherent_2020,King2021}. HCI clocks enable clock-comparisons with $\Delta K \approx 100$ \cite{2022QSNET}.

HCIs are excellent clock candidates, since they offer narrow transitions and long-lived metastable states together with a small sensitivity to systematic frequency shifts \citep{Schiller2007}. 
Many isoelectronic sequences have such states, and forbidden transitions from magnetic-dipole (M1) type are well known in the optical and ultraviolet range (see references in \cite{Crespo2008}), even up to charge states as high as Bi$^{82+}$ and other hydrogen-like systems \cite{Klaft1994,Crespo1996,Crespo1998,Beiersdorfer2001,Ullmann_2015,Skripnikov2018} which were already identified as appropriate for use in searches for a temporal variation of $\alpha$ \citep{Schiller2007,Oreshkina2017,Debierre2021}. For these reasons, many proposals for HCI clocks with high sensitivity to a change in $\alpha$ have been made \cite{berengut2012,berengut2012a,dzuba_ion_2012,safronova_highly_2014-1,ong2014optical,safronova_highly_2014,safronova_atomic_2014,Dzuba2015,Dzuba2015b,Porsev2015,nandy_highly_2016,Ir,2019MSreview,Kimura2020,Liang2021}, and several experimental groups are searching for them using both electron beam ion traps (see references cited in \cite{kozlov_highly_2018}), Penning traps \cite{Egl2019} and Paul traps \cite{micke_coherent_2020}. Searching for fifth forces by studying the $g$ factor of electrons bound in HCIs has also been suggested \cite{Debierre2020}. The low polarizability of HCIs is an additional, important advantage, since it strongly suppresses systematic frequency shifts due to, for example, black-body radiation and laser interactions. Furthermore, extant orbital crossings \cite{Berengut2010,Berengut2011,Porsev2015,Dzuba2015,Dzuba2015b,Windberger2015,Bekker2019,Ir,Kimura2020} can deliver an exquisite sensitivity to a time variation of $\alpha$, due to the near-degeneracy of energy levels in different configurations brought close by the choice of charge state in certain isoelectronic sequences.
Rapid progress has been made recently towards the realization of HCI-based optical clocks. Slowing and trapping of HCIs in Paul traps using laser-cooled beryllium \cite{schmoger_coulomb_2015} was the first step towards the first coherent laser spectroscopy using quantum logic \cite{micke_coherent_2020}. Algorithmic ground state cooling enabled full control over the motional state and a suppression of Doppler shifts below the $10^{-18}$ level \cite{King2021}. Recently, the first optical clock based on a HCI has been demonstrated with atom-related systematic uncertainties at a level of $10^{-18}$ and below \cite{king_private_2022}. The employed quantum logic spectroscopy technique  \cite{schmidt_spectroscopy_2005} is versatile and can be employed for most potential HCI clock candidates (see also Section~\ref{sec:toolbox}). Most of the interesting HCI candidates with a high sensitivity to a change in $\alpha$ and UDM mentioned above exhibit hyperfine structure. As a consequence of their high charge state, hfs splittings are several 10s to 100s of gigahertz, requiring the development of suitable (quantum logic) techniques for repumping and state preparation. 

Since the demonstration of extreme-ultraviolet frequency combs (see, e.g., \cite{cingoz2012direct}), atomic clocks at wavelengths shorter than optical became feasible. 
However, at high photon energies photoionization 
couples the bound electron with the continuum if the ionization threshold is surpassed. 
For this reason, only HCIs with sufficiently high ionization potential and nuclei are appropriate as frequency references in the extreme ultraviolet spectral range and beyond it \cite{CrespoHCI2016,nauta_towards_2017}.
Forbidden transitions of higher multipolarities up to magnetic octupole (M3) have been observed up in the X-ray domain. 
Very recently, a metastable state with an excitation energy of 202\,eV has been detected in $^{187}$Re$^{29+}$ ions through ultra-high-resolution mass measurements \cite{Schuessler2020}. 
A sizable fraction of the ion population was excited to the metastable state by electron impact in an electron beam ion trap, and transferred together with ground-state ions to a Penning trap \cite{Repp2012}. There, both species survived for week-long periods, allowing for ultra-precise mass measurements with 1\,eV resolution that revealed the difference in their rest masses. 
This proved that a high excitation energy does not preclude the very long metastable lifetimes needed for frequency metrology beyond the optical range. The excited state decays with a predicted lifetime of 130 days through a triacontadipole (E5) transition, induced by a weak hyperfine coupling with the nucleus. It has a frequency of $4.96\times 10^{16}$\,Hz, a linewidth of only $5\times 10^{-8}$\,Hz, and the highest quality factor hitherto found in an excited atomic system.
Such HCI species will be used in combination with extreme-ultraviolet frequency combs \cite{Jones2005,Bernhardt2009,Nauta2021,Pupeza2021} for laser stabilization and frequency metrology using schemes that have been theoretically analysed for their feasibility \cite{Lyu2020}, and could offer much higher sensitivities to hypothetical particles and fields. 

Another advantage of HCI-based frequency references is the strong overlap of the wave function of the `optically' active electron  with the nucleus caused by the high charge state. This shrinking of the wave function not only greatly enhances QED effects, but also isotopic shifts related to finite nuclear size, and thus the sensitivity of electronic transitions to Yukawa-type interactions of hypothetical particles \cite{Delaunay:2016brc,2021Ca,Debierre2021}. The existence of many stable isotopes and a sufficient number of forbidden transition for a given element, even in different charge states, enable the application of generalized King-plot methods \cite{Berengut:2020itu} also in HCI. This can remove systematic effects due to unknown nuclear structure parameters that affect the analysis of the experimental results. Summarizing, HCI-based frequency metrology offers many useful clock references up to the soft X-ray domain, which due to the great variety of available isoelectronic sequences provide choices of favorable properties for the search of new physics at the electron-nucleus boundary.
 
 \subsubsection{Development of a nuclear clock} 
 \label{Sec:Nuclock}
The transition frequencies of nuclear energy levels are generally outside of the laser-accessible range by many orders of magnitude.
A single exception is a  nuclear transition that occurs between the long-lived (isomeric) first excited state of the $^{229}$Th isotope and the corresponding nuclear ground state, with a wavelength near 150\,nm, within reach of modern lasers.
The transition energy was recently measured to be 8.19$\pm$0.12\,eV~\cite{2019nuclear,Sikorsky2020}. Note that also the excitation of the nuclear resonance in   a highly charged ion has been proposed \cite{Bilous2020}.

Two different approaches have been proposed for the realization of a nuclear clock: one based on trapped ions and another one using doped solid-state crystals. The first approach starts from individually trapped Th ions, e.g., in a Paul trap, comparable to trap-based optical atomic clocks. This approach promises an unprecedented suppression of systematic shifts of the clock frequency and leads to an expected nuclear clock uncertainty of about 1$\times 10^{-19}$~\cite{Campbell2012}. The other approach relies on embedding $^{229}$Th in vacuum-ultraviolet (VUV) transparent crystals (e.g. CaF$_2$, MgF$_2$, LiSrAlF$_6$, LiCaAlF$_6$)~\cite{Peik2003, Rellergert2010, Jackson2009, Dessovic2014}. This bears the advantage of the large number ($\geq$ 10$^{18}$\,cm$^{-3}$) of Th nuclei included in the crystal, leading to a considerably higher signal-to-noise ratio and a higher stability of the nuclear clock~\cite{Kazakov2012}. However, a precise characterization of the thorium isomer’s properties (especially the nuclear $^{229m}$Th resonance) with laser-spectroscopic precision remains a mandatory prerequisite for any kind of nuclear clock.
So far, the identification and characterization of the thorium isomer was largely nuclear physics driven, paving the road towards the realization of a nuclear clock. Now it remains to refine our knowledge of the isomeric excitation energy with laser-spectroscopic precision in order to enable optical control of the nuclear-clock transition with a suitable laser system. This requires bridging the gap of about 12 orders of magnitude in the precision of the $^{229m}$Th excitation energy from the present $\approx$0.1\,eV down to the (sub-)kHz regime. In a first step, existing broad-band laser technology can be used to localize the nuclear resonance with an accuracy of about 1\,GHz. In a second step, using VUV frequency-comb spectroscopy, presently under development via higher-harmonic generation in a noble gas, it is envisaged to improve the accuracy into the (sub-)kHz range. Another prerequisite for setting up a high-precision ion-trap based nuclear clock is the generation of thermally decoupled, ultra-cold $^{229}$Th$^{3+}$ ions via laser cooling. $^{229}$Th$^{3+}$ is particularly suited due to its electronic level structure with only one valence electron. Due to the high chemical reactivity of thorium, a cryogenic Paul trap is the ideal environment for laser cooling, since almost all residual gas atoms freeze out at 4\,K, increasing the trapping time into the region of a few hours. This will form the basis for direct laser excitation of $^{229m}$Th and will also enable a measurement of the not yet experimentally determined isomeric lifetime of $^{229}$Th ions. For the alternative development of a compact solid-state nuclear clock~\cite{Rellergert2010}, it will be necessary to suppress the $^{229m}$Th decay via internal conversion in a large band-gap, VUV transparent crystal (like CaF$_2$ or MgF$_2$) and to detect the $\gamma$ decay of the excited nuclear state. Here we also mention recent work on trapping and sympathetically cooling of single thorium ions in a Coulomb crystal of calcium ions \cite{Groot-Berning2019,Stopp2019Catching,Haas2020Recoil}, as well as proposals to study the nuclear transitions in relativistic highly charged ions in storage rings \cite{Budker2022_GF_Nuclear}.

Designing a clock based on this nuclear transition~\cite{Peik2021} is particularly attractive due to the suppression of the field-induced frequency shifts, as the nucleus interacts only via the relatively small nuclear moments and is highly isolated from the environment by the electron cloud.
 The nuclear clock sensitivity to the variation of $\alpha$ is expected to exceed the sensitivity of present clocks by about four orders of magnitude \cite{2020nuc}.
 In addition, nuclear clocks will be sensitive to a DM coupling to the hadronic sector of the SM.
 The expected extreme sensitivity of the nuclear clock to the variation of the fundamental constants and related new physics is based on the fact that the energy scales of the internal nuclear interactions are several orders of magnitude higher than the actual nuclear transition energy, which can be measured with a precision that is unprecedented in nuclear physics.
 In Figure~\ref{fig:scalar_de_plot}, we show the projected sensitivity of a nuclear clock.

  At the projected $10^{-19}$ fractional frequency precision level and strongly enhanced sensitivity, the nuclear clock can improve the ability to probe scalar dark matter by 5-6 orders of magnitude for a wide range of DM mass ranges in comparison with present limits~\cite{Kennedy:2020DM_atom-cavity}.

\subsection{Atom interferometers}
\label{Sec:3_AI}

The wide range of physics potential of long-baseline quantum sensors has already been recognized by the high-energy physics community \cite{osti_1358078, doe_quantum_2015, 2017arXiv170704591B}. The ever-improving precision and accuracy of atom interferometers enable a new variety of exciting fundamental physics experiments including, among many others, searches for UDM candidates. In particular, atom interferometers continue to set records in precise measurements of the fine structure constant \cite{PhysRevLett.106.080801, doi:10.1126/science.aap7706} and in tests for new fundamental forces \cite{Rosi2017, PhysRevLett.112.203002, PhysRevLett.125.191101, Hartwig_2015}, which can be utilized in experimental searches for dark matter. In this section, we introduce long-baseline clock atom interferometers 
and their potential in searches for UDM candidates with MAGIS-100, one such experiment currently under construction in the US, as a baseline.

\begin{figure}[t]
    \centering
    \includegraphics[width=0.5\textwidth]{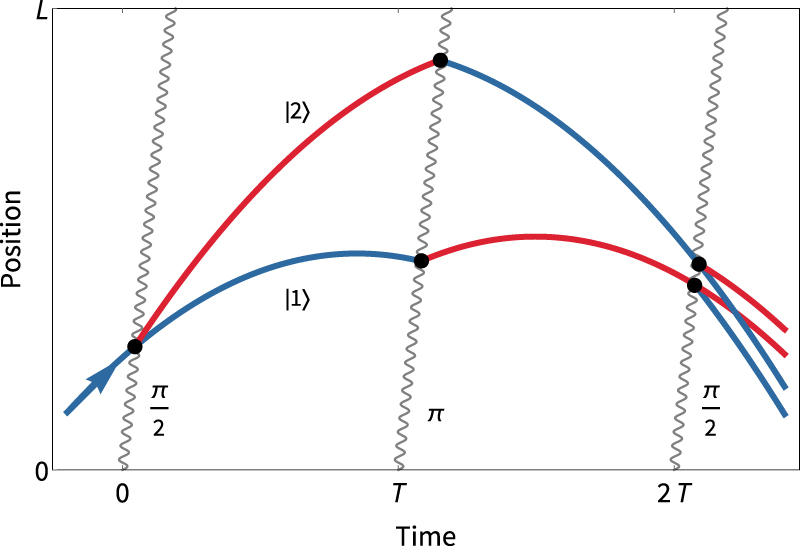}
    \caption{
    Adapted from Ref.\,\cite{magis-100_2021}. Space-time diagram for a clock atom interferometer. The atom cloud is launched upwards and falls freely under gravity, with a series of laser pulses applied (wavy lines). Blue (red) lines are atom trajectories in the ground (excited) state.
    }
    \label{fig:clock_atom_interferometer_schematic}
\end{figure}

\begin{figure}
    \centering
    \includegraphics[width=0.7\textwidth]{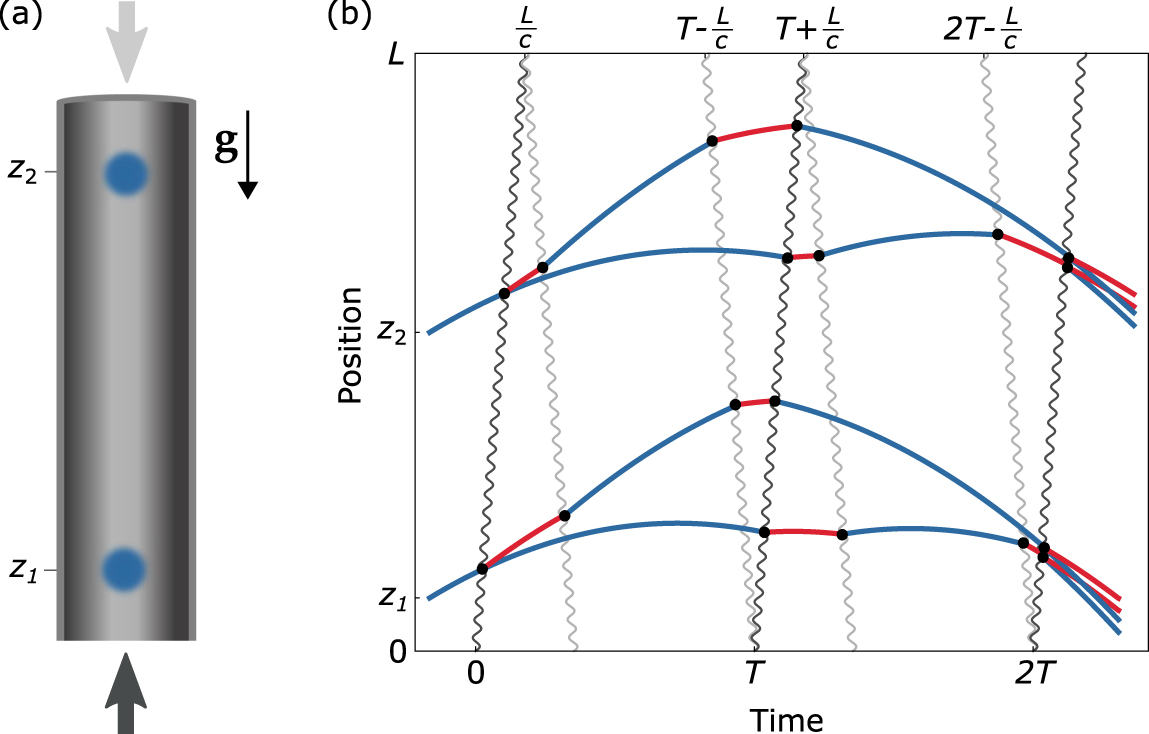}
    \caption{Adapted from Ref.\,\cite{magis-100_2021}. A clock gradiometer has two atom clouds manipulated with common laser pulses, creating a symmetric pair of clock atom interferometers across the baseline of the experiment. (a) Schematic showing two atom cloud positions; (b) Space-time diagram detailing the laser pulse sequence. Dark (light) gray lines are laser pulses traveling upwards (downwards). Blue (red) lines are atom trajectories in the ground (excited) state.
    }
    \label{fig:clock_gradiometer_schematic}
\end{figure}

At its core, an atom interferometer is an experiment that compares the phase accumulated by delocalized atom clouds (Fig.~\ref{fig:clock_atom_interferometer_schematic}). The phase $\phi$ of an atom cloud in a clock atom interferometer is proportional to $\omega_A L / c$, where $\hbar \omega_A$ is the excitation energy required for the clock transition\footnote{This is the energy required for the transition between (quasi-)stable atomic energy states.} and $L$ is the interferometer baseline length. Two clock atom interferometers sharing common laser pulses from a single laser operating together form a clock gradiometer, measuring the phase difference across the two interferometers (Fig.~\ref{fig:clock_gradiometer_schematic}). This approach has a major advantage in that it allows for common-mode rejection of the laser noise.

UDM has a large number density and can be described as a classical field that oscillates at a frequency determined by the mass of the dark matter particle as described in Section~\ref{theory}. The fact that the UDM signal oscillates at a frequency set by the mass of the dark matter serves as a powerful discriminant against a variety of noise sources, enabling high-precision searches for the ultra-weak effects of UDM. In atom interferometry, the presence of UDM can contribute to the phase of the atom cloud either: (1) by affecting the value of the internal energy splitting, or (2) by exerting additional forces upon the atom clouds. 

\begin{figure}[t]
    \centering
    \includegraphics[width=0.25\textwidth]{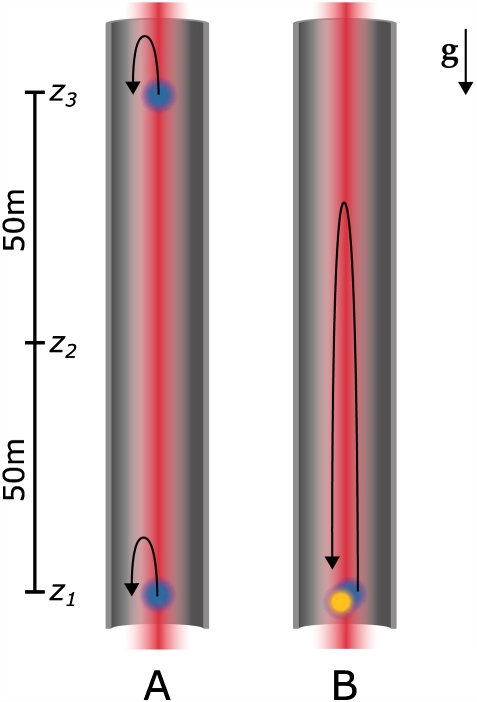}
    \caption{Adapted from Ref.~\cite{magis-100_2021}. MAGIS-100 detector operating modes. Mode A: two simultaneous atom interferometers separated across the baseline. Mode B: Dual-species atom interferometer with two atom clouds (orange and blue circles) launched from the same position.}
    \label{fig:magis_operation_modes}
\end{figure}

Firstly, UDM that affects fundamental constants (such as the electron mass or the fine structure constant) will change the internal energy levels of the atoms as  described in Section~\ref{Sec:ULDM_clocks}, causing them to oscillate at the Compton frequency of the UDM candidate. This effect can be searched for by comparing two simultaneous atom interferometers separated across the baseline of the experiment (Fig.~\ref{fig:magis_operation_modes} (A)). The sensitivity to such scalar UDM candidates using the MAGIS-100 experiment is shown in Figs.~\ref{fig:scalar_de_plot} and \ref{fig:scalar_dme_plot}. Secondly, UDM that causes accelerations can be searched for by comparing the accelerometer signals from two simultaneous atom interferometers run with different isotopes ($^{88}$Sr and $^{87}$Sr for example) \cite{graham2016dark}. This requires running a dual-species atom interferometer (Fig.~\ref{fig:magis_operation_modes} (B)), which is well established \cite{Kuhn_2014, PhysRevLett.113.023005, PhysRevA.88.043615, PhysRevLett.120.183604}. The potential sensitivity of MAGIS-100 to one such UDM candidate, a $B - L$ coupled new vector boson, is shown in Fig.~\ref{fig:vector_B-L_plot}. Potential sensitivities using more general accelerometer signals to dark matter candidates are shown in \cite{graham2016dark}.  Note that, compared to existing bounds, MAGIS-100 has the potential to improve the sensitivities to both scalar and vector UDM candidates with mass (frequency) range below approximately $\unit[10^{-15}]{eV}$ ($\unit[0.1]{Hz}$) by about an order of magnitude.

The sensitivity of atom interferometers is set by several parameters. The sensitivity is proportional to the baseline length $L$. The sensitivity also depends on the large-momentum-transfer (LMT) atom optics which refers to the number $n$ of photon momentum kicks provided to the atom clouds.  Lastly, the sensitivity is improved by reducing phase noise in the interference fringes by utilizing higher-flux atom sources and quantum entanglement. Atom interferometers of $\unit[10]{m}$ with Rb atom clouds were demonstrated with $n \approx 100$ LMT optics \cite{Kovachy2015}, and Sr clock transitions, to be used in MAGIS, have the potential to gain orders of magnitude increase in LMT $n$ \cite{PhysRevLett.119.263601, PhysRevLett.124.083604}. Also, entangled atom clouds were shown to have phase resolution $100$ times below the standard quantum limit \cite{Hosten2016}. Current and projected parameters of experiments based on the MAGIS concept are summarized in Table.~\ref{tab:magis_parameters}. The projected sensitivity of MAGIS detectors is shown on the summary plots in Section\,\ref{summary}. 

\begin{table}
\footnotesize
\centering
\begin{tabular}{cccccc}
    \hline
    Experiment &   (Proposed) Site &   \makecell{Baseline\\$L$ ($\unit{m}$)}   &   \makecell{LMT $n$} &   \makecell{Atom\\Sources} &   \makecell{Phase Noise\\$\delta\phi$ ($\unit{rad/\sqrt{Hz}}$}
    \\
    \hline
    \rule{0pt}{2.5ex}
    Sr prototype tower  &   Stanford    &   $10$            &   $10^2$  &   $2$ &   $10^{-3}$
    \\
    MAGIS-100 (initial) &   Fermilab (MINOS Shaft) & $100$ & $10^2$ & 3 & $10^{-3}$
    \\
    MAGIS-100 (final)   &   Fermilab (MINOS Shaft) & $100$ &  $4 \times 10^4$ & 3 & $10^{-5}$
    \\
    MAGIS-km    &   Homestake mine (SURF)   &   $2000$    &   $4 \times 10^4$   &   $40$    &   $10^{-5}$
    \\
    MAGIS-space &   Medium Earth Orbit  &   $4 \times 10^7$ &   $10^3$  &   $2$   &   $10^{-4}$
    \\
    \hline
\end{tabular}
\caption{Design parameters for MAGIS-concept experiments.}
\label{tab:magis_parameters}
\end{table}

Over the past several years, there has been widespread, growing international interest in pursuing long-baseline atomic sensors for UDM searches and gravitational-wave detection. This has sparked a number of proposals for both large-scale terrestrial and space-based instruments, some of which are already under construction today. These include MAGIS \cite{magis-100_2021}, AION \cite{AION_2020}, MIGA \cite{MIGA_2018}, ELGAR \cite{ELGAR_2020, canuel2020technologies}, ZAIGA \cite{ZAIGA_2020}, and AEDGE \cite{AEDGE_2020}. The ambitious scope of these endeavors from around the world is evidence of the widespread enthusiasm for the scientific prospects of long-baseline atom interferometry. These numerous projects complement each other through the diversity of approaches, allowing for the development of alternate atomic sensing technologies in parallel. The ultimate synergy of this global effort would be to realize a network of detectors.

MAGIS-100 is the first detector facility in this family of proposed experiments based on the MAGIS concept. The instrument features a 100-meter vertical baseline and is now under construction at the Fermi National Accelerator Laboratory (Fermilab). The state-of-the-art atom interferometers are currently operating at the 10-meter scale \cite{Hartwig_2015, PhysRevLett.125.191101, Kovachy2015, PhysRevLett.118.183602, PhysRevLett.111.083001, Zhou2011}. While already pushing limits for UDM sensitivity, MAGIS-100 also serves as a demonstrator to push the limits of atom interferometry beyond the lab-scale and bridge the gap to future kilometer-scale experiments. It is designed to operate as a full-scale detector, aims to achieve the high up-time required from such a facility, and explore a wide variety of systematic effects and background sources to serve as a technology demonstrator for future experiments.

\subsection{Optical cavities} 
\label{Sec:ULDM_cavities+LIFO}

Temporal variations of $\alpha$ and particle masses alter the geometric sizes of solid objects. 
In the non-relativistic limit, the length of a solid scales as $L \propto a_\textrm{B}$, where $a_\textrm{B} = 1/(m_e \alpha)$ is the atomic Bohr radius \cite{Stadnik:2015DM-laser,Stadnik:2016DM-cavity}. 
In the adiabatic limit (when sound-wave propagation through the solid occurs sufficiently fast for a solid to fully respond to changes in the fundamental constants), the size of a solid body therefore changes according to: 
\begin{equation}
    \label{adiabatic_solid_length_change}
    \frac{\delta L}{L} \approx \frac{\delta a_\textrm{B}}{a_\textrm{B}} = - \frac{\delta \alpha}{\alpha} - \frac{\delta m_e}{m_e} \, . 
\end{equation}
Relativistic corrections associated with electromagnetic processes and finite-nuclear-mass effects typically give only small corrections to the relation in Eq.\,(\ref{adiabatic_solid_length_change}) \cite{Stadnik:2015DM-laser,Pasteka:2019solids}. 

Let us specifically consider quasi-monochromatic oscillations in the fundamental constants due to an oscillating UDM field. 
If the oscillation frequency of the fundamental constants matches the frequency of a fundamental vibrational mode of the solid, then size changes of the solid are enhanced by the factor $Q = \textrm{min}(Q_\textrm{mech},Q_\textrm{DM})$ compared to the adiabatic case in Eq.\,(\ref{adiabatic_solid_length_change}), where $Q_\textrm{mech}$ is the relevant mechanical quality factor associated with the solid and $Q_\textrm{DM} \sim 10^6$ is the quality factor expected to be associated with the oscillating UDM field in our local Galactic region. 
If the oscillation frequency is much greater than the fundamental frequency (and different from the higher-harmonic frequencies), then size changes of the solid are suppressed compared to Eq.\,(\ref{adiabatic_solid_length_change}). 
See Refs.~\cite{Arvanitaki:2016DM-resonator,Geraci:2019DM-cavity,Savalle:2019DAMNED,Grote:2019DM-LIFO} for details. 

Optical cavities of various types can be used as sensitive probes of scalar UDM. 
In a cavity whose length depends on the length of the solid spacer between the mirrors, the cavity reference frequency $\nu_\textrm{cavity} \propto 1/L$ can freely respond to changes in the fundamental constants according to \cite{Stadnik:2015DM-laser,Stadnik:2016DM-cavity} 
\begin{equation}
    \label{free_cavity_length_changes}
\frac{\delta \nu_\textrm{cavity}}{\nu_\textrm{cavity}} = - \frac{\delta L}{L} \approx -\frac{\delta a_\textrm{B}}{a_\textrm{B}} = \frac{\delta \alpha}{\alpha} + \frac{\delta m_e}{m_e} \, , 
\end{equation}
where in the second (approximate) equality we have assumed the adiabatic relation, as in Eq.\,(\ref{adiabatic_solid_length_change}). 
If fluctuations in the cavity length are restricted (e.g., through the use of a multiple-pendulum suspension mirror system or if the DM oscillation frequency is too high), then the cavity reference frequency becomes practically insensitive to changes in the fundamental constants \cite{Stadnik:2016DM-cavity,Geraci:2019DM-cavity}. 

\subsubsection{Atom-cavity comparisons}
\label{Sec:clocks_cav}
Cavity-based reference frequencies can be compared against atomic or molecular transition frequencies. 
An optical atomic transition frequency scales as $\nu_{\rm{atom}} \propto \, m_{\rm{e}}\alpha^{2+K}$,  where $K$  of  Eq.\,(\ref{K}) accounts for enhanced sensitivity to $\alpha$-variation due to relativistic effects \cite{FlaDzu09}.\footnote{We note that $m_e$ dependence is the same for all optical clocks, making a ratio of two optical clocks insensitive to the variation of $m_e$.} Therefore, the comparison of an optical atomic clock against a cavity whose length is allowed to vary freely is sensitive to changes in $\alpha$, whereas the comparison of an optical atomic clock against a cavity whose length is ``fixed'' is sensitive to changes in both $\alpha$ and $m_e$ \cite{Stadnik:2016DM-cavity}. 
In contrast, comparisons of two co-located optical atomic clocks have no sensitivity to variations of the fundamental constants in the non-relativistic limit; such comparisons are sensitive to $\alpha$ variation due to relativistic corrections \cite{Dzuba:1999relativistic-A,Dzuba:1999relativistic-B}. 

 In clock-cavity comparisons, one measures the frequency of an atomic transition $\nu_{\rm{atom}}$ with respect to the resonance frequency of a reference optical cavity  ($\nu_{\rm{cavity}} \approx \nu_{\rm{atom}}$). 
The basic idea and detection signal in UDM clock-cavity searches is similar to the clock-clock comparisons described in Section~\ref{Sec:ULDM_clocks}.

Clock-cavity comparisons allow one to use high-precision clocks with small relativistic factors $K \approx 0$, as 
$\Delta K  \approx 1$ for such clock-cavity experiments, enabling searches for UDM of higher masses than clock-clock comparisons, and in some cases also providing sensitivity to $m_e$. 
The cavity could be represented by the laser internal resonator \cite{Antypas:2019DM_atom-cavity, AntypasQST2021} or some external optical cavity.
 A variety of clock-cavity comparisons searching for UDM via oscillations of the fundamental constants have been performed in the past several years \cite{Wcislo:2018DM_atom-cavity,Antypas:2019DM_atom-cavity,Kennedy:2020DM_atom-cavity,Aharony:2021DM_atom-cavity,Campbell:2021DM_atom-cavity,Oswald:2021vtc,Tretiak:2022ndx}.

Recently, new bounds on the coupling of UDM to SM fields were set by conducting  frequency comparisons between a state-of-the-art strontium optical lattice clock, a cryogenic crystalline silicon cavity, and a hydrogen maser \cite{Kennedy:2020DM_atom-cavity}.
In that frequency comparison, the most competitive bound on the electromagnetic gauge modulus $d_e$ is set by the strontium optical clock versus the cryogenic crystalline silicon optical resonator, leveraging the phenomenal long-term frequency stability of both resonator and clock. The unparalleled precision of this comparison improves the $d_e$ limits for a candidate UDM mass in the range $1 \times 10^{-19} - 10^{-17}$~eV. The frequency comparison of the hydrogen maser versus optical resonator is also sensitive to $\alpha$ variations, allowing the authors to set an additional bound on $d_e$. Despite the higher frequency instability of the hydrogen maser, the stronger sensitivity to $\alpha$ variation makes the bound competitive in a lower candidate mass range around $10^{-20}$ eV.

\begin{figure}[t]
    \centering
    \includegraphics[width=\textwidth]{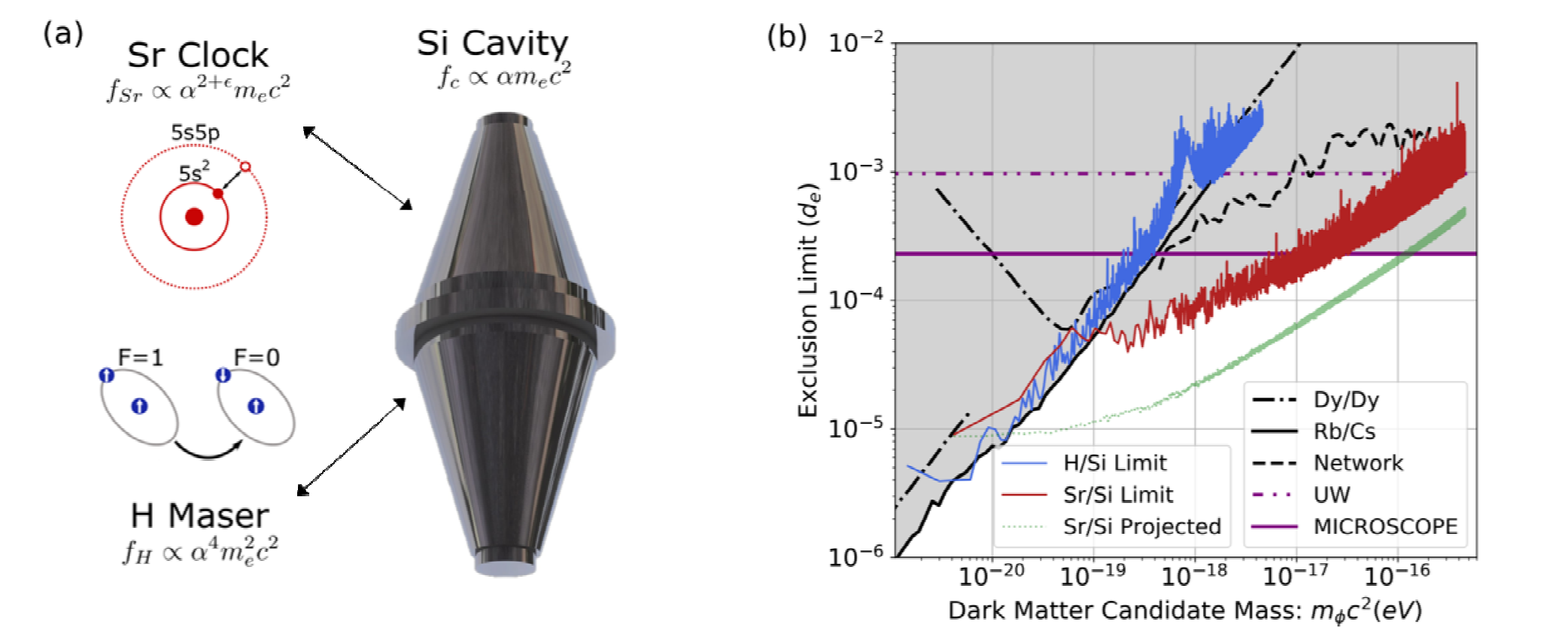}
    \caption{
    Adapted from Ref.\,\cite{Kennedy:2020DM_atom-cavity}. Simplified cartoon (a) of $\alpha$ and $m_e$ dependences of the Sr optical lattice clock, monocrystalline silicon cavity, and Hydrogen maser. Exclusion limits from these intercomparisons for $d_e$ are illustrated in (b) as blue traces for the $f_H/f_{Si}$ comparisons, and red traces for the $f_{Sr}/f_{Si}$ comparison. The $d_{m_e}$ exclusion limits obtained from this experiment are included in Fig.~\ref{fig:scalar_dme_plot}.
    }
    \label{fig:Sr_Si_H}
\end{figure}

There are various paths forward to improving the $d_e$ exclusion via the optical clock versus optical resonator comparison. The simplest would be to take more data with fewer time gaps, the potential for which is shown as the green curve in Fig.\,\ref{fig:Sr_Si_H} (b). A more fundamental improvement would come from reducing the noise of the optical resonator which is largely limited by the Brownian thermal noise of the mirror coatings. This noise is currently set by the operating temperature of the silicon cavity (124\,K), and the mechanical losses of the dielectric mirror coatings \cite{Matei}. Therefore one could improve the noise by using alternative mirror coatings that have lower mechanical losses at cryogenic temperatures  or by cooling the cavity to even lower temperature of 16\,K or 4\,K \cite{Robinson:21, Robinson:19}.

Improved bounds on the electron mass modulus $d_{m_e}$ were also placed by analyzing the optical resonator versus hydrogen-maser frequency record. The hydrogen maser operates on the magnetically insensitive hyperfine transition of $^1H$, which has the scaling $f_H \propto m_e^2$, whereas the strontium optical clock transition and optical resonator length are both linearly dependent on $m_e$, leading to an overall sensitivity for $f_H/f_c \propto m_e$. The authors placed a bound on $d_{m_e}$ that improved on previous equivalence principle tests for masses between $2\times 10^{-21} - 6\times10^{-19}$\,eV. This new bound is primarily limited by the noise performance of the hydrogen maser, a limit that has been stagnant for decades. Microwave transfer noise was observed at higher frequencies, giving some potential for marginal improvement. Improvements in $d_{m_e}$ from such a comparison would have to utilize a microwave clock that operates on a hyperfine transition with improved noise performance in this frequency band in comparison to a hydrogen maser.

The experiments~\cite{Antypas:2019DM_atom-cavity,Tretiak:2022ndx} search for fast apparent oscillations of the fundamental constants in the frequency range of 20\,kHz to 100\,MHz (corresponding to the UDM mass range $8 \times 10^{-11}\,$eV to $4 \times 10^{-7}$\,eV) using  the Cs D2 transition with $K = 0.26$. 
The relation of the difference between the Cs D2 atomic and optical cavity resonance frequencies, $\delta \nu = \nu_{\rm{Cs}} - \nu_{\rm{cavity}}$, to its average value $\nu$ is described by~\cite{KozlovADP2019} [see Eqs.\,(\ref{K}) and \eqref{free_cavity_length_changes}]:
\begin{equation}
\label{eq:dff2}
\frac{\delta \nu}{\nu} \approx \Big[2.26 h_{\rm{Cs}}(f_{\rm \phi})- h_{\rm{cavity}}(f_{\rm \phi})\Big]\frac{\delta \alpha}{\alpha}+\Big[ h_{\rm{Cs}}(f_{\rm \phi})-h_{\rm{cavity}}(f_{\rm \phi})\Big]\frac{\delta m_e}{m_e} \, ,
\end{equation}%

\noindent where $h_{\rm{Cs}}(f_{\rm \phi})$ and $h_{\rm{cavity}}(f_{\rm \phi})$ are the atomic Cs and optical cavity frequency-response functions that depend on the UDM frequency $f_{\rm \phi} \approx m_\phi c^2 / h$.

\begin{figure}
    \centering
    \includegraphics[width=.6\columnwidth]{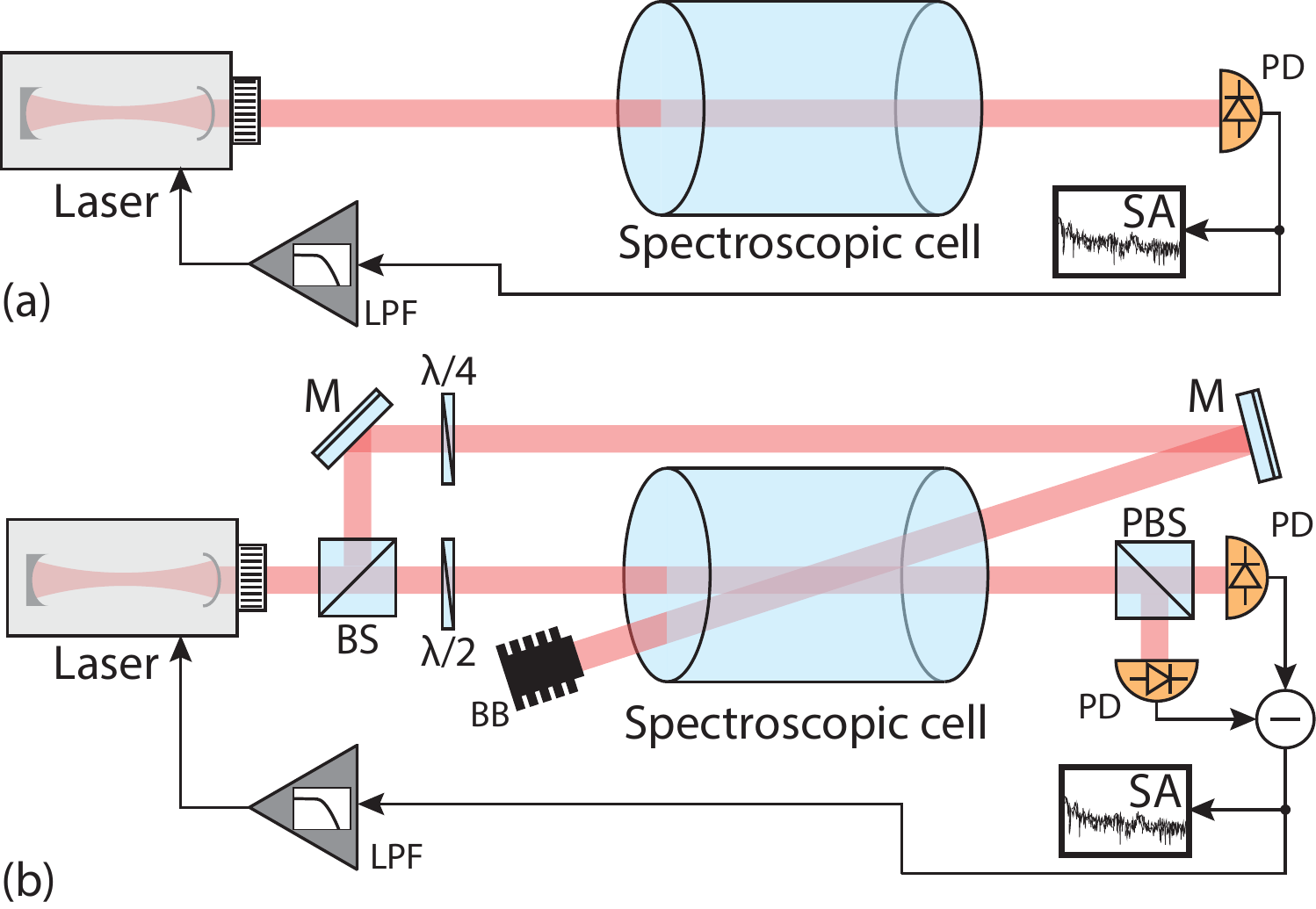}
    \caption{Simplified schematic of the Doppler-broadened (a) and Doppler-free (b) spectroscopy setups for the atomic transition-frequency modulation search. BB -- beam block, BS -- beam splitter, LPF -- electronic feedback loop with low-pass filter,  M -- mirror, PBS -- polarizing beam splitter, PD -- photodetector, SA -- spectrum analyser,   $\lambda/2 $ -- half-wave plate, $\lambda/4 $ -- quarter-wave plate. The internal laser resonator is used as a reference optical cavity. }
    \label{fig:oleg_spectroscopy_setups}
\end{figure}

Typical simplified spectroscopy setups are presented in Fig.~\ref{fig:oleg_spectroscopy_setups}. 
To detect the atomic transition frequency modulation the atomic resonance slope is used as a discriminator.  The spurious peak in the spectrum of the detected photo-current represent the possible candidates of this modulation. The $\nu_{\rm{Cs}}$ is sensitive to fundamental constants oscillations up to observed transition linewidth $\Gamma$.

Using an optical transition for atoms in vapor cells, the linewidth is dominated by the Doppler effect due to the thermal motion of the atoms, see Fig.~\ref{fig:oleg_spectroscopy_setups}\,(a). In this case the limitation is hundreds of MHz.
When desired, it is possible to eliminate Doppler broadening by using nonlinear-spectroscopy techniques employing counterpropagating pump and probe laser beams as shown in Fig.\,\ref{fig:oleg_spectroscopy_setups}\,(b). Then the cut-off frequency is about a few MHz. The advantage of the Doppler-free setups is a sharp spectroscopic features, giving a better sensitivity at low frequencies. In practice, $h_{at}(f_{\rm \phi})$ is determined through apparatus calibration.

The coupling of the reference optical cavity (laser resonator in Fig.~\ref{fig:oleg_spectroscopy_setups}) to oscillations of the fundamental constants is limited by its acoustic cut-off frequency $f_*$. 
Since the spacer between the mirrors of the resonator is a solid body, the maximum speed at which size changes can efficiently propagate is limited by the speed of sound, which is specific to the material of the spacer. 
In the experiments \cite{Antypas:2019DM_atom-cavity,Tretiak:2022ndx}, $f_* \approx 50~\textrm{kHz}$. 
For $f_\phi \gg f_*$, $h_\textrm{cavity} \approx 0$, while for $f_\phi \ll f_*$, $h_\textrm{cavity} \approx 1$. 

Finally, both setups depicted in Fig.~\ref{fig:oleg_spectroscopy_setups} have a feedback loop to keep the cavity centered on the atomic transition over time intervals much longer than the inverse frequency of the UDM oscillations. 
Using a high-performance graphical adaptor for the calculation and averaging of the photo-current spectra allows one to measure the relative variation $\delta \nu/\nu$ below $10^{-17}$ in less then 200 hours \cite{Tretiak:2022ndx}. 
In the absence of an observation of an oscillation of fundamental constants, Eq.\,\eqref{eq:dff2} can be used to place bounds on the interaction parameters $g_{e}$ and $g_{\rm{\gamma}}$ in Eq.\,(\ref{linear_scalar_interactions}), as was done in \cite{Antypas:2019DM_atom-cavity,Tretiak:2022ndx}.

Apart from experiments comparing atomic transition frequencies with those of a cavity, one can also perform similar comparisons with molecular transitions (see, e.g., \cite{2021Hanneke,AntypasQST2021,Oswald:2021vtc}). A key difference with atom-cavity comparisons is that such experiments are (additionally) sensitive to variations in nuclear masses.

\subsubsection{Cavity-cavity comparisons}
\label{cav}
\begin{figure}
    \centering
    \includegraphics[width=\columnwidth]{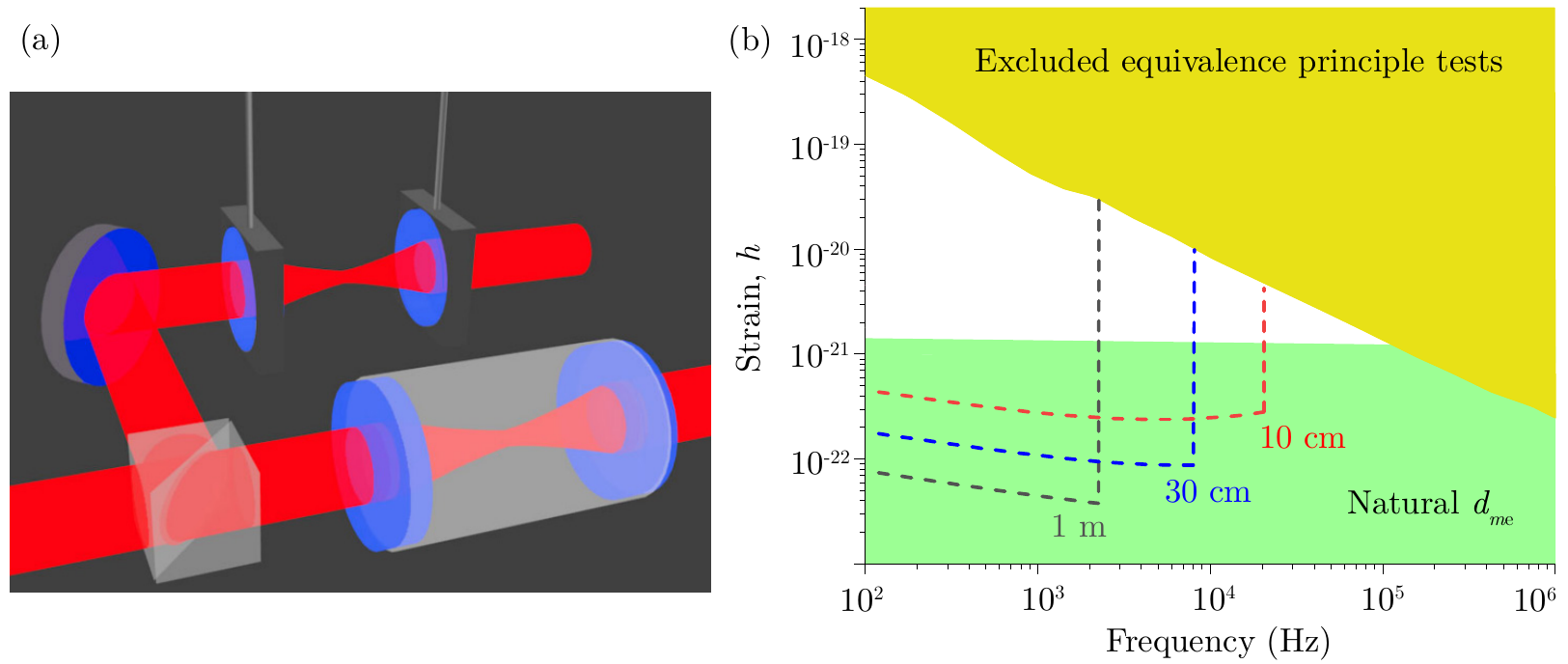}
    \caption{Adapted from Ref.~\cite{Geraci:2019DM-cavity}. (a) Configuration for measuring differential length change between the rigid and non-rigid cavities. (b) Sensitivity of the apparatus in (a) for different cavity lengths. The strain $h \equiv \delta L/L$, which is defined in Eq.\,(\ref{adiabatic_solid_length_change}), is plotted as a function of the DM Compton frequency. The region in yellow is excluded by tests of the equivalence principle. 
    }
    \label{fig:cavity_strain}
\end{figure}

The comparison of a cavity whose length is allowed to vary freely against a cavity whose length is ``fixed'' is sensitive to oscillations in both $\alpha$ and $m_e$ \cite{Stadnik:2016DM-cavity,Geraci:2019DM-cavity}. 
Figure\,\ref{fig:cavity_strain} (a) shows the configuration of the two cavities being compared. One of them has its mirrors 
hanging from a non-rigid suspension and the other one has its mirrors separated by a rigid spacer. 

All rigid objects in the apparatus will oscillate at the DM Compton frequency $f_{\phi}$. This includes the spacer of the rigid cavity and the platform from which the non-rigid cavity is suspended. 
However, DM-induced oscillations will tend to  
be suppressed in the non-rigid cavity due to the vibration isolation provided by the suspension. With a carefully designed vibration-isolation system, the change in the non-rigid cavity length can be suppressed by almost 10 orders of magnitude compared to its rigid counterpart. Moreover, any DM-induced changes to the interrogating laser frequency as a result of the laser-cavity length fluctuations will be cancelled in the common mode.  To reduce background technical noise, both cavities can be isolated from seismic and other external vibrations in the science frequency band ($100$~Hz to $10$~kHz). 

Figure\,\ref{fig:cavity_strain} (b) shows the projected ``strain'' [$h \equiv \delta L/L$ according to Eq.\,(\ref{adiabatic_solid_length_change})], as a function of $f_{\phi}$, in the apparatus for optical cavities of different lengths for a room-temperature experiment using state-of-the-art mirror coatings and substrates \cite{Geraci:2019DM-cavity}. These levels of (spacetime) strain sensitivity have already been demonstrated in large-scale gravitational wave detectors like LIGO, albeit at large length scales (e.g. 4\,km). 
With the sensitivity primarily expected to be thermal noise limited, cryogenic operation in a 4\,K environment could yield an additional order-of-magnitude improvement.

\subsection{Optical interferometers (including gravitational-wave detectors)}
\label{Sec:LIFO-GW_detectors}

Optical interferometers are directly sensitive to scalar and vector ultralight DM. 
In the case of a Michelson-type interferometer with equal arm lengths and identical end mirrors, scalar UDM primarily imprints its effects via the geometric asymmetry created by the central beam-splitter \cite{Grote:2019DM-LIFO}. 
Changes in the optical-path-length difference between the two arms of the interferometer arise due to oscillations in the thickness of the beam-splitter, $l$, as well as oscillations in the refractive index of the beam-splitter material, $n$: $\delta (L_x - L_y) \sim \delta (nl)$. 
For typical laser frequencies and common dielectric materials, changes in $n$ are generally sub-dominant compared to adiabatic changes in $l$ \cite{Savalle:2019DAMNED,Grote:2019DM-LIFO}, but can become dominant at frequencies far above the fundamental frequency associated with the relevant mechanical resonance of the material \cite{Grote:2019DM-LIFO}. 
Searches for scalar UDM with Michelson interferometers have been recently performed using older datasets from GEO600 \cite{Vermeulen:2021DM-GEO600} and the Fermilab holometer \cite{Aiello:2021DM-holometer}; see Figs.~\ref{fig:scalar_de_plot} and \ref{fig:scalar_dme_plot} for limits on the linear scalar-electron interaction. 
Small-scale Michelson interferometers (such as the Fermilab holometer) operating in the resonant narrowband regime can deliver a significant improvement in sensitivity to scalar UDM due to up to a $\sim 10^6$ enhancement in the UDM signal \cite{Grote:2019DM-LIFO}. 
In a Fabry-Perot-Michelson-type interferometer (Michelson-type interferometer with Fabry-Perot resonators in the two arms), the sensitivity to scalar UDM (in units of arm-length strain) is generally suppressed compared to a Michelson interferometer, but the sensitivity can be improved through the use of freely-suspended Fabry-Perot arm mirrors with different thicknesses in the two arms \cite{Grote:2019DM-LIFO}. 
Another approach involves using a three-arm Mach-Zender-type interferometer with unequal arm lengths to perform time-delay comparisons \cite{Savalle:2019DAMNED}. 
A recent search for ultralight scalar DM using such an approach was performed in the DAMNED experiment \cite{Savalle:2021DAMNED}; see Figs.~\ref{fig:scalar_de_plot} and \ref{fig:scalar_dme_plot} for limits on the linear scalar-electron interaction.

Ultralight DM may also exert time-varying forces on test bodies \cite{graham2016dark,Hees:2018DM_EP}, with $\boldsymbol{F} \propto \boldsymbol{\nabla} \phi$ in the case of ultralight scalar DM and $\boldsymbol{F} \propto \partial_t \boldsymbol{A}'$ in the case of ultralight vector DM. 
In the case of scalar UDM, such forces are suppressed, since the only vector quantity associated with a spinless DM field is its momentum, which results in a velocity suppression factor of order $v/c \sim 10^{-3}$; in the case of the electromagnetic and electron couplings, the force is further suppressed by the smallness of the electromagnetic and electron-mass contributions to the overall mass of a test body. 
On the other hand, there are no such suppression factors for vector UDM. 

In a two-arm interferometer with compositionally-identical 
test masses (including the beam-splitter and various mirrors), the common-mode suppression of forces exerted on different test masses is partially lifted due to motional gradients associated with the DM field \cite{Grote:2019DM-LIFO,Pierce:2018DM-LIFO_vector} and finite photon propagation speed through the interferometer \cite{Michimura:2021DM-LIFO_propagation}. 
The DM signal can further be boosted by using test masses of different materials in the interferometer \cite{Grote:2019DM-LIFO,Michimura:2020DM-LIFO_materials}. 
A search for vector UDM using data from LIGO and VIRGO's third observing run has recently been performed \cite{LIGO:2021DM_vector}; see Fig.~\ref{fig:vector_B-L_plot} for limits on the $B-L$ vector coupling. 
For related analyses using data from LIGO's first observing run, we refer the reader to Refs.~\cite{Michimura:2021DM-LIFO_propagation,LIGO:2019DM_vector}.

\subsection{Torsion balances}
\label{Sec:3_TB}
Torsion balances are exquisite sensors for differential forces on macroscopic test masses. The torsion balance instruments built at the University of Washington (UW) can resolve differential accelerations perpendicular to the torsion fiber that are as small as $10^{-15}$ m/s$^2$ on $\sim10-20$ g objects, after integrating over one day of data taking. Such sensitivity is sufficient to set limits on UDM over a wide range of DM masses \cite{graham2016dark}.
In static torsion-balance experiments that were designed to search for equivalence principle violations, one can derive limits by assuming that UDM bosonic particles create a differential force. 
For UDM masses $m_\phi<10^{-13}$~eV 
these experiments predominantly use the Earth as the source for such interactions, since at the latitude of the experiment 0.17\% more of the Earth's mass lies to the north than to the south. Because the noise in most precision measurements, including torsion balances, has 1/$f$-character the signal must be modulated; higher modulation frequencies lead to improvements in statistical noise. The team at UW pioneered a technology where the entire torsion balance instrument is placed on a continuously-rotating turntable to achieve a sinusoidal signal modulation at about two cycles per hour, with the noise content of the torsion balance being limited by the KT-torque noise in the torsion fiber.
These measurements set some of the strictest limits on equivalence principle violations and thereby competitive limits on scalar (e.g., Higgs-portal, or electron-mass coupled) and vector (e.g., B, L, and B-L coupled) UDM \cite{EotWash:2008, Wagner_2012_torsion, shaw2022torsion}. While the continuous rotation results in substantial noise improvement, systematic effects that can occur at the rotation frequency have to be well understood and ultimately dominate the uncertainty budget. New developments with fused quartz torsion fibers that have much lower KT-torque noise and techniques to reduce systematic uncertainty promise up to a factor of $5$ improvement in sensitivity.

Torsion-balance systems are also sensitive to time-varying UDM fields \cite{graham2016dark,shaw2022torsion}. Here the signal arises purely from the bosonic field and the signal modulation arises naturally because the field strength oscillates, while the torsion balance apparatus remains stationary, i.e., the latter does not need to be rotated.  
Such a measurement has much smaller systematic uncertainty since temperature effects, gravity gradients, etc., do not vary at the Compton frequency. Systematic effects at, or near, the daily frequency are averaged over as the direction of the signal is space-fixed. Therefore these measurements are only limited in practice by the thermal noise in the torsion fiber. The balances are run with fused quartz torsion fibers that deliver $Q$-values which can be as high as $10^6$.  The data analysis is more complex since the search for a signal must be performed along the three principal axes of the rest-frame of the fixed stars. 

Figure\,\ref{fig:vector_B-L_plot} includes preliminary limits on time-varying UDM $B-L$ coupling, as well as limits derived from torsion-balance equivalence-principle tests \cite{shaw2022torsion} and from the Microscope satellite \cite{MICROSCOPE:2017}. 
Torsion-balance experiments that are up to a factor of 10 more sensitive are in preparation at UW.  

\subsection{Mechanical resonators}  
\label{Sec:3_MR}
Both torsion balances and optical-cavity-based UDM searches involve the measurement of the deformation of an elastic body (the mechanical response) produced by a weak force. If the body has internal resonances at the UDM Compton frequency, the deformation can be amplified by a factor as large as the resonance $Q$ factor. This effect can yield massive sensitivity enhancement over a narrow bandwidth. 

A variety of proposals have been put forth in the last several years for mechanical resonator-based UDM detection, as summarized by Carney \emph{et al.}~\cite{carney2021mechanical}.  A key motivation is the diversity of well-studied nanoscale to centimeter-scale mechanical resonators in the field of cavity optomechanics \cite{aspelmeyer2014cavity}, spanning a range of frequencies from 1\,Hz to 1\,GHz, corresponding to a UDM mass from $10^{-14}$\,eV to $10^{-5}$\,eV.  These resonators can have $Q$ factors as high as $10^{10}$~\cite{maccabe2020nano,galliou2013extremely}, and can be read out at the thermal noise limit even in deep cryogenic environments~\cite{hauer2018two,Goryachev2014}. Development of ultra-sensitive mechanical UDM sensors is well-motivated, as these devices have also been proposed or already utilized in searches for new physics beyond UDM, including both light \cite{Afek:2021vjy} and heavy \cite{Montiero2020,Carney:2019pza} particle dark matter, dark energy \cite{Rider2016, Betz2022}, neutrinos \cite{smith1983coherent,domcke2017detection}, tests of quantum gravity \cite{Bushev2019,Bourhill2015}, and high-frequency gravitational waves \cite{goryachev2021,aggarwal2021challenges}. Several experimental searches for UDM have already produced preliminary results, based on levitated optomechanical systems \cite{Montiero2020}, cryogenic bulk acoustic mode resonators \cite{Campbell:2021DM_atom-cavity}, and re-purposing of gravitational wave detectors \cite{branca2017search}. Below we provide a brief summary of major proposed and nascent searches.

\begin{figure}
    \centering
    \includegraphics[width=1\columnwidth]{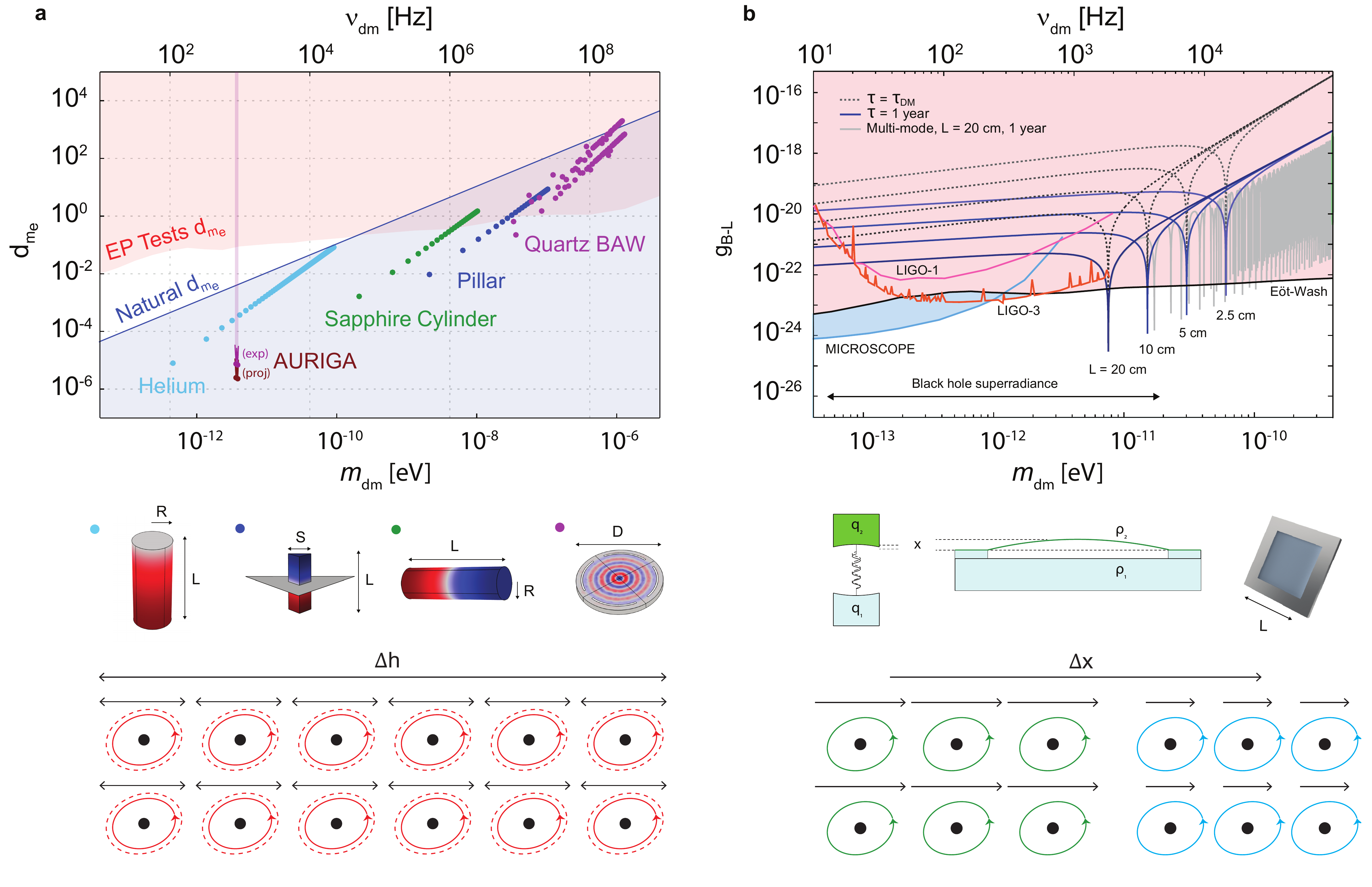}
    \caption{\textbf{Mechanical resonator based ultralight dark matter detection.} (a) Adapted from Ref.~\cite{manley2020searching}. Proposals to search for scalar UDM with breathing-mode mechanical resonators \cite{manley2020searching}, including a $^4$He bar resonator (light blue)\cite{de2017ultra}, a micropillar resonator (dark blue) \cite{neuhaus2016cooling}, a sapphire cylindrical test mass (green) \cite{rowan2000investigation}, and bulk acoustic wave resonators (BAW, purple)\cite{galliou2013extremely,Goryachev2013,Arvanitaki:2016DM-resonator}. (b) Adapted from Ref. \cite{manley2021searching}. Proposal to search for vector UDM with center-of-mass-mode mechanical resonators \cite{manley2021searching}: specifically, a silicon nitride membrane read out with an Fabry-P\'{e}rot interferometer. Below: Illustration of mechanical effects produced by scalar (left) and vector (right) UDM; specifically a homogeneous strain and a material dependent differential acceleration, respectively.
    }
    \label{fig:MechanicalResonatorExperiments}
\end{figure}

\textbf{Scalar UDM} -- In the Hz to GHz UDM frequency range, the E\"{o}t-Wash torsion balance EP tests have set strong constraints on scalar UDM coupling strength, imposing stringent requirements on the strain sensitivity of next generation mechanical detectors: 
\begin{subequations}
\begin{equation}
    d_e\lesssim 10^{-3}\left(\frac {f}{1\text{ Hz}}\right)^{1/4}  \,\,\,\, 
    \rightarrow \,\,\,\,\,
    \frac{\delta L}{L_0} \lesssim 10^{-21} \left(\frac{1 \text{ kHz}}{f}\right)^{3/4}
\end{equation}
\begin{equation}
    d_{m_e}\lesssim 10^{-2}\left(\frac {f}{1\text{ Hz}}\right)^{1/4} \,\,\, 
    \rightarrow   \,\,\,\,\,
    \frac{\delta L}{L_0} \lesssim 10^{-20} \left(\frac{1 \text{ kHz}}{f}\right)^{3/4}.
\end{equation}
\end{subequations}
Such small strain signals can be resonantly enhanced by Weber-bar-type detectors with longitudinal acoustic modes, enabling laboratory-scale detectors to probe for scalar UDM below the E\"{o}t-Wash torsion constraints \cite{Arvanitaki:2016DM-resonator,manley2020searching}. Of the experiments included in Figs. \ref{fig:scalar_de_plot} and \ref{fig:scalar_dme_plot}, AURIGA, DAMNED, and the Holometer make use of acoustic modes to resonantly amplify the UDM-induced strain signal and achieve the strongest constraints around their mechanical resonance frequencies. AURIGA is a resonant mass GW detector, whose data has been reanalysed to perform a retroactive search for scalar UDM and set some of the strongest constraints at $\sim 1$ kHz \cite{branca2017search}. In the DAMNED experiment \cite{Savalle:2021DAMNED}, an ultrastable optical cavity behaves effectively as a multimode resonant mass detector, where broadband readout is accomplished with an optical fiber interferometer. The Holometer, a Michelson interferometer, achieves its best UDM-strain sensitivity at the mechanical resonance frequency of its beamsplitter \cite{Aiello:2021DM-holometer}. 

A disadvantage of resonant mass detectors is that the high-$Q$ resonant enhancement goes hand-in-hand with a limitation on the bandwidth. To address bandwidth limitations, future resonant mass detectors for scalar UDM are seeking frequency tunability and array-based detection. Superfluid helium resonators are an attractive option \cite{vadakkumbatt2021prototype}, as their resonance frequencies can be tuned by up to 50\% via pressurization \cite{abraham1970velocity, souris2017tuning}.  As discussed in \cite{manley2020searching}, a variety of established table-top cavity optomechanical experiments may also be co-opted into scalar UDM searches (Fig.~\ref{fig:MechanicalResonatorExperiments} (a)). Quartz bulk-acoustic-wave resonators, for example, can have exceptionally high $Q$-$m$ products\footnote{Specifically, we refer to the product of the $Q$ factor and effective mass of the resonator mode, $m$, which characterizes the thermal force sensitivity.} at dilution refrigerator temperatures, can be read out with high precision using superconducting microwave cavities, and support overtones which enable ``xylophone"-type multi-mode resonant detection in the relatively unexplored parameter space between 1 MHz and 1 GHz~\cite{goryachev2014gravitational,goryachev2021,Arvanitaki:2016DM-resonator}.


\textbf{Vector UDM} --
Vector dark matter would produce an oscillating center-of-mass (CoM) acceleration of a test mass dependent on its material composition \cite{graham2016dark}. A typical vector model which is used as a benchmark comes from a gauged $B-L$ symmetry, as described in Section~\ref{theory}. If dark matter consists primarily of this field, it would accelerate free-falling bodies in proportion to the their neutron to nucleon ratio,
\begin{equation}
a_{\rm DM}(t) \approx g_{B-L} a_0 \frac{A-Z}{A}\sin(m_{\phi} t),
\label{eq:DMacceleration}
\end{equation}
where $A$ is the mass number, $Z$ is the atomic number, $g_{B-L}$ is the dimensionless coupling strength, and $a_0\propto \sqrt{\rho_\text{DM}}$\footnote{Explicitly, $a_0 =\sqrt{2\frac{e^2 \rho_{\text{DM}}}{{m_{\rm n}}^2 \epsilon_0}} \approx 3.7 \times 10^{11}~{\rm m/s^2}$, where $m_n$ is the neutron mass.}. From E\"{o}t-Wash constraints in the Hz to GHz range, the amplitude of the UDM-induced acceleration has an upper bound of approximately
\begin{equation}
    g_{B-L}a_0 \lesssim 10^{-11} \left(\frac{f}{1 \text{ kHz}}\right)^{1/4} \frac{\text{m}}{{\rm s}^2} .
\end{equation}
Equation  (\ref{eq:DMacceleration}) describes a material-dependent acceleration signal, as the neutron-nucleon ratio $\left(A-Z\right)/A$ is generally a material-dependent quantity. The experimentally accessible signal is a differential acceleration between two bodies, which would take the form $a_{\rm diff}(t) \approx g_{B-L} \Delta a_0 \sin(m_{\phi} t)$, where $\Delta = Z_1/A_1 - Z_2/A_2$ is the difference in the neutron-nucleon ratio of the bodies. 

The differential acceleration $a_\text{diff}(t)$ describes a force capable of deforming compositionally inhomogeneous elastic bodies. Resonant enhancement of this deformation can be achieved by the simple mechanical dimer arrangement shown in Fig. \ref{fig:MechanicalResonatorExperiments} (b), in which two masses made of different materials are bound by a spring.  The two masses may be fashioned into mirrors, forming an optical cavity for displacement-based readout.  This arrangement can be generalized to a cavity optomechanical system in which a mechanical resonator is coupled to an optical cavity made of a different material \cite{manley2021searching}.    

Contemporary cavity optomechanical systems offer a diversity of platforms for vector UDM detection based on the heterogeneous dimer model.  Manley \emph{et al.} \cite{manley2021searching} proposed a specific design based on a membrane optomechanical system that features ultra-high mechanical $Q$, frequency tuning via stress, quantum-limited cavity-based readout, and compatibility deep cryogenics.  Promising alternative platforms include high $Q$-$m$ levitated microspheres and optically trapped gram-scale mirrors, as considered in \cite{carney2021ultralight}. Arrays of membrane optomechanical system are also under exploration and construction to enhance the sensitivity to UDM~\cite{Windchime:2022whs}.

Levitated optomechanical systems, as high-sensitivity force \cite{Ranjit2016} and acceleration \cite{Moore2017} sensors, present another promising route towards high-sensitivity detection of scalar and vector UDM, as well as heavier-mass DM \cite{Moore_2021,Montiero2020, afek2021millicharges}. Recent constraints for a class of composite dark matter models with feeble, long-range interactions with normal matter are provided in Ref. \cite{Montiero2020}. Levitated optomechanical sensors have also been identified as a means to search for high-frequency gravitational waves \cite{ArvanitakiGeraci2013,Aggarwal2020}, where several sources may be dark matter related. The Levitated Sensor Detector (LSD), a  compact resonant high-frequency gravitational wave (GW) detector, is under construction \cite{Aggarwal2020}. Particularly well-motivated sources in this frequency range are gravitationally-bound states of UDM with decay constants near the grand unified theory scale that form through black hole superradiance and annihilate to GWs.

\subsection{ALP detectors constraining dark photons}
\label{sec:ALP}

Searches for dark photons which kinetically mix with the SM photon can benefit from many of the same technologies as searches for axions and ALPs. In the case of the axion, the local oscillating dark matter field sources an oscillating effective current in the presence of static magnetic field. The experimental signatures of dark photon-SM photon mixing lead in some cases to identical signals to axions, except that they are present even in the absence of a $B$-field. In principle, this means that dark photons can be searched for more readily than axions, and occasionally in the same experiment at the same time. Dark photon measurements can also be performed in new axion detectors before the magnetic field is turned on so that the detector can be tested without the added complications caused by the magnetic field. For example, in some proposed single-photon detection based axion searches, superconducting qubits are used as the single photon detectors; however, large magnetic fields are deleterious for superconductivity so it is useful to have a proof-of-concept measurement that also yields novel DM constraints~\cite{Dixit2020single}. This idea therefore motivates the use of many experiments as dual-purpose axion and dark photon detectors. However, the signals are not all identical, and there are several important considerations that should be made to ensure that experiments are sensitive to both particles at once.

There have been several dedicated searches for ultralight dark photon-SM photon kinetic mixing over the last few years. One concept with sensitivity across the MHz--THz range is the Dark $E$-field Radio~\cite{Godfrey:2021tvs}, which consists of an antenna placed inside a shielded room much larger than the dark matter Compton wavelength. Another popular strategy is to use broadband reflector-based experiments~\cite{Horns:2012jf,Jaeckel:2013sqa} which are suitable for searches over several different mass ranges~\cite{Andrianavalomahefa:2020ucg,Suzuki:2015sza,Knirck:2018ojz,Tomita:2020usq}. Constraints from dedicated searches for dark photons in haloscopes that are \emph{not} designed with magnetic fields are shown in red in Fig.\,\ref{fig:DarkPhotonBounds}. Searches will continue across a much wider mass range in the near future thanks to the proliferation of many novel proposals to search for axions and ALPs. In particular, the sensitivity required to reach QCD axions should allow future constraints to supersede competing bounds from cosmology and stellar cooling by several orders of magnitude; see the blue and green regions of Fig.\,\ref{fig:DarkPhotonBounds}, respectively. 

In addition to the LC circuits like DM-Radio described in 
Section~\ref{Sec:3_LC}, another notable example of a future dual axion/dark photon experiment is MADMAX~\cite{TheMADMAXWorkingGroup:2016hpc} which is currently being prototyped at DESY. MADMAX is an elaboration on the original dish-antenna concept that circumvents the volume restrictions of cavities at high frequencies by constructively interfering axion or dark-photon-induced radiation at a series of precisely spaced dielectric disks. The more recently proposed broadband reflector experiment, BREAD~\cite{BREAD:2021tpx}, is another re-imagining of the dish antenna, where the reflecting dish would be constructed with a specialised cone-like design so that the THz photons reflecting off of it would all be focused down to a small region containing a sensor or antenna. LAMPOST~\cite{Baryakhtar:2018doz,Chiles:2021gxk} is a higher-frequency dielectric haloscope with single-photon-detector readout allowing for sensitivity in the sub-eV mass range, with demonstrated constraints on dark photons~\cite{Chiles:2021gxk}. Finally, ALPHA, is a planned `plasma haloscope'~\cite{Lawson:2019brd,Gelmini:2020kcu} whose sensitivity relies on resonant conversion when the dark matter mass matches the plasma frequency of a thin wire metamaterial. DM-Radio, MADMAX, BREAD, LAMPOST, and ALPHA are all experiments that can set stringent limits on dark photons over a wide range of masses, even in the absence of a magnetic field. Their projections are displayed in Fig.\,\ref{fig:DarkPhotonBounds}. Future runs of haloscopes using cavity resonances~\cite{McAllister:2017lkb,Stern:2016bbw,Melcon:2018dba,Alesini:2017ifp,Jeong:2017hqs} will also be able set limits in the 1--100 $\upmu$eV range in the near future.

Due to the similarity between their signals, even previous runs of cavity-based axion haloscopes can be reinterpreted in the context of dark photons, as was done in Refs.~\cite{Arias:2012az,Ghosh2021searching,Caputo2021dark}. These bounds typically require no additional data analysis, however due to the subtle differences between the two particles, they are not necessarily as accurate, or as well-optimised, as they could be. For instance, a notable caveat applies to experiments that use the fact that the axion signal power depends on the presence of a $B$-field as a basis upon which to veto potential signals. Since the dark photon signal power would not be sensitive to the $B$-field at all, a positive detection could be discarded as noise if a candidate was interrogated this way. This prohibits the retroactive recasting of certain historical haloscope data, however it does not preclude the use of $B$-field veto entirely. Instead, this issue should encourage experiments to test signals with multiple techniques, as is already undertaken in collaborations like ADMX~\cite{Braine:2019fqb}. Since magnetic fields are not required for dark photon searches, the lack of geometric limitations associated with magnetic bores allows for distinct experimental configurations compared to axion-focused setups, as discussed for example in Ref.\,\cite{Gelmini:2020kcu}.

Another subtle issue that arises when searching for dark photons is the possibility for nontrivial time-dependence. Any dark photon signal is inherently directional as a result of its polarization. Many cavities, for instance, would measure an induced electric field strength that depends on the cosine of the angle between the dark photon polarization and their $B$-field. As a result, if even a subpopulation of dark matter is coherently polarized over a long period of time, then the rotation of the Earth would generate a daily modulation in the signal power with a period of one sidereal day, in addition to modulation originating from the coherence length of the UDM~\cite{Knirck:2018knd}.

The time dependence of the dark photon signal is one way in which reinterpretation of data originally taken and analysed in the context of an axion signal could be fraught with ambiguity. Central to this ambiguity is the fact that the dark photon polarization state in our local galactic region is unknown. This is a fundamental theoretical uncertainty, specific to vector DM, that does not have a clear answer at present. It has also received little attention in the literature, as it does not appear to impact the dark photon's viability as a dark matter candidate, or even any other observables apart from direct detection signals. At the extreme end, some production mechanisms like the simplest case of the misalignment mechanism~\cite{Arias:2012az, AlonsoAlvarez:2019cgw}, would imply that a fixed polarization is generated for the dark photon field within our cosmological horizon. This scenario would entail a single preferred direction over entire measurement campaigns. On the other hand, altogether different scenarios, including those in which dark photons are generated by the decay of topological defects~\cite{Long:2019lwl}, suggest a more randomised distribution of polarizations. The extreme case from the aforementioned fixed-polarization scenario would be the case when an effectively-random polarization vector were drawn in every coherence time. To make the situation even less clear, as discussed in Ref.~\cite{Caputo2021dark}, gravity may also play a role in precession of the polarization. Dedicated numerical simulations will likely be needed to assess the effect of structure formation on this important unresolved issue.

The randomized-polarization scenario is the simplest to set bounds on, however this is not the most conservative approach. In contrast, the \emph{fixed}-polarization scenario generally leads to a suppression in sensitivity overall, since there are certain experiments that could be `blind' to the dark photon if their orientations happened to be badly aligned during a measurement. Ref.~\cite{Caputo2021dark} suggested a statistical methodology for circumventing these issues, by essentially marginalising over the uncertainty. Unfortunately, applying the techniques as written is not always possible to do accurately for several past experiments, since precise timing data is usually not made public. Nevertheless an approximate recasting can still be done, as shown by the purple regions in Fig.~\ref{fig:DarkPhotonBounds}. This exercise can also be reversed to ask the question: what changes to axion/dark photon measurement campaigns should be made to maximise this sensitivity or, in other words, minimize the impact of this marginalization. Some of these recommendations are relatively simple, like altering the orientation of an antenna. However, most remarkably, with a careful planning of the data-taking to align with certain periods of the day, an experiment could boost sensitivity by around an order of magnitude. Since these changes would not alter the sensitivity to axions, these strategies motivate the planning of using haloscopes to act as dual-purpose axion/dark photon detectors in the future.

\subsection{Spin-based sensors: magnetometers and comagnetometers} 
\label{Sec:3_MN}

As noted in the previous section, techniques used to search for axions and axionlike particles (ALPs) can also be applied to search for vector bosons. One way in which signals can manifest in experimental searches is through a coupling between the spin-0 or spin-1 boson and the intrinsic spin of elementary particles. In the case of axions and ALPs, for example, interactions with standard model fermions can result from the Lagrangian density given by the coupling of the spacetime derivative of the ALP field $a$ to fermion axial-vector currents,
\begin{align}
{\mathcal L} \propto \frac{1}{f_{ai}}\partial_\mu a \times \bar \psi_{i} \gamma_\mu\gamma_5 \psi_{i}~,
\label{Eq:lin-ALP-spin-Lagrangian}
\end{align}
where $f_{ai}$ parameterizes the axion-fermion coupling strength, $\psi_i$ represents the field of fermion $i$, and $\gamma_\mu$ and $\gamma_5$ are Dirac matrices. This results in a non-relativistic Hamiltonian analogous to the Zeeman Hamiltonian governing magnetic interactions with spins $\boldsymbol{S}_i$ \cite{Safronova:2018RMP}:
\begin{align}
{\mathcal H} = -\frac{2\left(\hbar c\right)^{3/2}}{f_{ai}} \boldsymbol{S}_i \cdot {\boldsymbol{\nabla}} a(\boldsymbol{r},t)~.
\label{Eq:lin-ALP-spin-Hamiltonian}
\end{align}
In principle, a similar coupling between exotic vector fields $\boldsymbol{\chi}$ and spins could also exist \cite{garcon2019constraints}, ${\mathcal H} \propto \boldsymbol{\chi}\cdot\boldsymbol{S}_i$. In such cases, spin-based quantum sensors, such as atomic magnetometers \cite{budker2007optical,Bud13} and comagnetometers (see, for example, Refs.~\cite{rosenberry2001atomic,Kor02,Kor05,wu2018nuclear,wang2020single}), can be used to search for interactions with axions, ALPs, and vector bosons \cite{Safronova:2018RMP,graham2018spin,terrano2021comagnetometer}. 

Exchange of light spin-0 and spin-1 bosons between standard model particles generate long-range potentials \cite{moody1984new,dobrescu2006spin,fadeev2019revisiting} that can be searched for in experiments with spin-based sensors. A wide variety of experiments have been conducted to look for such effects (see, for example, Refs.~\cite{venema1992search,youdin1996limits,Vas09,hunter2013using,ledbetter2013constraints,terrano2015short,kotler2015constraints,Kim17GDM,lee2018improved,ding2020constraints}). 

A number of searches for ultralight bosonic dark matter, such as the Cosmic Axion Spin Precession Experiments (CASPEr), have been carried out using individual spin-based quantum sensors \cite{abel2017search,terrano2019constraints,wu2019search,garcon2019constraints,bloch2020axion,Crescini2020QUAX,aybas2021search,jiang2021search,Bloch2022ALPs}. Another modality is to employ a network of spin-based quantum sensors \cite{Pos13}. This is the strategy of the Global Network of Optical Magnetometers for Exotic physics searches (GNOME) \cite{Pus13}, an array of more than a dozen optical atomic magnetometers operating within magnetically-shielded environments located at stations all over the world \cite{afach2018characterization}. By analyzing the correlation between the signals from multiple, geographically-separated magnetometers, GNOME can be used to search for transient spin-dependent interactions that might arise, for example, if Earth passes through a compact, composite dark-matter object made up of ultralight bosons, such as a boson star \cite{Kim18AxionStars} or topological defect \cite{Pos13,afach2021search} (see Section~\ref{sec2}). While a single atomic-magnetometer system could in principle detect such transient events, in practice it is difficult to confidently distinguish a true signal heralding new physics from ``false positives'' induced by occasional abrupt changes of magnetometer operational conditions or local noise sources. The geographically distributed array of GNOME magnetometers enables consistency checks based on the relative timing and amplitudes of signals, enabling vetoing of false-positive events and suppressing uncorrelated noise \cite{Mas20}. GNOME can also be used to search for correlated signals from stochastic fluctuations of bosonic dark matter fields \cite{mas2022intensity} and bursts of exotic fields emanating from cataclysmic astrophysical events \cite{dailey2021quantum}. Importantly, correlated network searches offer the possibility to hunt for the unexpected.

Another interesting scenario is the case of kinetically-mixed hidden-photon dark matter. It turns out that the Earth itself can act as a transducer to convert hidden photon dark matter into a monochromatic oscillating magnetic field at the surface of the Earth \cite{fedderke2021earth}. The induced magnetic field from the hidden photons has a characteristic global vectorial pattern that can be searched for with unshielded magnetometers dispersed over the surface of the Earth. GNOME is insensitive to such kinetically-mixed hidden-photon dark matter because of the magnetic shields enclosing the magnetometers \cite{Chaudhuri2015Radio,Kim16}. Instead, a network of unshielded magnetometers is required. Data from the SuperMAG network used for geophysical field measurements has been used to place constraints on hidden photon dark matter with masses between $2 \times 10^{-18}$~eV and $7 \times 10^{-17}$~eV \cite{fedderke2021search}, and a dedicated unshielded magnetometer network targeting hidden photon dark matter may be able to extend the probed parameter space.

\subsection{Other ideas and searches for scalar and vector UDM} 
\label{Sec:3_Other}

\subsubsection{Rydberg atoms and superconducting transmons as qubit-based single photon detection}
\label{Sec:3_RA}
As discussed in Sec.\,\ref{sec:ALP}, haloscope axion detectors can be re-purposed to search for vector bosons. A typical haloscope configuration contains both a tunable microwave cavity to capture the photons generated by a dark matter signal and a low-noise amplifier to bring the tiny power fluctuations out of the cavity and into a detectable signal. 
Experiments looking for dark matter particles that interact with standard model photons above 10 GHz would require resonant microwave cavities smaller than 15 mm in each dimension. Because the signal power scales inversely with the fourth power of the frequency \cite{Sikivie1998},
higher frequency searches result in a decrease to the signal power induced by the dark matter signal. Additionally, cavity $Q$-factors decrease with increasing frequency due to the anomalous skin effect, resulting in a further reduction to the signal power. Measuring the small resulting signals requires reduction of system noise. However, efforts to reduce this noise in quadrature measurements inevitably meet a lower limit that stems from the Heisenberg uncertainty principle \cite{Caves:1981hw, lamoreaux2013analysis} and the unavoidable presence of quantum noise. 

Single-photon detection offers a promising way to eliminate quantum noise from the detection of low-mass dark matter candidates \cite{lamoreaux2013analysis}. 
In the case of single photon detection, the ability to measure the amplitude of a dark matter signal is limited by the shot noise on the number of detected background photons. The use of single photon detection offers a significant reduction in the detected noise power. Single-photon detection can be achieved through the use of two-level quantum systems, i.e., qubits, such as superconducting transmons and Rydberg atoms.\footnote{Two-level systems used in quantum information processing are referred to as qubits (quantum bits).}

In an attempt towards a single-photon-detection-based axion experiment, CARRACK successfully demonstrated the power of the Rydberg technique by measuring the blackbody spectrum of their system using Rydberg atoms {\it in situ} at cryogenic temperatures \cite{carrackiibbpaper}. One important feature of these systems is their tunability: the Rydberg atom transition frequency can be Zeeman shifted using a magnetic field $\sim10$\,mT; the resonant frequency of a transmon can be tuned with a bias current; and cavities can be tuned both mechanically and electrically.

Detecting a single photon with a high confidence requires the ability to monitor it repeatedly without destroying it. This can be achieved via a sequence of so-called quantum non-demolition (QND) photon-number measurements. One way to realize such a QND single-photon detection exploits off-resonant Rabi oscillations whereby the energy initially carried out by the photon oscillates repeatedly between the cavity field and a qubit whose transition frequency is far off-resonance relative to the photon frequency. This results in a rotation of the phase of the cavity field that can be then measured, providing an unambiguous signature of the presence of the single photon. 

\subsubsection{Molecular absorption DM detectors}

Atoms and molecules have been considered as a means to search for interactions with axions, for example through atomic and molecular transitions being induced from oscillating nuclear moments \cite{flambaum2020atomic}, via resonant production and absorption of axions with atoms \cite{flambaum2018resonant,tan2019interference}, or coherent axion-photon transformations in forward scattering on atoms \cite{flambaum2018coherent}. Similarly, bosonic DM with $\sim$eV mass could have a measurable cross-section for interaction with an ensemble of molecular absorbers. In fact, a molecular-absorption-based bosonic dark matter detection experiment is projected to be the most sensitive technique for searching for bosonic dark matter candidates in the eV mass range \cite{2018UDMwithMolecules}.
After molecular excitation by the dark matter wave, decay to the molecular ground state will typically result in the emission of a single real photon.  By tuning the molecular transition (e.g., using the Stark effect) to match the resonant absorption DM frequency, such photons can be distinguished from the background.  Maximizing the signal-to-noise ratio in the experiment requires large sample volume (high event rate), high single photon detection efficiency, and low dark-count rate (i.e. low detector area). 

\subsubsection{Quantum materials}

Solid-state quantum materials offer a promising target for UDM absorption in the meV-eV range \cite{Hochberg_et_al:2016,Bloch:2016sjj}, particularly the absorption of kinematically mixed dark photons and pseudoscalar dark matter. Low-energy quasiparticle excitations such as electrons and phonons in solid-state materials can both have good kinematic match with UDM, and have the advantage of `Avogadro scaling' with a large target volume owing to the large number of atoms in a crystal~\cite{Essig_et_al:2012}. Quantum materials, therefore, provide an excellent opportunity as dark photon and pseudoscalar absorbers, owing to the $\sim$ meV energy scales in their excitations, and the wealth of phases and order parameters that can be explored for optimal coupling \cite{Orenstein:2012, Kahn/Tongyan:2021}. Moreover, the growth of \textit{ab initio} calculations of these properties and materials-informatics approaches for the discovery of materials now enables the bespoke design of real materials with optimal properties for interactions with specific DM models \cite{Jain:2013,Inzani:2020szg, Geilhufe_et_al:2018}. 

In solid-state materials, the dark photon inherits the properties of the ordinary photon in the detector target, and as such, its interaction is determined by the optical response of the target given by its dielectric function. Several options for exploiting quasiparticle phenomena in solid-state materials for UDM absorption have been proposed including electronic, superconducting, and phononic excitations in solid-state systems \cite{Hochberg_et_al:2016, Bloch:2016sjj, Hochberg_et_al:2017}. 
Dirac materials provide an optimal test-bed for target development as meV-eV UDM absorbers: their gaps can be reliably predicted by treating spin-orbit coupling as a perturbation \cite{Inzani:2020szg}, and their electronic structure near the band edge is simple enough to allow analytic estimates of the scattering rate \cite{Hochberg:2017wce}. Crucially, in-medium dielectric effects are suppressed in Dirac materials, allowing for dark photon absorption in real systems right up to the theoretical freeze-in limit \cite{Hochberg_et_al:2017,Geilhufe:2019ndy}. Similar sensitivity to dark photon absorption is found for optical phonons in polar semiconductors \cite{Knapen_et_al:2018, Griffin_et_al:2018}. Importantly, for these phonon-based schemes, high quality detector-grade crystals are already available owing to their use in microelectronics and quantum computing such as GaAs \cite{Griffin_et_al:2018} and SiC \cite{Griffin/Hochberg:2021}.

More recent efforts have explored UDM interactions with magnetoelectric couplings in quantum materials, such as bulk multiferroics \cite{Roising_et_al:2021} and antiferromagnetic topological insulators \cite{Marsh_et_al:2019,Schutte-Engel:2021bqm}. The latter proposed low-noise THz photon detection as a readout scheme, with projected sensitivity to axions in the 0.7 to 3.5 meV range. With this wealth of possible DM-matter interactions in quantum materials, future challenges lie in the design and selection of high-quality target materials to maximize the DM-matter coupling, exploration of highly correlated and entangled quantum phases for improved or novel sensing, and read-out strategies for these low-energy excitations that typically lie at or beyond the limits of current sensing thresholds \cite{Fink_et_al:2020, Sochnikov2020}.

\subsubsection{LC oscillators}
\label{Sec:3_LC}
Lumped-element LC oscillators can be used to look for a dark photon signal. The LC oscillator resonance frequency $\omega_0 = 1/\sqrt{LC}$ determines the accessible dark photon Compton frequency. Due to kinetic mixing, the dark photon field produces an oscillating, circumferential magnetic field within a cylindrical superconducting shield when the Compton wavelength of the dark photon is much larger than the dimensions of the shield. Placing a toroidal pick-up inductor within the shield allows the magnetic flux signal to be sensed by a DC SQUID. By making the pick-up inductor part of an RLC circuit, the dark photon induced signal is enhanced to measurable levels.\cite{Chaudhuri2015Radio} The DM-Radio Pathfinder detector has used this technique to set a direct-detection limit on dark photons around a narrow mass range near 2 neV \cite{Phipps:2020HP}. The ADMX SLIC experiment \cite{Crisoto:2020SLIC} has also used the lumped-element resonator technique to search for lower-mass axions and their search data can also be used to set limits on dark photons near 175 neV. 

LC oscillators can also be used to detect UDM candidates of positive intrinsic
parity~\cite{Donohue:2021jbv}, thus probing the possibility of scalar and axial 
vector UDM --- noting, as in Sec.~\ref{theory}, that 
the cosmological mechanisms which permit 
the production of UDM can do so without regard to its intrinsic parity. The discussion of scalar UDM in this white paper has
focused on probes of the couplings of Eqs.~(\ref{scalar-field_Lagrangian_alternative}) 
and
(\ref{linear_scalar_interactions}), as 
developed in dilaton-like
models~\cite{Arvanitaki:2014faa,Damour:2010EP-A,Damour:2010EP-B,Hees:2018DM_EP}. These can come from 
modifications of fundamental constants, such
as $\alpha$ or $m_e$~\cite{Arvanitaki:2014faa,Stadnik:2015DM-laser,Stadnik:2015DM_VFCs}, or from 
the nonobservation of long-range forces, leading to 
EP 
violation~\cite{Damour:2010EP-A,Damour:2010EP-B,Hees:2018DM_EP}
and other effects. 
However, broader possibilities exist. 
In particular, the appearance of scalar UDM can be framed
as a modification of electrodynamics~\cite{Donohue:2021jbv}, 
in analogy to that 
of axion electrodynamics~\cite{Sikivie:1983ip,Wilczek:1987mv},
giving rise to the
interaction $g_\gamma \phi F^{\mu\nu}F_{\mu\nu}/4$.
The interactions between 
electromagnetic fields and UDM are thus 
quite different from the axion case, 
because parity symmetry 
plays a key differential role. Moreover, the 
electromagnetic couplings of such UDM candidates can
be probed directly~\cite{Donohue:2021jbv}. 
In this case, a resonant, superconducting LC circuit with
a large, static electric field 
can induce weak magnetic fields
in the presence of scalar UDM. 
The experimental setup 
proposed in Ref.~\cite{Donohue:2021jbv} 
yields sensitivity to 
$g_\gamma$ up to ${\cal O}(10^{-22}\,
{\rm eV}^{-1})$
in the $10^{-11} - \,4\times 10^{-8}$ eV mass
($2\, {\rm kHz} - 10\,{ \rm MHz}$ frequency) range~\cite{Donohue:2021jbv} 
using the large electric 
fields developed for use in neutron EDM experiments~\cite{nEDM:2019qgk}. This would 
improve upon previous sensitive searches for
scalar particles from scalar-photon couplings in 
``light shining
through a wall'' experiments~\cite{Ballou2015PhRvD..92i2002B} 
by some three orders of
magnitude. 
At these mass scales, more sensitive constraints, noting
Fig.~\ref{fig:scalar_de_plot}, come
from probes to which both terms in 
Eq.~(\ref{scalar-field_Lagrangian_alternative}) 
could contribute, 
admitting the possibility of cancellation. 

\subsubsection{Trapped ions}
\label{ions}
In this white paper, many examples are given of the use of 
magnetic field detection to probe, and thus far to constrain, 
the existence of UDM. Yet, broader possibilities exist. 
For example, as in Ref.~\cite{Gilmore2021Sci...373..673G}, the 
center-of-mass motion of a trapped-ion crystal can be laser 
cooled to its quantum-mechanical ground state and its motional 
excitations, in the absence of cooling, sensitively measured. 
This enables searches for weak electric fields, potentially from UDM candidates. 
The electric field measurement sensitivity improves as $N^{1/2}$ where N is the number of ions in the crystal. 
Recent work demonstrated the ability to measure excitations (or displacements) of the 1.6 MHz center-of-mass motion of single-plane crystals with $N\sim 150$ ions at a level of $\sim$ 9 dB below the size of the ground-state wave function size \cite{Gilmore2021Sci...373..673G} in a single measurement. 
The enhanced sensitivity is obtained by preparing an entangled state of 
the ion spin and the center-of-mass motional degrees of freedom. Reading out the displaced entangled state is accomplished by reversing the interaction that produced the initial entangled state, 
resulting in an unentangled state where the ion spins are rotated by an amount proportional to the displacement. 
The angle of the ion spin-rotation can then be measured. Improvements to the demonstration of Ref.~\cite{Gilmore2021Sci...373..673G} 
indicate the potential for measuring electric fields of 1\,nV/m in averaging times of a few minutes. 
A 1.6\,MHz center-of-mass frequency implies sensitivity to an UDM candidate of 1.6\,neV in mass. 
For a dark photon, the estimated sensitivity to a kinetic-mixing parameter $\epsilon$ is $\sim 10^{-9}$ 
after one day of averaging~\cite{Gilmore2021Sci...373..673G}, 
though anticipated improvements should allow to access the 
$\epsilon < 10^{-12}$ range not yet excluded by existing 
searches~\cite{Chaudhuri2015Radio}.

\subsubsection{Nonlinear optics with dark photons}

Nonlinear optics is a commonplace tool in classical and quantum optics. In a nonlinear optical medium, the material response leads to effective three-photon and four-photon vertices.
A canonical example is the process of spontaneous parametric down-conversion (SPDC) in which a pump photon down-converts to two photons, usually called a signal and an idler. Other nonlinear optics processes include $2\to 1$ interactions, such as frequency doubling and sum- or difference-frequency generation, as well as $2\to2$ interactions known as four-wave mixing. 

If a dark photon is present as a degree of freedom, then a nonlinear optics process can take place with a photon replaced by a dark photon. Recently a search concept for dark photons was proposed, known as dark SPDC (dSPDC)~\cite{Estrada:2020dpg}. In dSPDS, a pump photon down-converts in a nonlinear medium to a signal photon plus a dark photon, which replaces SPDC's idler. 
The presence of the signal photon may be used to infer the presence of a dark state, akin to missing energy searches that are commonplace in high-energy physics. 
Though conceptually similar to SM SPDC, the kinematics in dSPDC  can be very different because dark photons possess a different dispersion relation than that of photons -- a mass for one and an index of refraction for the other. This can be used to distinguish the dark photon signal from the SM background~\cite{Estrada:2020dpg}.

The dSPDC process, we note, is not a dark matter search per-se, but rather a search for the dark photon as a degree of freedom. One can also imagine nonlinear optics processes in which the dark photon is in the initial state~\cite{Blinov-in-prep}. 
In addition, nonlinear optics are a key component in the toolbox described in Section~\ref{sec:toolbox} for enhancing dark matter searches.

\section{Dark matter searches with networks of quantum sensors on Earth and in space }
\label{networks}

In Section~\ref{Sec:3_MN}, we discussed how a network of magnetometers and comagnetometers can be used to detect UDM. Here, we describe some general benefits of using networks of quantum sensors. 
Using a network of detectors can be beneficial in searches for UDM. 
In particular, if there are $N$ nodes located within a single coherence volume, then it is possible to improve the signal-to-noise ratio in searches for UDM by a factor of $N^{1/2}$ \cite{Derevianko:2018Stochastic}. Additionally, using a network of spatially-separated detectors, it is possible to probe the spatio-temporal correlation function for a range of UDM masses; see Ref.~\cite{2022QSNET} for an example of the capabilities of a surface-based clock network with a network size of $\sim 300~\textrm{km}$. In addition, for spatially extended networks of optical clocks, the use of differential comparisons between clocks with a shared clock laser can relax requirements on clock laser performance, obviate the need for frequency combs, and significantly enhance detector sensitivity \cite{2022JunSr,zheng2022differential,clements2020lifetime}.

A network of spatially-separated sensors can also provide opportunities in searches for structured dark objects that are made up of ultralight bosonic fields, but do not necessarily comprise oscillatory DM, such as topological defects. 
A number of proposals have been put forward to search for the passage of topological defects through a terrestrial or space-based network of detectors, including magnetometers \cite{Pos13}, atomic clocks \cite{Derevianko:2014TDM}, pulsars \cite{Stadnik:2014TDM_pulsars}, optical cavities and interferometers \cite{Stadnik:2015DM-laser,Stadnik:2016DM-cavity,Grote:2019DM-LIFO}, and gravimeters/accelerometers  \cite{McNally:2020TDM_gravimeters,Figueroa2021DMaccel}. 
Several searches for passing scalar-field topological defects have been performed in recent years \cite{Wcislo:2016TDM_clock-cavity,Roberts:2017TDM_clocks,Wcislo:2018DM_atom-cavity,roberts_search_2020}. 
Greater sensitivity in identical models of scalar-field topological defects has been attained via various types of quantum sensors searching for an environmental dependence and spatial variations of the fundamental constants \cite{Stadnik:2020TDM_enviro}. 
Networks of quantum sensors may also be useful in searches for bursts of relativistic bosonic waves \cite{dailey2021quantum,Eby:2021ece} (see also Refs.~\cite{Stadnik:2021comment,Derevianko:2021reply} for commentary related to the proposal in \cite{dailey2021quantum}), including those that arise in collapse of boson stars \cite{Eby:2016cnq,Levkov:2016rkk}. 

It has recently been realised that space-based quantum sensors can offer a significant advantage over analogous ground-based detectors in searches for ultralight scalar fields when the scalar field is strongly screened near the surface of the Earth \cite{Hees:2018DM_EP,Stadnik:2020TDM_enviro,Stadnik:2021comment} (see also Refs.~\cite{Damour-Esposito:1993Nonperturb,Olive:2008Enviro,Hinterbichler:2010Screening} for earlier work on screening in scalar-field models), in particular in scalar-field models with the $\phi^2$ interactions in Eq.~(\ref{quadratic_scalar_interactions}). 
When the scalar field is strongly screened inside Earth, the scalar-field amplitude can be suppressed by the factor of $\sim h/R_\oplus$ near the surface of Earth, where $h$ is the height above Earth's surface and $R_\oplus$ is the radius of Earth; for a typical height of a ground-based apparatus of $h \sim 1~\textrm{m}$, one has $h/R_\oplus \sim 10^{-7}$, which means that the utilisation of space-based detectors can provide an enormous advantage. 
See Refs.~\cite{Hees:2018DM_EP} and \cite{Stadnik:2020TDM_enviro} for results pertaining to models of scalar-field UDM and scalar-field topological defects, respectively. 
Future space-based missions, such as ACES/PHARAO \cite{ACES-PHARAO:2015}, FOCOS \cite{FOCOS:2021}, OACESS \cite{OACESS:2021}, and MAGIS-Space \cite{magis-100_2021}, should offer even more sensitive platforms. 

Recent atomic clocks placed in space \cite{Alonso:2022oot}, including NASA's deep space atomic clock (DSAC) \cite{DSAC2021} and other space clocks, provide new opportunities for UDM searches, especially if coupled with future missions similar to the Parker Solar Probe \cite{PhysRevLett.127.255101}. Space-based quantum sensors, including atomic, molecular, and nuclear clocks, can probe unexplored parameter space of solar-bound UDM \cite{Banerjee:2019epw,Banerjee:2019xuy}, covering theoretical relaxion targets motivated by naturalness and Higgs mixing \cite{Tsai:2021lly}.

\section{Quantum measurement toolbox}\label{sec:toolbox}
Multiple UDM sensors discussed here involve detection of weak forces or fields, down to the single quantum level. Quantum optics provides a rigorous formalism and multiple experimental platforms to investigate the statistical properties of bosonic fields and single-photon detection. A detailed analysis of these techniques and their extension to dark matter detection will add to the mutually symbiotic relationship that has existed between astronomy and quantum optics for over a century. From the explanation of the solar blackbody spectrum to using auto-correlation measurements to characterize quantum devices, several advances in quantum optics have been motivated by astronomical measurements. Efforts to understand the nature of quantum measurements were initially motivated by gravitational-wave detection \cite{Caves:1981hw}, and laid the foundation for the field of quantum metrology employing quantum-enhanced measurements \cite{pezze_quantum_2018, braun_quantum-enhanced_2018}. Many ideas and techniques that emerged from this work can also be applied to UDM detection.

In order to use the quantum measurement toolbox to improve precision one needs to start with understanding the SQL. The SQL arises in linear measurements of any quantum mechanical observable $A$ of a system that does not commute with itself at different times. If the observable $A$ is measured with some uncertainty $\Delta A$, then its conjugate variable $B$ will have an uncertainty $\Delta B$ given by the Heisenberg uncertainty relation $\Delta A \Delta B \geq  (i/2)[A,B]$. This uncertainty feeds back into the subsequent evolution of $A$, limiting the accuracy of successive measurements. The SQL is a function of the specific physical system and type of measurements under consideration, and it is now well understood that it is not a fundamental limit to measurement precision. 

One can tailor the Heisenberg uncertainty in one observable so the error is primarily in the conjugate variable, as done in the case of quantum squeezing \cite{Caves:1981hw, walls_squeezed_1983}. Alternatively, one can pick an observable that commutes with itself at later times, as is the case with quantum non-demolition (QND) measurements \cite{caves1980measurement}. Both techniques have been used very recently in the context of dark matter detection. Recent, proof-of-principle demonstrations  involve the use of squeezed state receivers to subvert the SQL, resulting in a factor of two increase in scan rate in the search of axions \cite{backes2021quantum}, and the use of superconducting transmon qubits in a dark photon search involving QND measurements, giving a factor of $10^3$ improvement in scanning rate \cite{Dixit2020single}.


More broadly, one can tailor the coupling between a quantum system and/or the “meter” used to detect its dynamics to minimize (or evade) measurement back-action effects. A detailed discussion of this topic can be found in Ref. \cite{braginsky1995quantum,  caves1980measurement, WisemanBook2009, meystre2021quantum, tsang2012, polzik2015}. Some experimentally-demonstrated techniques to evade measurement back-action include the use of QND measurement of photons \cite{guerlin_progressive_2007, brune_quantum_1990}, constructing quantum-mechanics-free subspaces to avoid the backaction of measurements \cite{wasilewski2010, moller2017, ockeloen2016}, and using quantum feedback control \cite{schmid_coherent_2022, tebbenjohanns_quantum_2021, magrini_real-time_2021, rossi_measurement-based_2018, wilson_measurement-based_2015}. These techniques can be implemented to dramatically improve the sensitivity and/or bandwidth of multiple quantum sensors searching for UDM, and warrant detailed theoretical study and experimental prototype demonstration.

\textbf{Coherent clock comparisons}
-- To date, frequency ratio measurements of optical quantum clocks based on different transitions have almost exclusively been carried out incoherently.  In incoherent frequency ratio measurements, each clock is operated completely independently and the ratio is derived by performing beat note measurements between the two clock lasers and a femtosecond frequency comb \cite{Beloy2021}.  The precision of such measurements is fundamentally limited by the quadrature sum of the QPN of each clock at a fractional frequency uncertainty
\begin{equation}\label{eq:SQL}
\sigma(\tau) = \frac{1}{2 \pi \nu \sqrt{N T \tau}} \ ,
\end{equation}
where $\tau$ is the measurement duration, $\nu$ is the clock transition frequency, $N$ is the number of unentangled atoms or molecules, and $T$ is the spectroscopy probe duration \cite{Itano1995QPN}.  In turn, the probe duration is limited by the shorter of the coherence time of the clock transition and the coherence time of the clock laser.  However, for many of the highest performance optical quantum clocks, the coherence of the clock transition is much longer than the coherence of the best current and anticipated future clock lasers, and the precision of frequency ratio measurements would be improved if the probe duration could be extended beyond the laser coherence time.

Using a technique called correlation spectroscopy, frequency difference measurements of clocks based on the same transition have been performed with probe durations beyond the laser coherence time, by using the same laser to synchronously probe both clocks and performing parity measurements of the transition probability of the two clocks.  Experimental demonstrations of optical frequency difference measurements between multiple ions \cite{Chwalla2007CorrSpec,Chou2011CorrSpec,Kaewuam2020ThreeIonCorrSpec} or atoms \cite{Young2020TweezerCorrSpec} in a single trap using correlation spectroscopy have been performed, culminating in a measurement of the difference in gravitational redshift for atoms spread across a mm-scale strontium optical lattice trap \cite{2022JunSr}.  Recently, correlation spectroscopy was used to improve the precision of frequency difference measurements between two independent aluminum-ion clocks \cite{clements2020lifetime} by more than on order of magnitude with respect to previous incoherent comparisons \cite{Chou2010AlAl}.  
In contrast to conventional clock comparisons that rely on coherence between the atoms or molecules and the clock laser within each clock, correlation spectroscopy requires quantum coherence between the two quantum clocks and can therefore be referred to as a coherent clock comparison.

Very recently, an extension of correlation spectroscopy called differential spectroscopy has been proposed \cite{Hume2016DiffSpec} and demonstrated \cite{Kim2022DiffSpec} that enables frequency difference measurements of optical quantum clocks based on different transitions with probe durations longer than the laser coherence time.  In Ref.~\cite{Kim2022DiffSpec}, the precision of a frequency difference measurement between an aluminum-ion clock and a ytterbium optical lattice clock was improved by a factor of 7 with respect to incoherent comparisons \cite{Beloy2021}.  With technical improvements, it is anticipated that the precision of future frequency ratio measurements between a wide variety of quantum clocks can be improved by orders of magnitude using differential spectroscopy, making coherent clock comparisons a very powerful technique for UDM searches.

\textbf{Squeezed states} -- While the standard quantum limit (SQL) given by Eq.~\ref{eq:SQL} is a fundamental limit for clocks based on $N$ unentangled atoms, entangled atoms or molecules offer the possibility of precision that scales more favorably with $N$.  In particular, clocks based on squeezed states \cite{Wineland1992Squeezing,Kitagawa1993Squeezing} with $N \gg 1$ have a stability that scales like $\sigma \propto N^{-2/3}$ when the probe duration is limited by laser decoherence \cite{Andre2004Squeezing}.  Coherent comparisons of squeezed clocks can circumvent both laser decoherence and Dick effect noise that may prevent this scaling from being reached \cite{2020PietSqueezing}. Experimental demonstrations of squeezing thousands of atoms at microwave transition frequencies have achieved quantum projection noise a factor of one hundred below the SQL \cite{Cox2016Squeezing,Hosten2016}.  A promising recent experiment has generated squeezing of hundreds of ytterbium atoms on an optical transition \cite{pedrozo-penafiel_entanglement_2020}.

\textbf{Heisenberg-limited spectroscopy} -- Beyond extending the probe duration and reducing the QPN of clocks based on unentangled atoms or molecules, correlation and differential spectroscopy also remove the key barrier to the use of entangled states in trapped-ion clocks.  While squeezed states offer modest metrological gain even in the presence of laser decoherence, this is not true for maximally entangled states that saturate the Heisenberg limit $\sigma \propto N^{-1}$, because the improved scaling of QPN with $N$ is negated by a reduction in coherence time between the atoms or molecules and the laser \cite{Huelga1997EntangledClocks}.  However, coherent clock comparisons do not require atom-laser coherence and thus offer the tantalizing prospect of clock comparisons at the Heisenberg limit.  Trapped ion platforms in particular offer exquisite quantum control capabilities including the generation of arbitrary entangled states optimized for minimal sensitivity to noise \cite{Marciniak2021OptimalQM} and maximally entangled GHZ states of up to $N = 24$ to-date \cite{Pogorelov2021GHZ24Ion}.  With a focused effort, it would be possible to achieve Heisenberg-limited UDM searches with tens of trapped ions within the next decade.

\textbf{Dynamical decoupling} -- Optical quantum clocks based on conventional spectroscopy have poor sensitivity to UDM with Compton frequencies greater than the reciprocal of typical probe durations $100~\textrm{ms} < T < 1~\textrm{s}$.  Dynamical decoupling quantum control techniques offer a route to maximize the sensitivity of clocks to high-frequency UDM signals \cite{Aharony:2021DM_atom-cavity} while simultaneously minimizing their sensitivity to laser decoherence and other noise sources \cite{Viola1999DynamicalDecoupling,Dorscher2020DynamicalDecoupling}.  In Ref.~\cite{Aharony:2021DM_atom-cavity}, a strontium-ion optical clock is operated using a dynamical decoupling pulse sequence that is highly sensitive to UDM with a Compton frequency of 1~kHz.  With higher-intensity lasers, there are no fundamental obstacles to extending UDM searches based on optical quantum clocks up to 1~MHz or beyond.

\textbf{Quantum logic spectroscopy}
-- Quantum optics not only tells us how to perform measurements in the most efficient way, but also enables measurements on previously inaccessible species with high sensitivity to UDM. A prominent example is quantum logic spectroscopy (QLS) \cite{wineland_quantum_2002, schmidt_spectroscopy_2005}, where techniques developed in the context of quantum information processing with trapped ions \cite{haffner_quantum_2008-1, blatt_entangled_2008, leibfried_quantum_2003} are employed to perform precision spectroscopy. The lack of a suitable transition for laser cooling and internal state detection for most spectroscopy ions is mitigated through a co-trapped logic ion providing sympathetic cooling, state preparation and state readout using quantum algorithms. The spectroscopy ion can therefore be chosen exclusively based on its sensitivity to new physics. The aluminium ion optical clock demonstrated the potential of the technique \cite{rosenband_frequency_2008} and has advanced since to the most accurate clock \cite{2019Alclock}. It has served as an almost unaffected anchor for the search in a variation of $\alpha$ \cite{rosenband_frequency_2008} and UDM \cite{Beloy2021} (see also Section~\ref{Sec:ULDM_clocks}). Through QLS new clock candidates with a high sensitivity to a change in $\alpha$ and UDM searches, such as HCI and molecules, become accessible and significant advances have been made in this direction. HCIs have been slowed, trapped, and sympathetically cooled in Paul traps using laser-cooled beryllium \cite{schmoger_coulomb_2015, King2021} and an optical clock based on an Ar$^{13+}$ HCI has been demonstrated \cite{micke_coherent_2020, king_private_2022}. There have also been significant advances in extending quantum optics control over molecular ions using QLS. Non-destructive internal state detection and simple spectroscopy \cite{wolf_non-destructive_2016, sinhal_quantum-nondemolition_2020}, coherent manipulation \cite{chou_preparation_2017} and high-resolution frequency-comb spectroscopy \cite{chou_frequency-comb_2020} of simple diatomic molecular ions have been demonstrated. QLS should also be possible with $^{229}$Th ions \cite{Groot-Berning2019}, see Sec.~\ref{Sec:Nuclock}.
QLS significantly extends our possibilities to search for new physics using atomic and molecular species with high sensitivity to UDM and other new physics.

\section{Towards dark energy}
Despite the apparent success of the $\Lambda$CDM model (Lambda Cold Dark Matter), outstanding theoretical issues with the cosmological constant explanation of cosmic acceleration have inspired a variety of alternative dark energy (DE) models, many of which utilize scalar fields similar to UDM models. For example, quintessence models \cite{Wetterich:1987fm,Ratra:1987rm,Tsujikawa2013,Brax2017} are built around a slowly-rolling scalar field which can behave as DE, explaining the observed accelerated expansion of the universe \cite{Planck2018}. Coupled DE models (i.e. those with scalar-SM interactions) assume either Yukawa-type couplings as discussed in Section \ref{theory} or universal, gravitational-strength couplings which arise from scalar-tensor theories of gravity \cite{Brax2017}. The small scalar field mass that is necessary for the field to drive cosmic acceleration proves troublesome for quintessence theories, as the corresponding long-range scalar-mediated fifth force is highly constrained \cite{Adelberger2009}, implying that the scalar-SM coupling would have to be negligibly small to have remained undetected. However, the discovery of screening mechanisms \cite{Khoury2004,Hinterbichler:2010Screening,Vainshtein1972}, which suppress the scalar-mediated force, allow the reconciliation of null results in detection experiments with the presently observed cosmic acceleration. For example, two common screening mechanisms utilize dynamic scalar-field mass \cite{Khoury2004} and scalar-SM couplings which depend on the local matter density \cite{Hinterbichler:2010Screening}. The result is an associated fifth force which is exceedingly sensitive to the composition and environment of the source and test masses. At present, the best constraints on screened scalar fields come from astrophysical observations \cite{Wilcox2015,Vikram2018,Jain2013}, atom interferometry \cite{Jaffe2017}, and torsion-balance experiments \cite{Upadhye2012,Adelberger2003} (see Ref. \cite{Burrage2018} for a review). Existing and proposed experiments have considered using levitated \cite{Rider2016,Betz2022} and cavity \cite{Qvarfort2021} optomechanical systems to probe the screened-scalar field parameter space. Despite the theoretical differences between scalar UDM and DE, several existing and future experiments discussed here can perform UDM and DE searches simultaneously.

\section{Combined exclusion plots and projections for future experiments} 
\label{summary}

\begin{figure}
\centering
\includegraphics[width=\columnwidth]{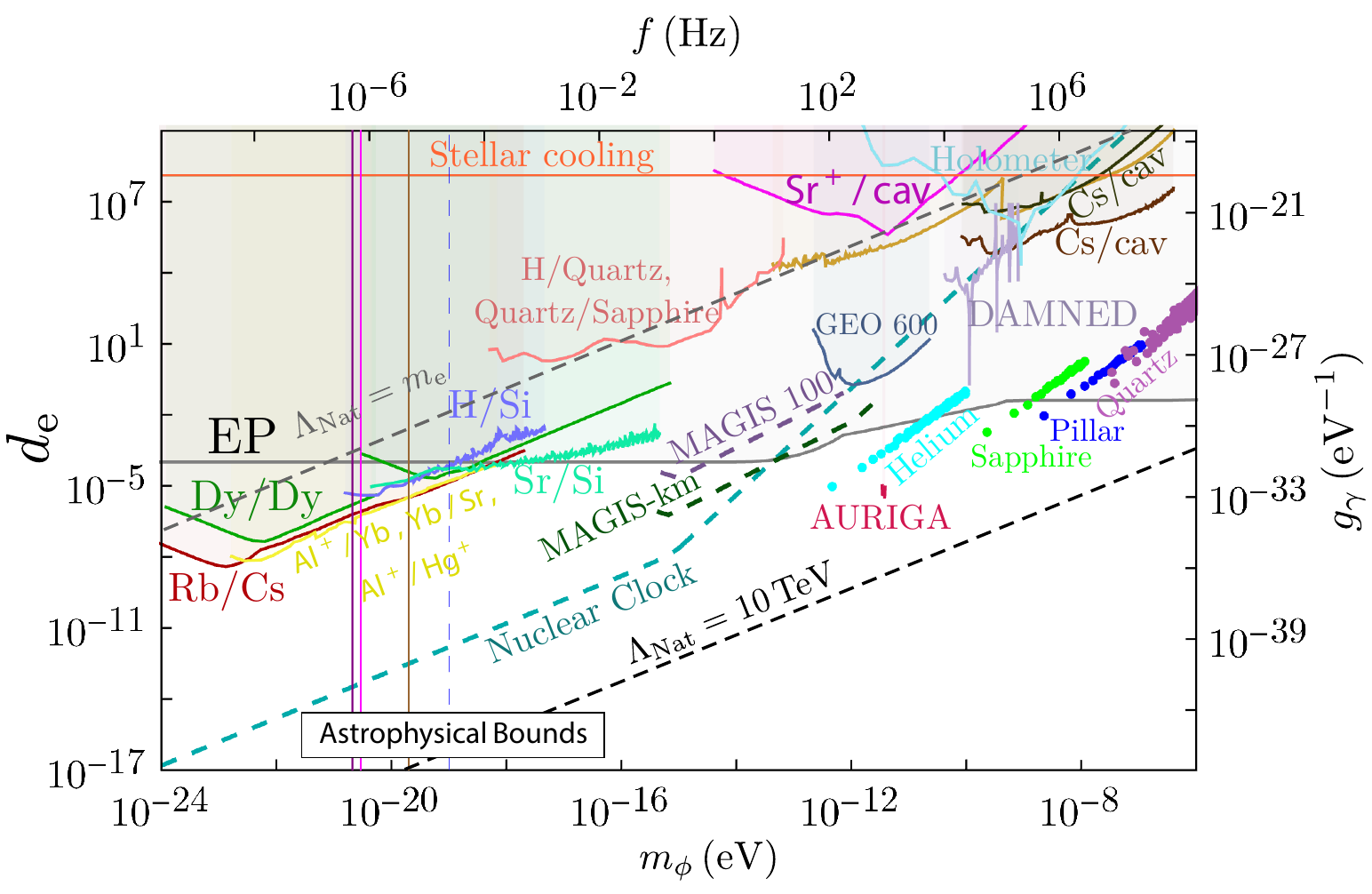}
\caption{ 
Summary plot for $d_e$, the scalar DM coupling to the electromagnetic field tensor, defined in Eq.~\eqref{scalar-field_Lagrangian_alternative}, assuming only $d_e\neq 0$. Experiments which search for EP violation, represented by the gray shaded region, provide stringent constraints over the mass scale shown in the figure~\cite{RotWash:1999,EotWash:2008,Wagner_2012_torsion,MICROSCOPE:2017,Berge:2017ovy,Hees:2018DM_EP}. 
Existing bounds from precision spectroscopy measurements in Dy/Dy (dark green)~\cite{PhysRevLett.115.011802}, Rb/Cs atomic clocks (dark red)~\cite{PhysRevLett.117.061301} and a combination of Al$^+$/Yb, Sr/Yb and Al$^+$/Hg$^+$ clocks  (light yellow)~\cite{Beloy2021} are also shown. 
Bounds from the comparison of a H-maser and Sr clock with a Si cavity (labelled H/Si and Sr/Si, respectively) are denoted by the lavender and aqua-green regions, respectively~\cite{Kennedy:2020DM_atom-cavity}, the comparison of a Sr$^+$ clock with a Si cavity is shown by the magenta region~\cite{Aharony:2021DM_atom-cavity}, comparisons of a bulk acoustic wave quartz oscillator with a H-maser and a cryogenic sapphire oscillator are shown by the peach-orange line~\cite{Campbell:2021DM_atom-cavity}, and comparisons of a Cs clock with a cavity are shown by the very dark green~\cite{Antypas:2019DM_atom-cavity} and brown~\cite{Tretiak:2022ndx} regions. 
The dark yellow line represents bounds from molecular I$_2$ experiments~\cite{Oswald:2021vtc}, whereas the cherry-red shaded region represents bounds from the AURIGA experiment~\cite{branca2017search}. 
Optical interferometry bounds from GEO600 (dark blue)~\cite{Vermeulen:2021DM-GEO600}, DAMNED (lilac)~\cite{Savalle:2021DAMNED} and Fermilab Holometer (light turquoise)~\cite{Aiello:2021DM-holometer} experiments are also shown. 
We show the projected sensitivities of the MAGIS experiments by dashed purple and dashed dark green lines~\cite{magis-100_2021}. 
We show the projected sensitivity of a nuclear clock as a dashed dark turquoise line. 
We also show projections from various proposed mechanical resonators by cyan (superfluid $^{4}$He), light green (sapphire), blue (pillar) and mauve (quartz bulk acoustic wave) circles~\cite{manley2020searching}. 
Note that, for masses $m_\phi\lesssim 10^{-21}$ eV, the scalar field cannot account for $100\%$ of the observed DM. 
The sensitivities of the considered experiments scale as the square root of the local DM density, meaning that e.g.~for a fractional abundance of $1\%$, the actual sensitivities weaken by a factor of about 10. 
We have shown the stellar cooling bounds by the horizontal orange line~\cite{Raffelt:1996wa}. 
The vertical solid (dashed) lines show various current (projected) lower bounds on the DM mass coming from astrophysical considerations~\cite{Schutz200105503, Nadler200800022, 2021PhRvL.126g1302R,Kobayashi:2017jcf,2017PhRvL.119c1302I,Armengaud:2017nkf,Drlica-Wagner190201055}. Natural values of $d_{\rm e}$ for the cut-offs $\Lambda_{\rm Nat} = 10 \, {\rm TeV}$ and $\Lambda_{\rm Nat} = m_{e}$ lie below the black and gray dashed lines, respectively. 
}
\label{fig:scalar_de_plot}
\end{figure}

\begin{figure}
\centering
\includegraphics[width=\columnwidth]{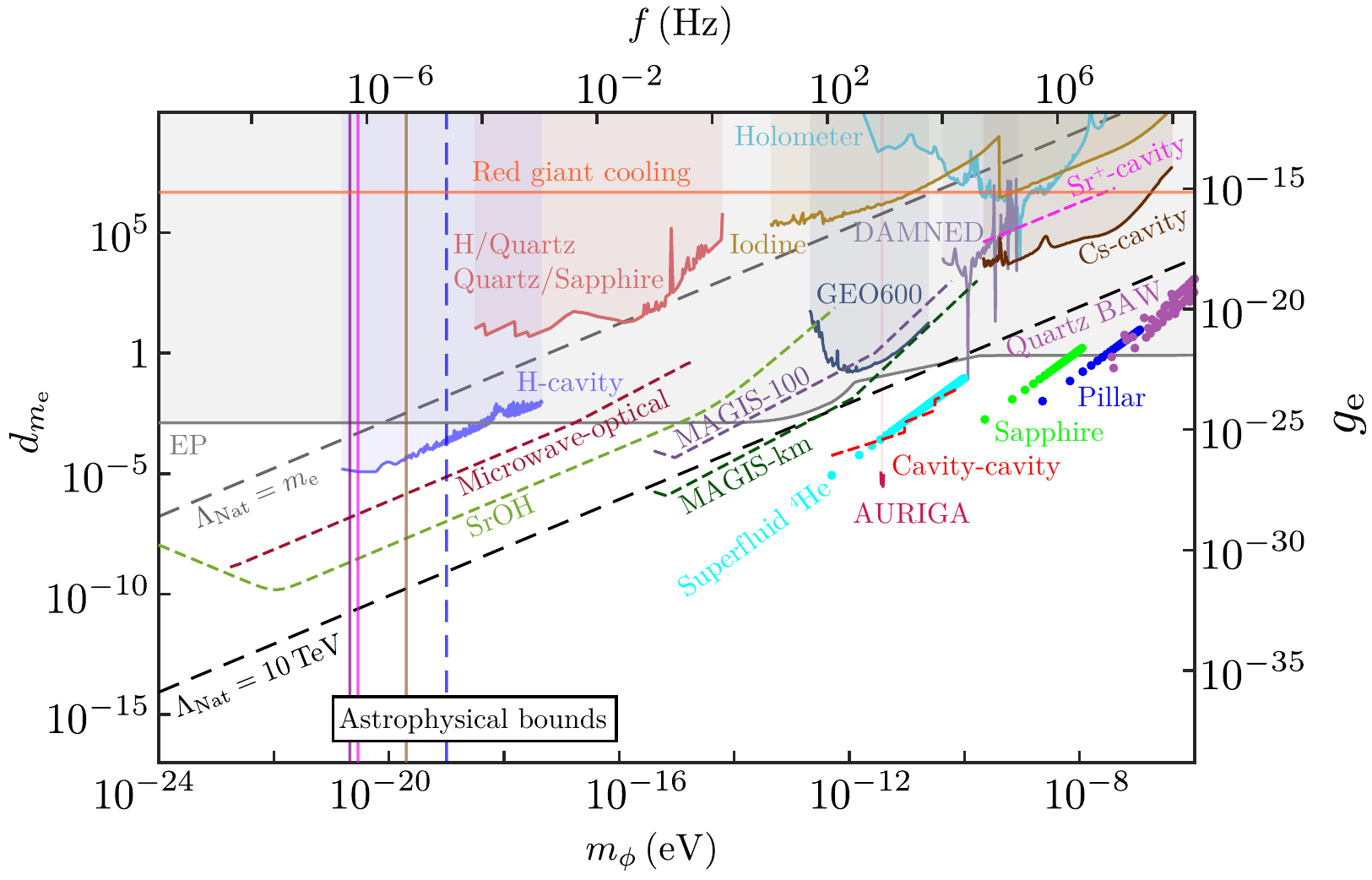}    
\caption{Summary plot for $d_{m_e}$, the scalar DM coupling to the electron, defined in Eq.~\eqref{scalar-field_Lagrangian_alternative}, assuming only $d_{m_e} \neq 0$. 
Experiments which search for EP violation, represented by the gray shaded region, provide stringent constraints over the mass scale shown in the figure~\cite{RotWash:1999,EotWash:2008,Wagner_2012_torsion,MICROSCOPE:2017,Berge:2017ovy,Hees:2018DM_EP}. 
Bounds from the comparison of a H-maser with a Si cavity are denoted by the lavender region~\cite{Kennedy:2020DM_atom-cavity}, comparisons of a bulk acoustic wave quartz oscillator with a H-maser and a cryogenic sapphire oscillator are shown by the peach-orange line~\cite{Campbell:2021DM_atom-cavity}, and comparison of a Cs clock with a cavity are shown by the brown~\cite{Tretiak:2022ndx} region. 
The dark yellow line represents bounds from molecular I$_2$ experiments~\cite{Oswald:2021vtc}, whereas the cherry-red shaded region represents bounds from the AURIGA experiment~\cite{branca2017search}. 
Optical interferometry bounds from GEO600 (dark blue)~\cite{Vermeulen:2021DM-GEO600}, DAMNED (lilac)~\cite{Savalle:2021DAMNED} and Fermilab Holometer (light turquoise)~\cite{Aiello:2021DM-holometer} experiments are also shown. 
We show the projected sensitivities of the MAGIS experiments by dashed purple and dashed dark green lines~\cite{magis-100_2021}, as well as the projected sensitivities of microwave-optical atomic clock comparison (dashed dark red line)~\cite{Arvanitaki:2014faa},  SrOH molecular clock spectroscopy (dashed lime-green line)~\cite{Kozyryev2021}, Sr$^{+}$ clock-cavity comparison (dashed magenta line) \cite{Aharony:2021DM_atom-cavity}, and cavity-cavity comparison (dashed red line)~\cite{Geraci:2019DM-cavity} experiments. 
We also show projections from various proposed mechanical resonators by cyan (superfluid $^{4}$He), light green (sapphire), blue (pillar) and mauve (quartz bulk acoustic wave) circles~\cite{manley2020searching}. 
We have shown the red giant cooling bounds by the horizontal orange line~\cite{Raffelt:1996wa}. 
The vertical solid (dashed) lines show various current (projected) lower bounds on the DM mass coming from astrophysical considerations~\cite{Schutz200105503, Nadler200800022, 2021PhRvL.126g1302R,Kobayashi:2017jcf,2017PhRvL.119c1302I,Armengaud:2017nkf,Drlica-Wagner190201055}. Natural values of $d_{m_{\rm e}}$ for the cut-offs $\Lambda_{\rm Nat} = 10 \, {\rm TeV}$ and $\Lambda_{\rm Nat} = m_{e}$ lie below the black and gray dashed lines, respectively.
}
\label{fig:scalar_dme_plot}
\end{figure}

\textbf{Scalar DM current limits and future perspectives} -- Summary plots for the current and projected constraints on the scalar DM couplings to the electromagnetic field tensor and the electron field in Eq.~\eqref{scalar-field_Lagrangian_alternative} are given in  Figs.~\ref{fig:scalar_de_plot} and \ref{fig:scalar_dme_plot}, respectively. 
The dimensionless parameters $d_e$ and $d_{m_e}$ encode the strengths of the interactions between the scalar field and SM fields relative to the strength of gravity. 
The dashed black lines correspond to upper bounds on natural values of $d_{\rm e}$ and $d_{m_{\rm e}}$ for a 10 TeV cutoff of the Yukawa modulus \cite{Arvanitaki:2016DM-resonator}. 
For convenience, the right-hand axes give the same limits in a common alternative notation: $g_\gamma \leftrightarrow \kappa d_e$ and $g_e \leftrightarrow \kappa m_e d_{m_e}$, where $\kappa = (\sqrt{2}M_\textrm{Pl})^{-1}$ with $M_\textrm{Pl}$ being the reduced Planck mass. 
We note that another often used convention in the literature involves the new-physics energy scales $\Lambda_\gamma$ and $\Lambda_e$ explicitly via the identifications $g_\gamma \leftrightarrow 1/\Lambda_\gamma$ and $g_e \leftrightarrow m_e/\Lambda_e$, respectively. 
Both graphs have the DM mass on the bottom axis and the corresponding DM Compton frequency at the top axis. 

Figs.~\ref{fig:scalar_de_plot} and \ref{fig:scalar_dme_plot} show limits from searches for EP violation~\cite{RotWash:1999,EotWash:2008,Wagner_2012_torsion,MICROSCOPE:2017,Berge:2017ovy,Hees:2018DM_EP}, atom-cavity experiments \cite{Kennedy:2020DM_atom-cavity,Aharony:2021DM_atom-cavity,Campbell:2021DM_atom-cavity,Antypas:2019DM_atom-cavity,Tretiak:2022ndx},
molecular iodine (I$_2$) experiments~\cite{Oswald:2021vtc}, the AURIGA experiment~\cite{branca2017search}, optical interferometry experiments~\cite{Vermeulen:2021DM-GEO600,Savalle:2021DAMNED,Aiello:2021DM-holometer}, atom interferometry experiments~\cite{magis-100_2021}, the stellar cooling bounds~\cite{Raffelt:1996wa}, and astrophysical constraints ~\cite{Schutz200105503, Nadler200800022, 2021PhRvL.126g1302R,Kobayashi:2017jcf,2017PhRvL.119c1302I,Armengaud:2017nkf,Drlica-Wagner190201055}.
Fig.~\ref{fig:scalar_de_plot} also show $d_e$ bounds from 
precision spectroscopy  \cite{PhysRevLett.115.011802} and atomic clocks  ~\cite{PhysRevLett.117.061301,Beloy2021}.
We note that the frequency ratio of two optical atomic clocks is insensitive to $d_{m_e}$, as discussed in Sec.~\ref{Sec:ULDM_clocks}. 
Fig.~\ref{fig:scalar_dme_plot} also shows projected limits for a microwave-optical atomic clock comparison \cite{Arvanitaki:2014faa} and a SrOH molecular spectroscopy experiment~\cite{Kozyryev2021}.

The EP limits come from torsion balance experiments described in Section~\ref{Sec:3_TB} \cite{RotWash:1999,EotWash:2008,Wagner_2012_torsion} and the MICROSCOPE satellite \cite{MICROSCOPE:2017,Berge:2017ovy,Hees:2018DM_EP}. 
MICROSCOPE aims to tests the EP in 
orbit to an ultimate precision
of $10^{-15}$ using 
electrostatic accelerometers on board a drag-free micro-satellite~\cite{Rodrigues:2021csn}, with current results
 based on a fraction of their data~\cite{MICROSCOPE:2017,Berge:2017ovy}.
Torsion balance experiments that are up to a factor of 10 more sensitive are in preparation at the University of Washington. 

Limits from precision spectroscopy (Dy/Dy)~\cite{PhysRevLett.115.011802}, Rb/Cs microwave clocks~\cite{PhysRevLett.117.061301}, and  Al$^+$/Hg$^+$ optical clocks~\cite{Beloy2021} come from the re-analyses of $\alpha$ drift data in terms of an oscillating DM field. 
Such experiments have previously looked for a drift of $\alpha$ on a time scale of months and years. 
Therefore, re-analyzing such prior experiments can give limits on DM with long oscillation periods corresponding to small DM masses. 
New clock experiments included clock-comparison experiments with Yb/Al$^+$ and Yb/Sr  clock pairs~\cite{Beloy2021}.
UDM experiments involving atomic clocks described in Section~\ref{Sec:ULDM_clocks} are broadband and are naturally sensitive to smaller DM masses due to the details of the clock operation. 
The use of dynamic decoupling that involves applying an addition sequence of laser pulses during the clock probe time \cite{Aharony:2021DM_atom-cavity} can be used to enhance clocks' sensitivity to UDM of specific and/or higher masses.

Clock-based limits are expected to significantly improve with the increased precision of current clocks and the development of highly charged ion (HCI) and nuclear clocks which have much higher sensitivities to UDM. 
Recently, the first optical clock based on a HCI has been demonstrated with atom-related systematic uncertainties at a level of $10^{-18}$ and below \cite{king_private_2022}. 
HCI clocks will enable clock-comparison DM searches with $\Delta K \approx 100$ \cite{2022QSNET}, see Section~\ref{HCI}.
The nuclear clock sensitivity to the variation of $\alpha$ is expected to exceed the sensitivity of present clocks by $\sim 4$ orders of magnitude \cite{2020nuc}. 
In Figure~\ref{fig:scalar_de_plot}, we show the projected sensitivity of a nuclear clock  plotted for $\rm SNR=1$ with an averaging time of 1\,s, integration time of $10^6$\,s, and the time interval between two $\pi/2$-pulses in the Ramsey method of $0.5$\,s as discussed in~\cite{Banerjee:2020kww}. The sensitivity enhancement factor of $K=10^4$ was used for the projection plot.

As discussed in Section~\ref{Sec:clocks_cav}, one can also compare an atomic clock frequency to that of a reference cavity that could be represented by the laser internal resonator (which can be a part of the clock) or some external optical cavity. 
The bounds set by conducting  frequency comparisons between a state-of-the-art strontium optical lattice clock and a cryogenic crystalline silicon cavity, and a hydrogen maser and the cavity are represented by the Sr/Si and H/Si lines, respectively~\cite{Kennedy:2020DM_atom-cavity}. 
We note that ``Si'' here just refers to a material inside the cavity. Future projected Sr/Si cavity limits are given in Fig.~\ref{fig:Sr_Si_H}.

Other atom-cavity experiments discussed in Section~\ref{Sec:clocks_cav} are sensitive to higher DM masses. Present limits include  the comparison of a Sr$^+$ clock with a Si cavity~\cite{Aharony:2021DM_atom-cavity} that demonstrated dynamic decoupling, comparisons of a bulk acoustic wave quartz oscillator with a H-maser and a cryogenic sapphire oscillator~\cite{Campbell:2021DM_atom-cavity}, and comparisons of a Cs clock with a cavity~\cite{Antypas:2019DM_atom-cavity,Tretiak:2022ndx}. 
Significant orders-of-magnitide improvements are expected with atom-cavity experiments; future Cs-cavity experiments will allow one to measure the relative variation $\delta \nu/\nu$ below $10^{-17}$ in less than 200 hours \cite{Tretiak:2022ndx}. 
Projected bounds on $d_{m_e}$ for cavity-cavity experiment  \cite{Geraci:2019DM-cavity} described in Section~\ref{cav}.

The bounds from optical interferometry experiments described in Section~\ref{Sec:LIFO-GW_detectors} come from the GEO600~\cite{Vermeulen:2021DM-GEO600}, DAMNED~\cite{Savalle:2021DAMNED} and Fermilab Holometer~\cite{Aiello:2021DM-holometer} experiments. 
Large-scale Fabry-Perot-Michelson interferometers (such as LIGO, VIRGO or KAGRA) can improve over current optical-interferometry bounds at lower frequencies by at least an order of magnitude if freely-suspended Fabry-Perot arm mirrors of different thicknesses are used in the two arms, while small-scale Michelson interferometers (such as the Fermilab holometer) operating in the resonant narrowband regime can deliver a significant improvement in sensitivity over current optical-interferometry bounds at mid-to-high frequencies due to up to a $\sim 10^6$ enhancement in the UDM signal \cite{Grote:2019DM-LIFO}.

\begin{figure}
    \centering
    \includegraphics[width=\columnwidth]{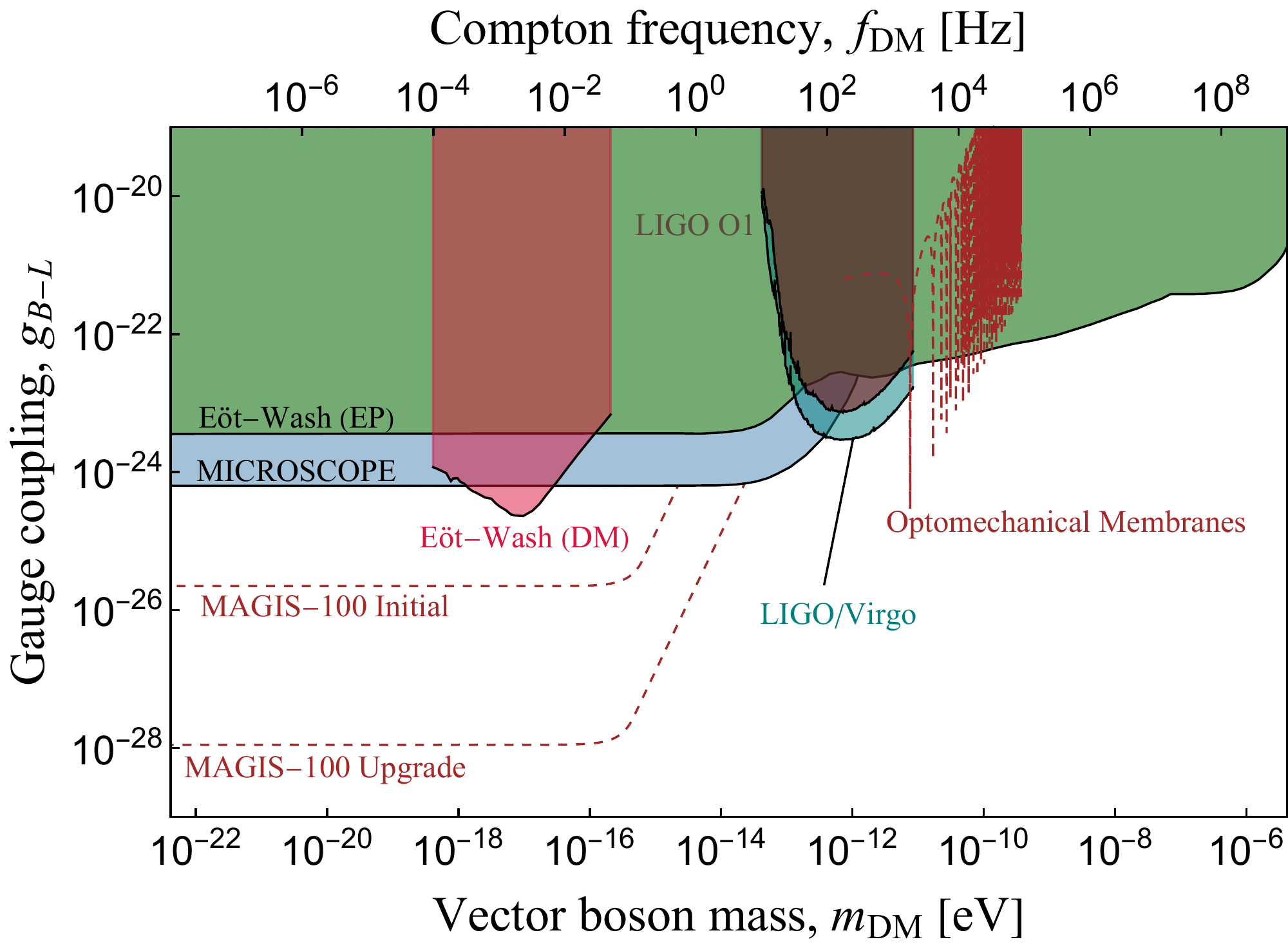}
    \caption{Summary plot for $g_{B-L}$, the vector DM gauge coupling parameter, defined in Eq.~\eqref{vector_gauge_coupling}. 
    Current constraints: MICROSCOPE EP test~\cite{Berge:2017ovy,MICROSCOPE:2017}, E\"{o}t-Wash torsion balance EP tests~\cite{Wagner_2012_torsion}, E\"{o}t-Wash torsion balance DM search~\cite{shaw2022torsion}, LIGO's first observing run~\cite{Michimura:2021DM-LIFO_propagation}, combined constraints from the third observing run of LIGO and VIRGO~\cite{LIGO:2021DM_vector}. Proposed experiments: MAGIS-100 and its upgrade~\cite{magis-100_2021}, optomechanical membranes~\cite{manley2021searching}.}
    \label{fig:vector_B-L_plot}
\end{figure}

The projected sensitivities of the atom interferometry MAGIS-100 and MAGIS-km experiments are shown by dashed lines. 
The MAGIS program \cite{magis-100_2021} has the potential to improve sensitivity in a variety of UDM models (see Figs.~\ref{fig:scalar_de_plot}, \ref{fig:scalar_dme_plot}, and \ref{fig:vector_B-L_plot}) and can do so in the near future, as MAGIS-100 is already under construction and scheduled to start its physics run in 2023. Looking into the further future, it should be noted that MAGIS-100 is just the first experiment of its type---long baseline atom interferometers for fundamental physics---and that there are many more to come. The technology developed and the experiences obtained from MAGIS-100 will be crucial for enabling the next-generation experiments of the MAGIS program such as MAGIS-km, as well as other international long baseline atom interferometers such as AION \cite{AION_2020}. These experiments will not only greatly extend the reach of UDM searches, but also lead to new developments and discoveries in quantum science and gravitational wave physics \cite{magis-100_2021}.

We also show projections from various proposed mechanical resonators described in Section~\ref{Sec:3_MR}, including superfluid $^{4}$He, sapphire, pillar, and  quartz bulk acoustic wave by circles~\cite{manley2020searching}. 

For masses $m_\phi < 2 \times 10^{-20}$ eV, the scalar field cannot account for $100\%$ of the observed DM owing to bounds from small-scale cosmic structure \cite{2021PhRvL.126g1302R}.
The sensitivities of the considered UDM experiments scale as the square root of the local DM density, meaning that e.g.~for a fractional abundance of $1\%$, the actual sensitivities weaken by a factor of about 10. 
On the other hand, the sensitivities of the more traditional EP tests shown in Figs.~\ref{fig:scalar_de_plot} and \ref{fig:scalar_dme_plot} do not depend on the local DM density, since the effects in those types of experiments arise due to the exchange of virtual scalar bosons. 

The vertical solid (dashed) lines show various current (projected) lower bounds on the DM mass coming from astrophysical considerations~\cite{Schutz200105503, Nadler200800022, 2021PhRvL.126g1302R,Kobayashi:2017jcf,2017PhRvL.119c1302I,Armengaud:2017nkf,Drlica-Wagner190201055}. There is important complementarity between astrophysical and terrestrial probes, where a future detection of a small-scale cut-off in the matter clustering or halo mass function would motivate terrestrial searches (and \textit{vice versa}). Further, null detections in astrophysical data set an important lower limit on the allowed scalar mass to be all of the dark matter (with sub-dominant contributions also probed).

\begin{figure}
    \centering
    \includegraphics[width=\columnwidth]{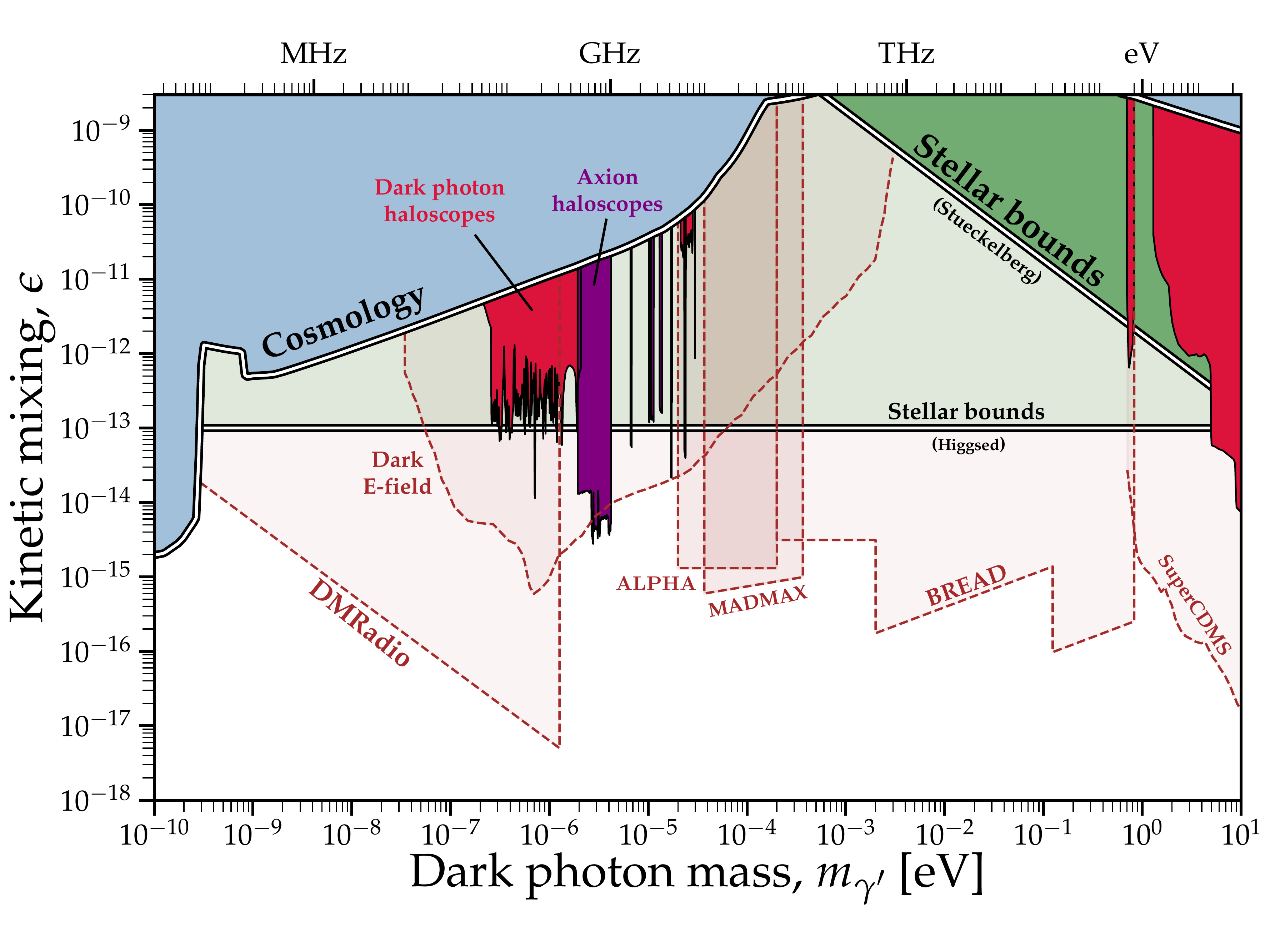}
    \caption{
    Current bounds and projections on the strength of the kinetic mixing parameter $\epsilon$ between the SM photon and a dark photon, defined in Eq.~(\ref{qa}). 
    In blue we show the bounds on dark photons as dark matter, set using cosmological data~\cite{Arias:2012az,McDermott:2019lch,Witte:2020rvb,Caputo:2020rnx,Caputo:2020bdy}, whereas the bounds from stellar cooling arguments are shown in green~\cite{Vinyoles:2015aba}. The stellar bounds in particular are sensitive to the mass generation mechanism for the dark photon---the standard Stueckelberg case is shown in dark green, and the alternative Higgs mechanism case is shown in lighter green~\cite{An:2013yua,An:2020bxd}. All bounds with the exception of these stellar bounds assume that the dark photon comprises all of the dark matter. In red we show experimental limits set by dedicated dark-photon searches, namely: SHUKET~\cite{Brun:2019kak}, WISPDMX~\cite{Nguyen:2019xuh}, SQuAD~\cite{Dixit2020single}, Dark E-field Radio~\cite{Godfrey:2021tvs}, LAMPOST~\cite{Chiles:2021gxk}, MuDHI~\cite{Manenti:2021whp}, FUNK~\cite{Andrianavalomahefa:2020ucg}, and three Tokyo-based dish antennae~\cite{Suzuki:2015sza,Knirck:2018ojz,Tomita:2020usq}. In purple we show reinterpreted bounds from axion haloscopes, namely ADMX~\cite{Asztalos:2001jk,Asztalos:2009yp,Du:2018uak,Boutan:2018uoc,Braine:2019fqb}, HAYSTAC~\cite{Zhong:2018rsr,backes2021quantum}, and CAPP~\cite{Lee:2020cfj,Jeong:2020cwz,CAPP:2020utb}. These have been recasted using the procedure outlined in Ref.~\cite{Caputo2021dark} that accounts for the dark photon's unknown polarization state. The transparent regions bounded by dashed lines are all projected limits for proposed experiments: DM-Radio~\cite{DMRADIO}, Dark E-field Radio~\cite{Godfrey:2021tvs}, ALPHA~\cite{Lawson:2019brd,Gelmini:2020kcu}, MADMAX~\cite{TheMADMAXWorkingGroup:2016hpc}, BREAD~\cite{BREAD:2021tpx}, LAMPOST~\cite{Baryakhtar:2018doz}, and SuperCDMS~\cite{Bloch:2016sjj}. Plots and limit data files available at Ref.~\cite{AxionLimits}.}
    \label{fig:DarkPhotonBounds}
\end{figure}

\textbf{Vector DM current limits and future perspectives} -- Fig.~\ref{fig:vector_B-L_plot} shows the parameter space for vector dark matter with the $B-L$ gauge coupling, in terms of the gauge coupling parameter $g_{B-L}$ as a function of the vector boson mass (see Section~\ref{Sec:vector_DM_theory}). 
Current constraints from the MICROSCOPE EP test~\cite{Berge:2017ovy,MICROSCOPE:2017}, E\"{o}t-Wash torsion balance EP tests~\cite{Wagner_2012_torsion}, E\"{o}t-Wash torsion balance DM search~\cite{shaw2022torsion}, LIGO's first observing run~\cite{Michimura:2021DM-LIFO_propagation}, and combined constraints from the third observing run of LIGO and VIRGO~\cite{LIGO:2021DM_vector} are shown.
The projected sensitivities of the MAGIS-100 experiment and its upgrade with improved LMT optics \& high-flux atom cloud sources~\cite{magis-100_2021} and optomechanical membranes~\cite{manley2021searching} are also shown. 

Figure~\ref{fig:DarkPhotonBounds} shows current bounds and projections on the strength of the kinetic mixing parameter $\epsilon$ between the SM photon and dark photon defined in Eq.~(\ref{qa}). 
Plots and limit data files are available at Ref.~\cite{AxionLimits}. 
The figure show the bounds on dark photons as dark matter, set using cosmological data~\cite{Arias:2012az,McDermott:2019lch,Witte:2020rvb,Caputo:2020rnx,Caputo:2020bdy}. The bounds from stellar cooling arguments~\cite{Vinyoles:2015aba} are sensitive to the mass generation mechanism for the dark photon---the standard Stueckelberg case is shown in dark green, and the alternative Higgs mechanism case is shown in lighter green~\cite{An:2013yua,An:2020bxd}. All bounds with the exception of these stellar bounds assume that the dark photon comprises all of the dark matter.  Fig.~\ref{fig:DarkPhotonBounds} shows  experimental limits set by dedicated dark-photon searches: SHUKET~\cite{Brun:2019kak}, WISPDMX~\cite{Nguyen:2019xuh}, SQuAD~\cite{Dixit2020single}, Dark E-field Radio~\cite{Godfrey:2021tvs}, LAMPOST~\cite{Chiles:2021gxk}, MuDHI~\cite{Manenti:2021whp}, FUNK~\cite{Andrianavalomahefa:2020ucg}, and three Tokyo-based dish antennae~\cite{Suzuki:2015sza,Knirck:2018ojz,Tomita:2020usq}. Reinterpreted bounds from axion haloscopes are also shown: ADMX~\cite{Asztalos:2001jk,Asztalos:2009yp,Du:2018uak,Boutan:2018uoc,Braine:2019fqb}, HAYSTAC~\cite{Zhong:2018rsr,backes2021quantum}, and CAPP~\cite{Lee:2020cfj,Jeong:2020cwz,CAPP:2020utb}. These have been recasted using the procedure outlined in Ref.~\cite{Caputo2021dark} that accounts for the dark photon's unknown polarization state. The transparent regions bounded by dashed lines are all projected limits for proposed experiments: DM-Radio~\cite{DMRADIO}, Dark E-field Radio~\cite{Godfrey:2021tvs}, ALPHA~\cite{Lawson:2019brd,Gelmini:2020kcu}, MADMAX~\cite{TheMADMAXWorkingGroup:2016hpc}, BREAD~\cite{BREAD:2021tpx}, LAMPOST~\cite{Baryakhtar:2018doz}, and SuperCDMS~\cite{Bloch:2016sjj}.

%

\providecommand{\href}[2]{#2}\begingroup\raggedright\endgroup

\end{document}